\documentclass[twocolumn]{svjour3} 

\usepackage[english]{babel}
\usepackage[utf8]{inputenc}
\usepackage[T1]{fontenc}
\usepackage{ae}
\usepackage{aecompl}

\usepackage{amsmath}
\usepackage{amssymb}
\usepackage{commath}
\usepackage{mathtools}
\usepackage{mathrsfs}
\usepackage[bbgreekl]{mathbbol}
\usepackage{dsfont}
\usepackage{bbold}
\usepackage{booktabs}
\usepackage{bm}

\usepackage[numbers]{natbib}
\usepackage{color}
\usepackage{graphicx}
\usepackage{subfigure}
\usepackage{booktabs}
\usepackage{bm}
\usepackage{cases}

\everymath={\displaystyle}

\newlength{\arrow}
\newcommand*{\myrightarrow}[1]{\xrightarrow{\mathmakebox[\arrow]{#1}}}

\newcommand{\R}{\ensuremath{\mathbb{R}}}           
\newcommand{\N}{\ensuremath{\mathbb{N}}}           

\newcommand{\FS}{\textnormal{\tiny FS}}
\newcommand{\BR}{\textnormal{\tiny BR}}
\newcommand{\SE}{\textnormal{\tiny SE}}
\newcommand{\KE}{\textnormal{\tiny KE}}
\newcommand{\G}{\textnormal{\tiny G}}

\newcommand{\domain}[1]{\mathcal{#1}}

\newcommand{\referential}[1]{\mathcal{#1}}

\newcommand{\functional}[1]{\mathcal{#1}}

\newcommand{\EffOp}{\functional{E}}

\newcommand{\PowerOp}{\functional{P}}

\newcommand{\KEOp}{\functional{T}}

\newcommand{\SEOp}{\functional{V}}

\newcommand{\ExtWorkOp}{\functional{W}}

\newcommand{\DampPotOp}{\functional{D}}

\newcommand{\MassOp}{\functional{M}}

\newcommand{\DampOp}{\functional{C}}

\newcommand{\StiffOp}{\functional{K}}

\newcommand{\ForceOp}{\functional{F}}

\newcommand{\BoundOp}{\mathscr{B}}

\newcommand{\varOp}[1]{\delta #1}

\newcommand{\wfunc}[1]{\psi_{#1}}

\newcommand{\intfunc}[3]{\functional{#1}_{#2}^{#3}}

\renewcommand{\vec}[1]{\bm{#1}}

\newcommand{\mat}[1]{\left[ #1 \right]}

\newcommand{\transp}[1]{#1^{\scriptscriptstyle T}}

\newcommand{\integraldef}[3]{\int_{#1}^{#2} #3 dx}

\newcommand{\tensorprod}[2]{#1\otimes #2}

\newcommand{\tensor}[1]{\mathds{\bm{#1}}}

\newcommand{\dotprod}[2]{#1 \bm{\cdot} #2}

\newcommand{\ddotprod}[2]{#1 \, \textbf{:} \, #2}

\newcommand{\trace}[1]{\texttt{tr} \left(#1\right)}

\newcommand{\sgn}[1]{\texttt{sgn} \left(#1\right)}

\newcommand{\indfunc}[1]{\ensuremath{\mathds{1}_{#1} }}

\newcommand{\dimless}[1]{#1^{\ast}}

\renewcommand{\SS}{\Theta}

\newcommand{\SSpt}{\theta}

\newcommand{\SA}{\mathbb{\Sigma}}

\newcommand{\PM}{\mathbb{P}}

\newcommand{\randvar}[1]{\mathbb{#1}}

\newcommand{\randproc}[1]{\mathbb{#1}}

\newcommand{\pdf}[1]{p_{\tiny{#1}}}

\newcommand{\supp}[1]{\ensuremath{\texttt{Supp}\,{#1} }}

\newcommand{\expval}[1]{\ensuremath{\mathbb{E}\left[  #1 \right] }}

\newcommand{\mean}[1]{m_{#1}}

\newcommand{\var}[1]{\sigma^2_{#1}}

\newcommand{\stddev}[1]{\sigma_{#1}}

\newcommand{\shannon}[1]{\textsc{S} \left( #1 \right)}

\journalname{Computational Mechanics}

\begin{document}

\title{Computational modeling of the nonlinear stochastic dynamics of horizontal drillstrings}
\titlerunning{}

\author{Americo Cunha~Jr \and Christian Soize \and Rubens Sampaio}
\authorrunning{A. Cunha~Jr \and C. Soize \and R. Sampaio}

\institute{A. Cunha~Jr (corresponding author) \at
              Universidade do Estado do Rio de Janeiro, Instituto de Matem\'{a}tica e Estat\'{i}stica,
              Departamento de Matem\'{a}tica Aplicada,
			  Rua S\~{a}o Francisco Xavier, 524, Pav. Jo\~{a}o Lyra, Bl. B, Sala 6032,
			  Rio de Janeiro, 20550-900, Brasil\\
              \email{americo@ime.uerj.br}
              \and
              A. Cunha~Jr \and C. Soize \at
			  Universit\'{e} Paris-Est, Laboratoire Mod\'{e}lisation et Simulation Multi Echelle,
			  MSME UMR 8208 CNRS, 5, Boulevard Descartes 77454, Marne-la-Vall\'{e}e, France\\
              \email{christian.soize@univ-paris-est.fr}
              \and
              A. Cunha~Jr \and R. Sampaio \at
              PUC--Rio, Departamento de Engenharia Mec\^{a}nica,
			  Rua M. de S\~{a}o Vicente, 225 - Rio de Janeiro, 22451-900, Brasil\\
              \email{rsampaio@puc-rio.br}
}

\date{Received: date / Accepted: date}

\maketitle

\begin{abstract}
This work intends to analyze the nonlinear stochastic dynamics of drillstrings
in horizontal configuration. For this purpose, it considers a beam theory,
with effects of rotatory inertia and shear deformation, which is capable
of reproducing the large displacements that the beam undergoes. The friction
and shock effects, due to beam/borehole wall transversal impacts,
as well as the force and torque induced by bit-rock interaction,
are also considered in the model. Uncertainties of bit-rock interaction
model are taken into account using a parametric probabilistic approach.
Numerical simulations have shown that the mechanical system of interest
has a very rich nonlinear stochastic dynamics, which generate phenomena
such as bit-bounce, stick-slip, and transverse impacts.
A study aiming to maximize the drilling process efficiency, varying
drillstring velocities of translation and rotation is presented. Also,
the work presents the definition and solution of two optimizations
problems, one deterministic and one robust, where the objective is to 
maximize drillstring rate of penetration into the soil respecting its 
structural limits.

\end{abstract}

\section{Introduction}
\label{introd}

\begin{figure}[h]
	\centering
	\includegraphics[scale=0.4]{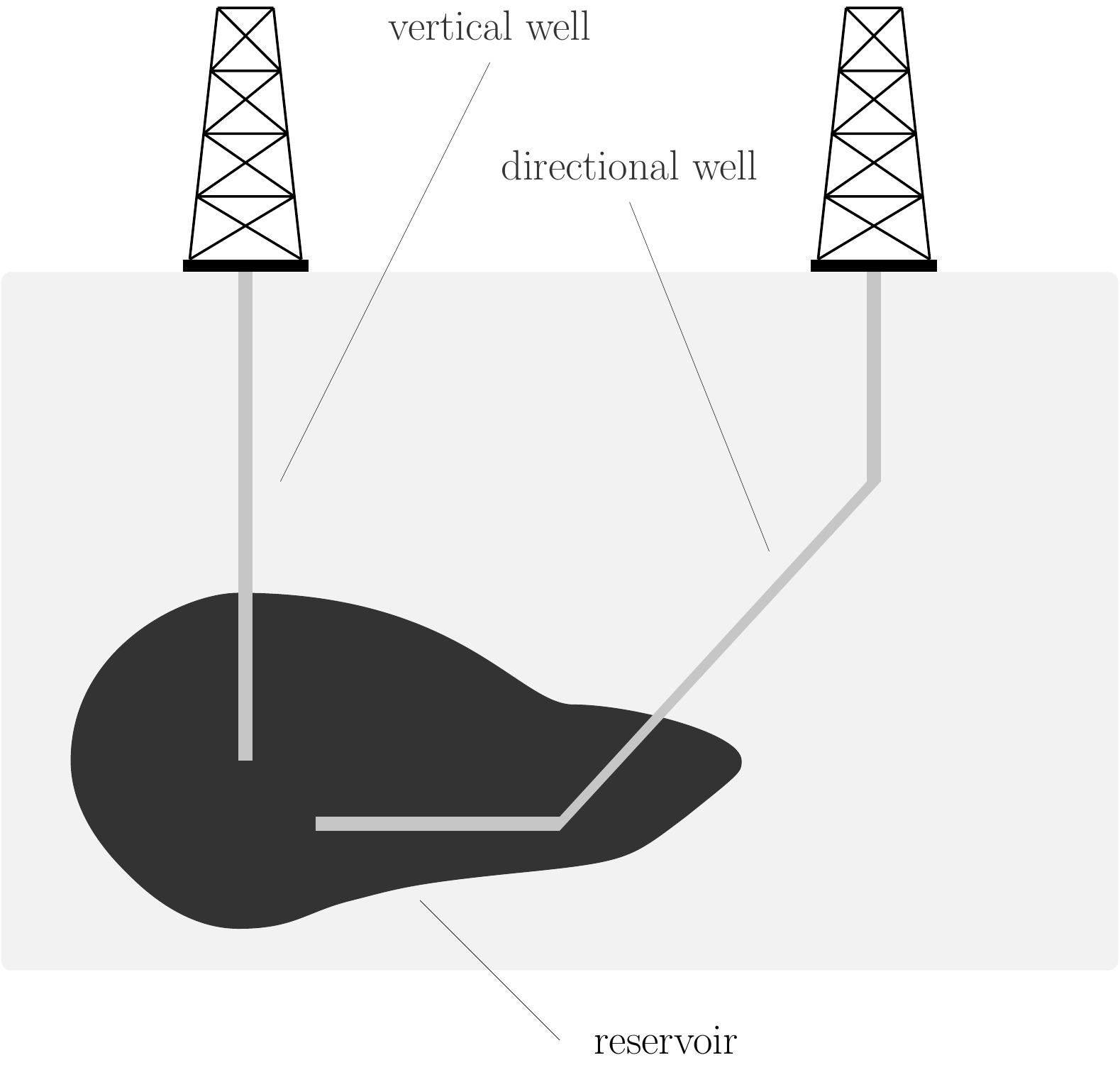}
	\caption{Schematic representation of two exploratory wells.
				The left well configuration is vertical while the right
				one is directional.}
	\label{oil_wells}
\end{figure}

High energy demands of 21st century make that fossil fuels, like oil and shale gas,
still have great importance in the energy matrix of several countries. Prospection
of these fossil fuels demands the creation of exploratory wells.
Traditionally, an exploratory well configuration is vertical, but directional or even horizontal
configurations, where the boreholes are drilled following a non-vertical way, are also possible
\cite{willoughby2005}. An illustration of the different types of configurations which an exploratory
well can take is presented in Figure~\ref{oil_wells}.

\begin{figure}[h]
	\centering
	\includegraphics[scale=0.4]{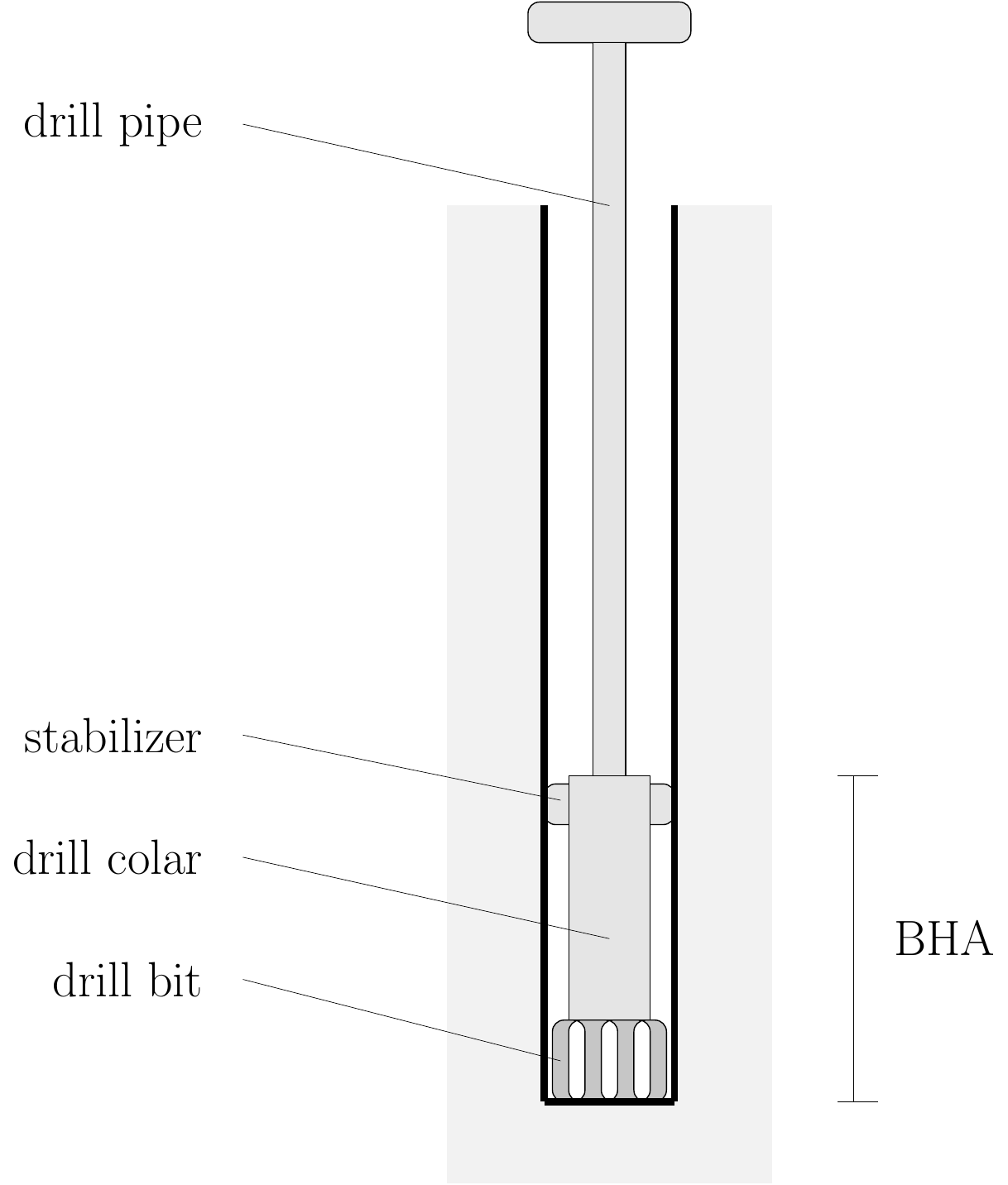}
	\caption{Schematic representation of a typical drillstring.}
	\label{drillstring_fig}
\end{figure}

The equipment used to drill the soil until the reservoir level
is called \emph{drillstring}.
This device is a long column, composed of a sequence of connected
drill-pipes and auxiliary equipment.
Furthermore, within the column flows drilling mud,
which is used to cool the drilling system and to remove
drilling cuttings from the borehole.
The bottom part of this column is called
\emph{bottom hole assembly} (BHA)
and consists of a pipe of greater thickness, named \emph{drill-colar},
and a tool used to stick the rock, the \emph{drill-bit}
\cite{freudenrich2001}.
The BHA presents stabilizers throughout
its length, whose function is to maintain structural integrity
of the borehole before cementation process.
A schematic representation of a typical vertical
drillstring and its components is presented in
Figure~\ref{drillstring_fig}, but a column in horizontal
configuration essentially has the same structure.

Since the axial dimension of a drillstring is orders of magnitude larger
than the characteristic dimension of its cross section area, the column
is a long flexible structure with a very complex flexural dynamics.
Furthermore, during drilling process, the drillstring is also subjected
to other two mechanisms of vibration (longitudinal and torsional),
which interact nonlinearly with the flexural mechanism, resulting in a
further complicated dynamics \cite{spanos2003p85}. The coupling between
these three mechanisms of vibration, which imposes severe complications on
the drillstring dynamics modeling, comes from the action of several
agents, such as: structure self weight (for a vertical column);
tensile and compressive loads due to the \emph{weight on bit} (WOB)
and soil reaction force; dry friction and impacts with borehole wall;
bit-rock interaction forces; internal flow pressure; forces induced
by internal flow instabilities; etc \cite{spanos2003p85}.

The dynamics of a drillstring is not a new subject in the
technical/scientific literature. Works on this subject, covering
experimental analysis, numerical and/or analytical modeling,
can be seen since the 1960s. Most of the numerical works
developed between 1960s and 1990s, have used lumped parameters
approach to gain insight about drillstrings dynamical behavior.
On the other hand, the analytical works focused on simple distributed
parameters models. Little has been done using finite element-based
approaches until the beginning of 1990s. A comprehensive literature survey
of the research work produced until 2000 can be found in
\cite{chevallier2000} and \cite{spanos2003p85}.

In recent studies, lumped parameters approach have been used,
for example, to seek configurations which reduce stick-slip
occurrence during drillstring operation \cite{silveira2009p012056};
to identify suitable values for drilling system operational parameters
\cite{liu2013p61};
to analyze the coupling between axial and torsional vibrations and its stability
\cite{franca2004p789,divenyi2012p1017,nandakumar2013p2575,depouhon2014p2019}.
On the other hand, approaches based on distributed parameters
models have been used to: investigate drillstring failure mechanisms
\cite{jansen1993}; better understand transversal impacts between
the column and borehole wall \cite{trindade2005p1015};
study the effects induced by the nonlinear coupling between longitudinal
and torsional dynamics of the drillstring \cite{sampaio2007p497};
describe the column dynamic behavior taking into account
the coupling between the three mechanisms of vibration
\cite{ritto2009p865,ritto2010}; investigate the chaotic regime which
drillstring transverse vibration mechanism is subjected
\cite{chatjigeorgiou2013p5592}.

Despite the fact that directional drilling has been used in practical engineering
for a few decades, and most of the exploratory wells drilled today be directional
in configuration, all the works mentioned above model vertical drillstrings only.
To the best of authors' knowledge, there are very few papers in open literature
which models drillstring in directional configurations
\cite{sahebkar2011p743,hu20121p66,ritto2013p145}. All of these works
use a distributed parameters approach, but while \cite{sahebkar2011p743,ritto2013p145}
only address the drillstring longitudinal dynamics, \cite{hu20121p66}
uses generalized Euler-Bernoulli beam theory to describe the drillstring
three-dimensional dynamics in a sloped directional well. In \cite{sahebkar2011p743},
the authors study a sloped configuration for the borehole and uses a perturbation technique
to discretize the model equations. Conversely, model equations are discretized by
finite element in \cite{ritto2013p145}.

In addition to the difficulties inherent to the nonlinear dynamics,
drillstrings are subjected to randomness on their
geometrical dimensions, physical properties, external forcing, etc.
The lack of knowledge on these parameters, known as
\emph{system-parameter uncertainty}, is a source of inaccuracies
in drillstring modeling, which may, in an extreme case,
completely compromise the model predictability
\cite{schueller1997p197,schueller2007p235}.
Furthermore, during the modeling process, hypotheses about the
drillstring physical behavior are made. These considerations may be or
not be in agreement with reality and should introduce additional 
inaccuracies in the model, known as \emph{model uncertainty} induced by modeling errors
\cite{soize2012,soize2013p2379}.
This source of uncertainty is essentially due to the use of simplified computational model 
for describing the phenomenon of interest and, usually, is the largest source
of inaccuracy in computational model responses \cite{soize2012,soize2013p2379}.

Therefore, for a better understanding of the drillstring dynamics,
these uncertainties must be modeled and quantified. In terms of
quantifying these uncertainties for vertical drillstrings, the reader
can see \cite{spanos2002p512}, where external forces
are modeled as random objects and the method of statistical
linearization is used along with the Monte Carlo (MC) method
to treat the stochastic equations of the model. Other works
in this line include: \cite{ritto2009p865,ritto2010}, where
system-parameter and model uncertainties are considered using a nonparametric
probabilistic approach; and \cite{ritto2010p250,ritto2012p111},
which use a standard parametric probabilistic approach to take into
account the uncertainties of the system parameters.
Regarding the works that model directional configurations,
only \cite{ritto2013p145} considers the uncertainties,
which, in this case, are related to the friction effects due to
drillstring/borehole wall contact.

From what is observed above, considering only the theoretical point of view,
the study of drillstring nonlinear dynamics is already a rich subject.
However, a good understanding of its dynamics also has 
significant importance in applications. For instance, it is fundamental to predict the fatigue
life of the structure \cite{macdonald2007p1641} and the drill-bit wear
\cite{zhu2014p64}; to analyze the structural integrity of an exploratory well
\cite{davies2014inpress}; to optimize the drill-bit \emph{rate of penetration} (ROP)
of into the soil \cite{ritto2010p415}, and the last is essential
to reduce cost of production of an exploratory well.

In this sense, this study aims to analyze the three-dimensional nonlinear dynamics
of a drillstring in horizontal configuration, taking into account the
system-parameter uncertainties. Through this study it is expected to gain a better understanding
of drillstring physics and, thus, to improve drilling process efficiency,
and maximize the column ROP accordingly. All results presented here were developed
in the thesis of \cite{cunhajr2015}.

The rest of this work is organized as follows. Section~\ref{phys_model}
presents the mechanical system of interest in this work, its parametrization
and modeling from the physical point of view. Mathematical formulation 
of initial/boundary value problem that describes the mechanical system 
behavior, as well as the conservative dynamics associated, is shown
in section~\ref{math_model}. The computational modeling of the problem, 
which involves model equations discretization, reduction of the discretized 
dynamics, the algorithms for numerical integration and solution of nonlinear 
system of algebraic equations, can be seen in section~\ref{comp_model}.
The probabilistic modeling of uncertainties is presented in 
section~\ref{prob_mod_data_uncert}.
Results of numerical simulations are presented and discussed in
section~\ref{num_results}. Finally, in section~\ref{concl_remarks},
the main conclusions are emphasized, and some paths to future works
are pointed out.


\section{Physical model for the problem}
\label{phys_model}

\subsection{Definition of the mechanical system}
\label{def_mech_sys}

\begin{figure}[h]
	\centering
	\includegraphics[scale=0.6]{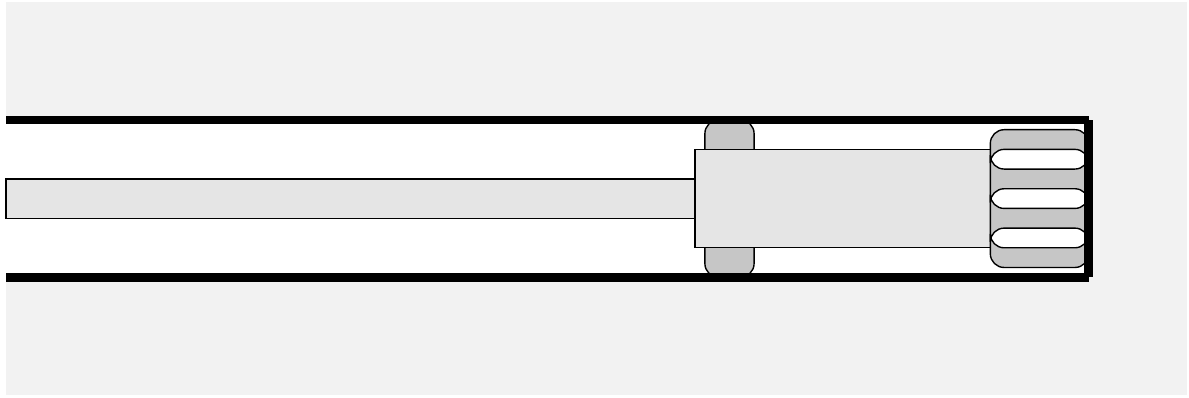}
	\caption{Schematic representation of the mechanical system under analysis.}
	\label{horiz_drillstring_fig}
\end{figure}

The mechanical system of interest in this work, which is schematically
represented  in Figure~\ref{horiz_drillstring_fig}, consists of a horizontal
rigid pipe, perpendicular to gravity, which contains in its interior a
deformable tube under rotation. This deformable tube is subjected to 
three-dimensional displacements, which induces 
longitudinal, lateral, and torsional vibrations of the structure. 
These mechanisms of vibration are able to generate slips and shocks 
in random areas of the rigid tube. Also, the contact between drill-bit, 
at the right extreme of the tube, with soil generates nonlinear forces 
and torques on drillstring right extreme, which may completely block 
the structure advance over the well.

\subsection{Nonlinear dynamical system parameterization}

For purposes of modeling, the only part of column considered
is the BHA. So, any variation of diameter along the column
is ignored. In this way, the bottom part of the deformable tube described,
in section~\ref{def_mech_sys}, is modeled as a rotating beam in horizontal
configuration, whose transverse displacement ($y$ and $z$) at both ends
is blocked, as well as transverse rotations on the left extreme.
It looks like the left end of the system is a stabilizer and the right one a support.
This beam is free to rotate around the $x$ axis, and to move longitudinally.
The rigid pipe is treated as a stationary cylindrical rigid wall in
horizontal configuration.

As the beam is confined within the borehole, it is reasonable to assume that it
undergoes small rotations in transverse directions.
On the other hand, large displacements are observed in $x$, $y$, and $z$, as well as large
rotations around the $x$-axis. Therefore, the analysis that follows uses a beam theory
which assumes large rotation in $x$, large displacements in the three spatial directions,
and small deformations \cite{bonet2008}.

Seeking not to make the mathematical model excessively complex, this work will
not model the fluid flow inside the beam, nor the dissipation effects induced by
the flow on the system dynamics.

Due to the horizontal configuration, the beam is under action
of the gravitational field, which induces an acceleration $g$.
The beam is made of an isotropic material with mass density $\rho$,
elastic modulus $E$, and Poisson's ratio $\nu$.
It has length $L$ and annular cross section,
with internal radius $R_{int}$ and external radius $R_{ext}$.

\begin{figure}[h]
	\centering
	\includegraphics[scale=0.41]{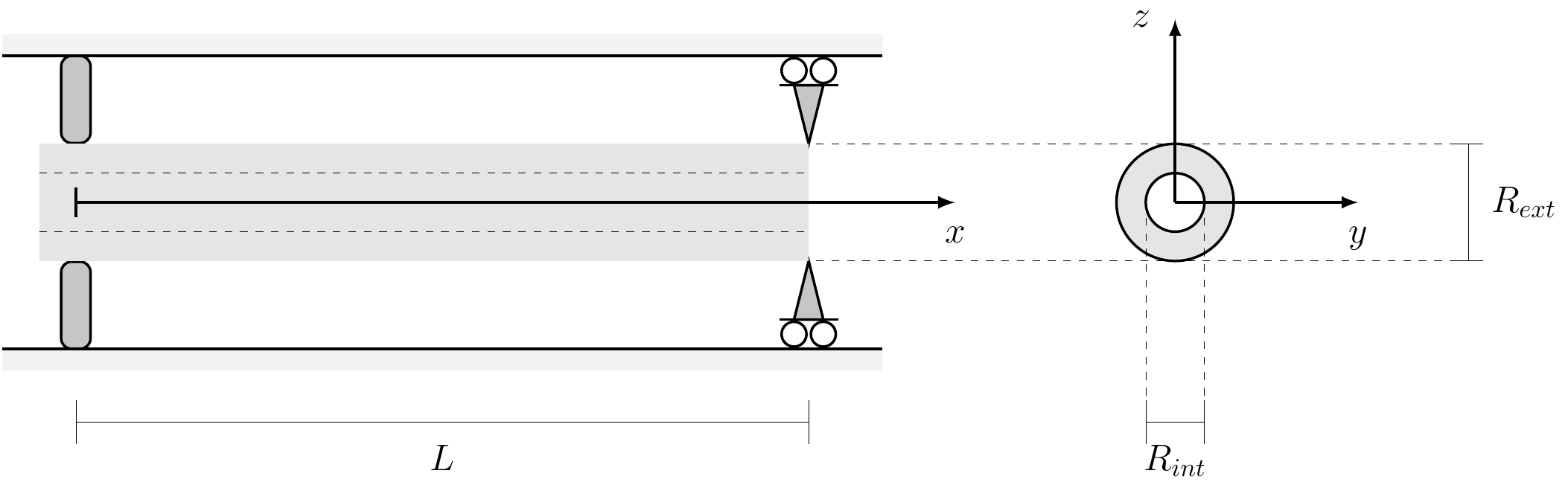}
	\caption{Schematic representation of the beam geometry
				used to model the deformable tube under rotation, and the
				inertial system of coordinates used.}
	\label{drillstring_geometry_fig}
\end{figure}

An illustration of beam geometric model
is presented in Figure~\ref{drillstring_geometry_fig}. It is important to note that
this model also ignores the mass of drill-bit and its geometric shape.

Using the cartesian coordinate system $(x,y,z)$, defined by orthonormal
basis $\{\vec{e}_{x},\vec{e}_{y},\vec{e}_{z}\}$, fixed in the inertial
frame of reference $\referential{R}$, and shown in Figure~\ref{drillstring_geometry_fig},
one can describe the beam undeformed configuration by

\begin{equation}
		\domain{B}_b =
		\left\lbrace (x,y,z) \in \R^3
		~~ \big\vert ~~
		0 \leq x \leq L,~
		(y,z) \in \domain{S}_b
		\right\rbrace,
		\label{def_domain}
\end{equation}

\noindent
where the beam cross section undeformed configuration is described by

\begin{equation}
		\domain{S}_b =
		\left\lbrace (y,z) \in \R^2
		~~ \big\vert ~~
		R^{2}_{int} \leq y^2+z^2 \leq R^{2}_{ext}
		\right\rbrace.
		\label{def_sec_transv}
\end{equation}

%
%
%
%

With the undeformed cross section configuration characterized,
one can define and compute the cross-sectional area,

\begin{equation}
		A = \iint_{\domain{S}_b} \,dy\,dz = \pi \left( R_{ext}^2 - R_{int}^2\right),
		\label{def_area}
\end{equation}

\noindent
the second moment of area around the $y$ axis

\begin{equation}
		I_{yy} = \iint_{\domain{S}_b} z^2 \,dy\,dz = I_{4},
		\label{def_Iyy}
\end{equation}

\noindent
the second moment of area around the $z$ axis

\begin{equation}
		I_{zz} = \iint_{\domain{S}_b} y^2 \,dy\,dz = I_{4},
		\label{def_Izz}
\end{equation}

\noindent
the polar moment of area

\begin{equation}
		I_{xx} = \iint_{\domain{S}_b} \left(y^2 + z^2 \right) \,dy\,dz = 2 \, I_{4},
		\label{def_Ixx}
\end{equation}

\noindent
the fourth moment of area around the $z$ axis

\begin{equation}
		I_{zzzz} = \iint_{\domain{S}_b} y^4 \,dy\,dz = 3 \, I_{6},
		\label{def_Izzzz}
\end{equation}

\noindent
and the fourth product of area

\begin{equation}
		I_{yyzz} = \iint_{\domain{S}_b} y^2 z^2 \,dy\,dz = I_{6},
		\label{def_Iyyzz}
\end{equation}

\noindent
where

\begin{equation}
		I_4 = \frac{\pi}{4} \left( R_{ext}^4 - R_{int}^4\right),
		\label{def_I4}
\end{equation}

\noindent
and

\begin{equation}
		I_6 = \frac{\pi}{24} \left( R_{ext}^6 - R_{int}^6 \right).
		\label{def_I6}
\end{equation}

%

In this work other three coordinate systems (all of then with the same origin
as the $(x,y,z)$ coordinate system) are also used,
each one fixed in a non-inertial frame of reference $\referential{R}_n$,
where $n=1,2,3$, and defined by an orthonormal basis of vectors of the form
$\{\vec{e}_{x_n},\vec{e}_{y_n},\vec{e}_{z_n}\}$.

These systems of coordinates are related by a sequence of elementary rotations,
such as follows

\begin{equation}
\scalebox{0.82}{
$
\begin{array}{ccccccc}
	\referential{R} & \myrightarrow{\displaystyle\theta_x} & \referential{R}_1 & \myrightarrow{\displaystyle\theta_y} & \referential{R}_2 & \myrightarrow{\displaystyle\theta_z} & \referential{R}_3, \\
		(x,y,z)         &                                         & (x_1,y_1,z_1)    &                                         &   (x_2,y_2,z_2)   &                                         & (x_3,y_3,z_3)\\
\end{array}
$
}
\label{syst_coord_relation}
\end{equation}

\noindent
where $\theta_x$ is the rotation around the $x$ axis,
$\theta_y$ is the rotation around the $y$ axis, and
$\theta_z$ is the rotation around the $z$ axis.
These rotations follow the right hand rule.


Thus, with respect to the non-inertial frame of reference,  the instantaneous angular velocity
of rotating beam is written as

\begin{equation}
		\vec{\omega} =
		\dot{\theta}_x \vec{e}_x +
		\dot{\theta}_y \vec{e}_{y_1} +
		\dot{\theta}_z \vec{e}_{z_2},
		\label{angular_velocity1}
\end{equation}

\noindent
where $\dot{\theta}_x$, $\dot{\theta}_y$, and $\dot{\theta}_z$
denote the rate of rotation around the $x$, $y$, and $z$ directions,
respectively. From now on, upper dot will be used as an abbreviation 
for time derivative, i.e., $\dot{\square} \equiv \pd{\square}{t}$.

Referencing vector $\vec{\omega}$ to the inertial frame of reference,
and using the assumption of small rotations in transversal directions,
one obtains

\begin{eqnarray}
		\vec{\omega} =
		\left( \begin{array}{c}
			 \dot{\theta}_x + \dot{\theta}_z \theta_y\\
			 \dot{\theta}_y \cos{\theta_x} - \dot{\theta}_z \sin{\theta_x} \\
			 \dot{\theta}_y \sin{\theta_x} + \dot{\theta}_z \cos{\theta_x} \\
		\end{array} \right).
		\label{angular_velocity3}
\end{eqnarray}

Regarding the kinematic hypothesis adopted for beam theory,
it is assumed that the three-dimensional displacement of a beam point,
occupying position $(x,y,z)$ at instant of time $t$, can be written as

\begin{eqnarray}
		\label{kinematic_hyp}
		u_{x} (x,y,z,t) & = & u - y \theta_z + z \theta_y, \\ \nonumber
		u_{y} (x,y,z,t) & = & v + y \left( \cos{\theta_x} -1\right) - z \sin{\theta_x}, \\
		u_{z} (x,y,z,t) & = & w + z \left( \cos{\theta_x} -1\right) + y \sin{\theta_x}, \nonumber
\end{eqnarray}

\noindent
where $u_{x}$, $u_{y}$, and $u_{z}$ respectively denote the displacement of a beam
point in $x$, $y$, and $z$ directions. Moreover, $u$, $v$, and $w$ are the
displacements of a beam neutral fiber point in $x$, $y$, and $z$ directions, respectively.

Finally, it is possible to define the vectors

\begin{eqnarray}
		\vec{r} =
		\left( \begin{array}{c}
			 x \\
			 y \\
			 z
		\end{array} \right),
		~~
		\vec{\mbox{v}} =
		\left( \begin{array}{c}
			 \dot{u} \\
			 \dot{v} \\
			 \dot{w}
		\end{array} \right),
		~~ \mbox{and} ~~
		\vec{\dot{\theta}} =
		\left( \begin{array}{c}
			 \dot{\theta}_x \\
			 \dot{\theta}_y \\
			 \dot{\theta}_z
		\end{array} \right),
\end{eqnarray}

\noindent
which, respectively, represent the position of a beam point,
the velocity of a neutral fiber point, and the rate of rotation
of a neutral fiber point.

Note that the kinematic hypothesis of Eq.(\ref{kinematic_hyp})  is expressed 
in terms of three spatial coordinates ($x$, $y$, and $z$) and six field variables 
($u$, $v$, $w$, $\theta_x$, $\theta_y$, and $\theta_z$).

It is important to mention that, as the analysis assumed small rotations in $y$ and $z$, 
this kinematic hypothesis presents nonlinearities, expressed by 
trigonometric functions, only in $\theta_{x}$.
Besides that, since a beam theory is employed, the field variables 
in Eq.(\ref{kinematic_hyp}) depend only on the spatial coordinate $x$ and time $t$.
Therefore, although the kinematic hypothesis of Eq.(\ref{kinematic_hyp}) is 
three-dimensional, the mathematical model used to describe the beam nonlinear dynamics 
is one-dimensional.


\subsection{Modeling of friction and shock effects}

\begin{figure}[h]
	\centering
	\includegraphics[scale=0.41]{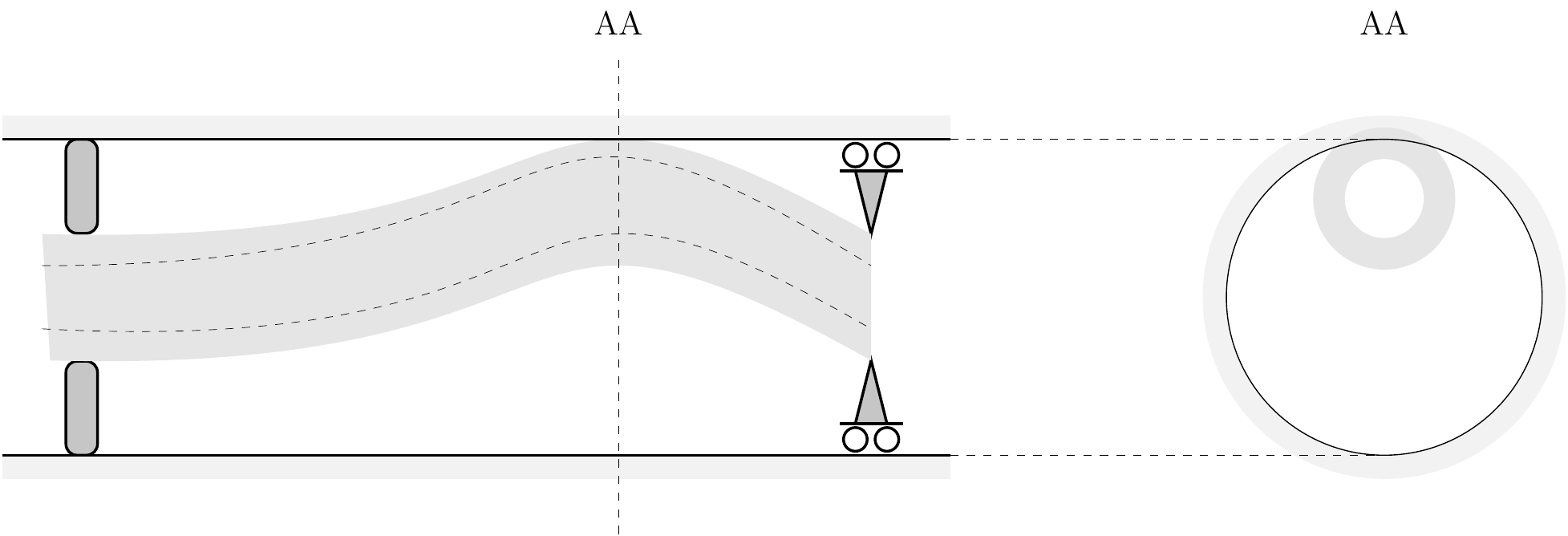}
	\caption{Schematic representation of the situation where there is a mechanical contact
	between a drillstring and the borehole wall.}
	\label{beam_wall_contact_fig}
\end{figure}

When a drillstring deforms laterally, there may occur a mechanical contact
between the rotating beam and the borehole wall, such as illustrated in
Figure~\ref{beam_wall_contact_fig}. This mechanical contact, which generally
take place via a strong impact, gives rise to friction and shock effects
\cite{gilardi2002p1213,wriggers2006,litewka2002p26}.

\begin{figure}[h]
	\centering
	\includegraphics[scale=0.45]{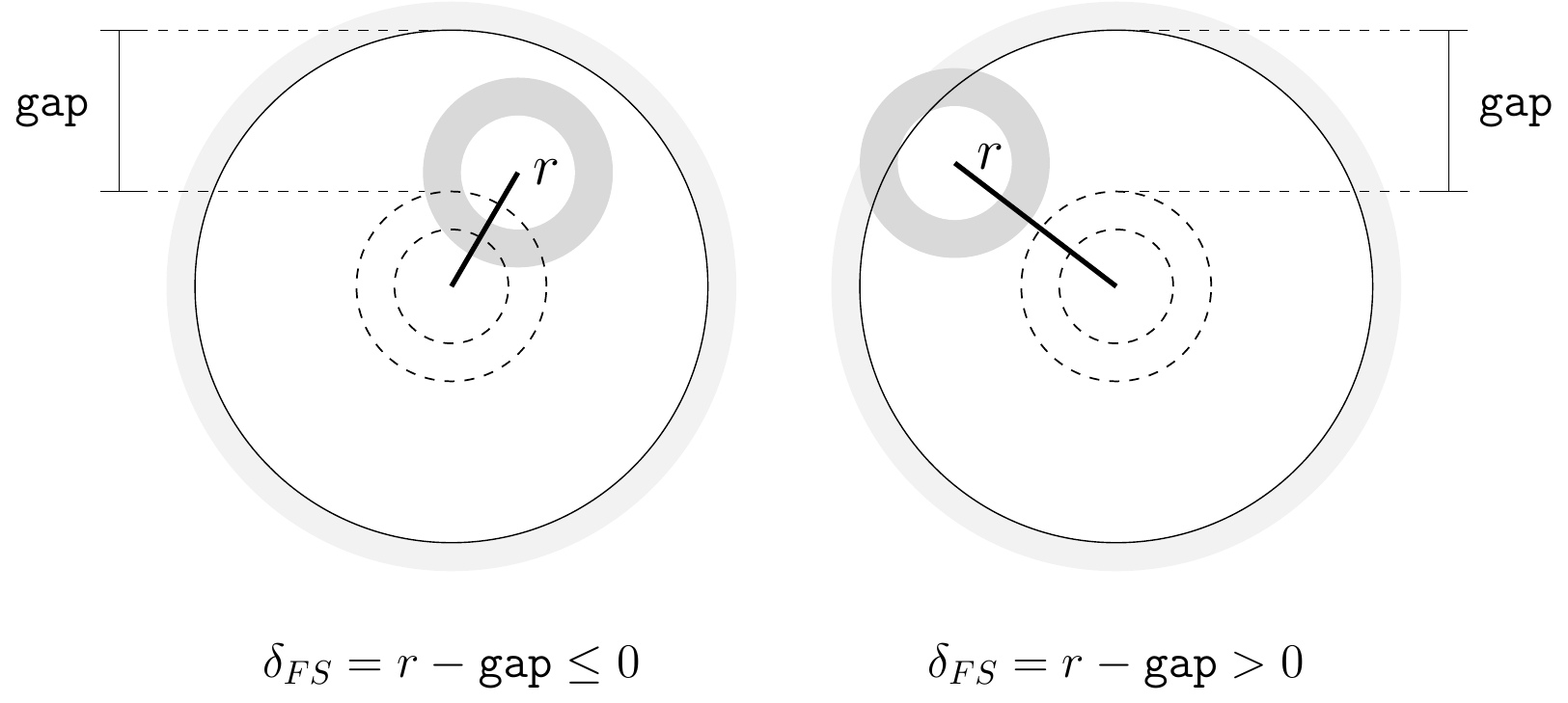}
	\caption{Illustration of the indentation parameter in a situation
	 without impact (left) or with impact (right).}
	\label{indentation_fig}
\end{figure}

The modeling of friction and shock phenomena is made in terms of a
geometric parameter dubbed \emph{indentation}, which is defined as

\begin{equation}
		\delta_{\FS} = r - \texttt{gap},
		\label{indentation_def}
\end{equation}

\noindent
where $r = \sqrt{v^2 + w^2}$ is the neutral fiber lateral displacement,
and $\texttt{gap}$ denotes the spacing between the undeformed beam
and the borehole wall. One has that $\delta_{\FS} > 0$ in case of an impact,
or $\delta_{\FS} \leq 0$ otherwise, as can be seen in Figure~\ref{indentation_fig}.
Note that the indentation corresponds to a measure of penetration in the wall
of a beam cross section \cite{gilardi2002p1213}.

When an impact occurs, a normal force of the form

\begin{equation}
		F_{\FS}^{n} = - k_{\FS_1} \, \delta_{\FS}
					  - k_{\FS_2} \, \delta_{\FS}^{3}
					  - c_{\FS} \, |\delta|^{3} \dot{\delta}_{\FS},
		\label{force_shock_eq1}
\end{equation}

\noindent
where $k_{\FS_{1}}$, $k_{\FS_{2}}$ and $c_{\FS}$ are constants of the shock model,
begins to act on the beam cross section.
In this nonlinear shock model, proposed by Hunt and Crossley \cite{hunt1975p440},
the first (linear spring) and the second (nonlinear spring) terms describe the elastic
deformation during an impact, while the third term (nonlinear damper) takes into
account the loss of energy during an impact.

Once the column is rotating and moving axially, an impact also induces
a frictional force in axial direction, $F_{\FS}^{a}$, and a torsional friction
torque, $T_{\FS}$. Both are modeled by Coulomb friction law  \cite{cull1999p2103},
so that the force is given by

\begin{equation}
		F_{\FS}^{a} = - \mu_{\FS} \, F_{\FS}^{n} \, \sgn{\dot{u}},
		\label{force_friction_eq1}
\end{equation}

\noindent
where the torque is described by

\begin{equation}
		T_{\FS} = - \mu_{\FS} \, F_{\FS}^{n} \, R_{bh} \, \sgn{\dot{\theta}_{x}},
		\label{torque_friction_eq1}
\end{equation}

\noindent
being $\mu_{\FS}$ the friction coefficient, $\sgn{\cdot}$ the sign function, and
the radius of the borehole is $R_{bh} = R_{ext} + \texttt{gap}$.

In order to find all points of contact between the beam and the borehole wall,
it is necessary to discover all values of $x$ where $\delta_{\FS} > 0$. This is
usually done by solving an optimization problem with constraints
\cite{wriggers1987p215,wriggers2004}.

Although the strategy of detection based on the optimization problem
is robust in terms of accuracy, it is extremely complex in terms of
implementation and computational cost. For this reason, this work
uses an approach that introduces the forces of Eqs.(\ref{force_shock_eq1})
and (\ref{force_friction_eq1}), and the torque of Eq.(\ref{torque_friction_eq1}),
as efforts concentrated on the nodes of finite element mesh,
defined in the section~\ref{discretization_dynamics}.
This procedure sacrifices accuracy, but simplifies the
friction and shock model implementation.


\subsection{Modeling of bit-rock interaction effects}

During the drilling process, in response to drillstring 
rotational advance, a force and a torque of reaction begin to act 
on the drill-bit, giving rise to the so-called bit-rock interaction effects
\cite{detournay2008p1347,franca2010p043101}.

In this work, the model proposed by \cite{ritto2013p145} is considered
to describe the bit-rock interaction force

\begin{figure}[h!]
	\centering
	\includegraphics[scale=0.5]{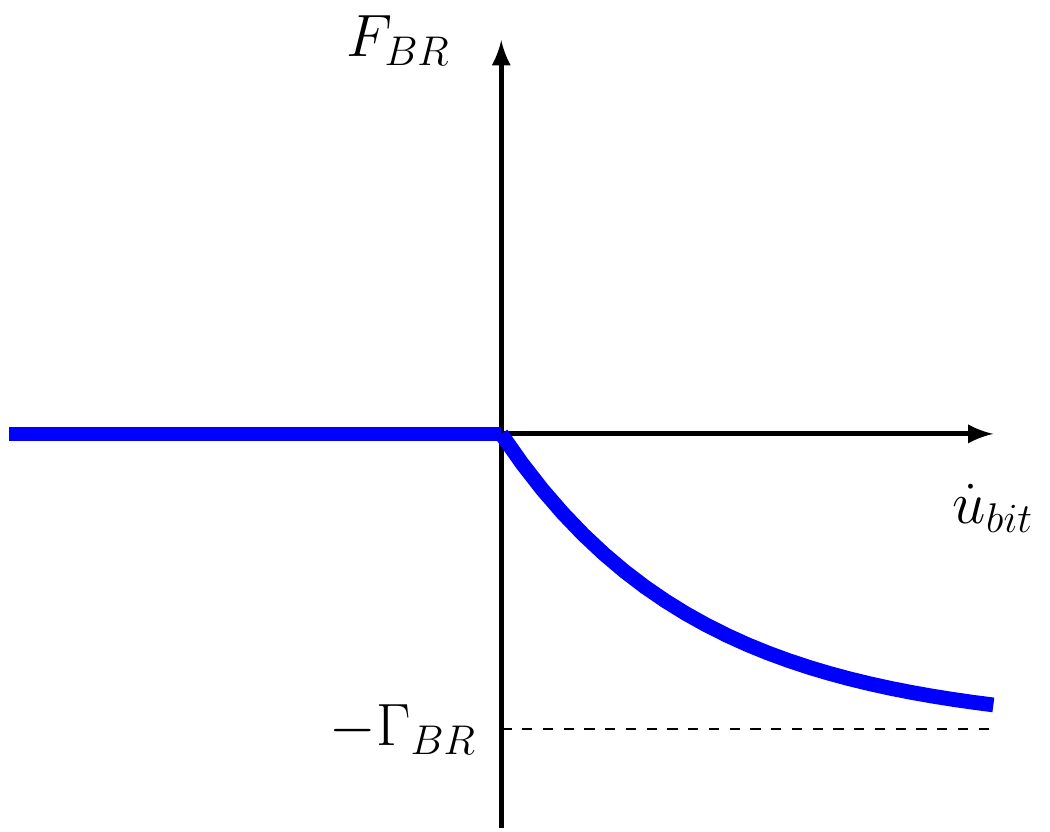}
	\caption{Illustration of the function used to describe the reaction force
	on drill-bit, due to bit-rock interaction effects.}
	\label{bit-rock_force_fig}
\end{figure}

\begin{numcases}{F_{\BR} =}
    \label{br_force_eq}
			\Gamma_{\BR} \left( e^{- \alpha_{\BR} \, \dot{u}_{bit}} -1 \right)  & for $\dot{u}_{bit} > 0$,\\ \nonumber
			                                                                                                    0 & for $\dot{u}_{bit} \leq 0$,
\end{numcases}

\noindent
where $\Gamma_{\BR}$ is the bit-rock limit force, $\alpha_{\BR}$ is the
rate of change of bit-rock force, and $\dot{u}_{bit} = \dot{u}(L,\cdot)$.
The graph of $F_{\BR}$ is illustrated in Figure~\ref{bit-rock_force_fig}.

\begin{figure}[h!]
	\centering
	\includegraphics[scale=0.5]{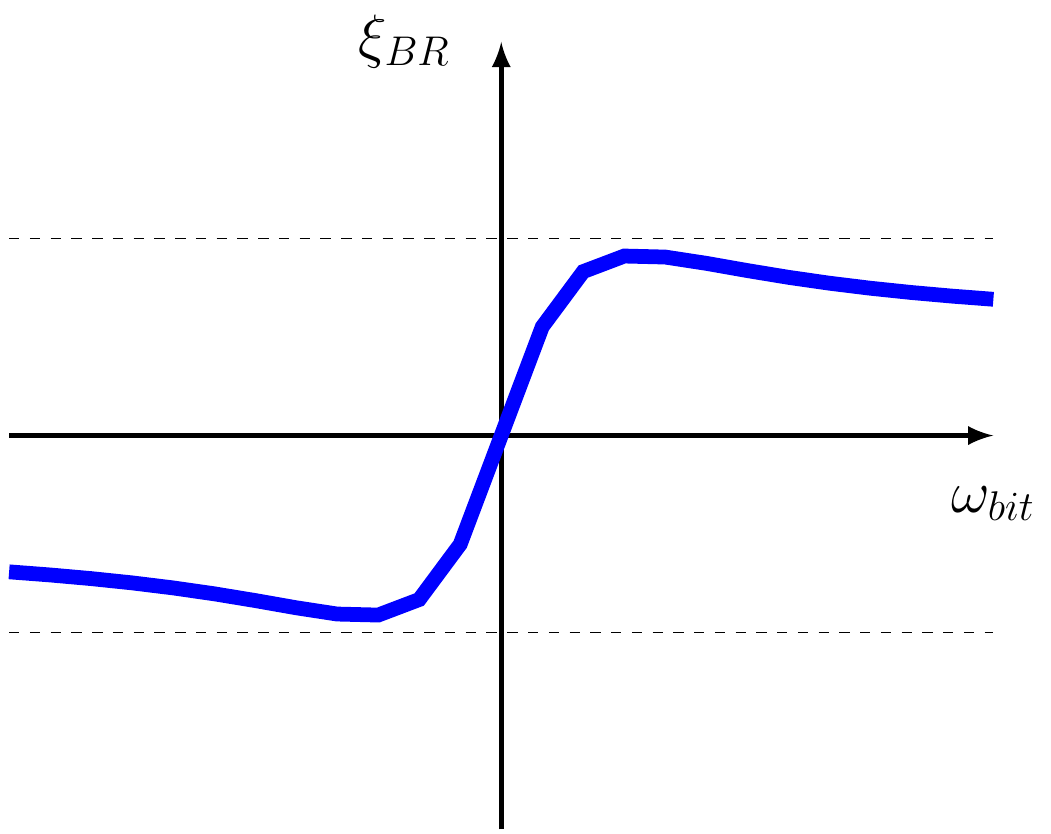}
	\caption{Illustration of the smooth function used to regularize the reaction torque
	on drill-bit, due to bit-rock interaction effects.}
	\label{bit-rock_torque_reg_fig}
\end{figure}

Also, for the bit-rock interaction torque, it is adopted the 
regularized Coulomb model used by \cite{khulief2007p540}, 
which is expressed as

\begin{equation}
		\label{br_torque_eq}
		T_{\BR} = - \mu_{\BR} \, F_{\BR} \, R_{bh} \, \xi_{\BR} \left( \omega_{bit} \right),
\end{equation}

\noindent
where $\omega_{bit} = \dot{\theta}_{x}(L,\cdot)$, the 
bit-rock friction coefficient is $\mu_{\BR}$, and

\begin{equation}
		\xi_{\BR} \left( \omega_{bit} \right) = \tanh\left( \omega_{bit} \right) +
																	  \frac{2 \, \omega_{bit}}{1 + \omega_{bit}^2},
		\label{reg_func_eq}
\end{equation}

\noindent
is a regularization function. The graph of $\xi_{\BR}$
is illustrated in Figure~\ref{bit-rock_torque_reg_fig}.

\subsection{Kinetic energy}

The kinetic energy of rotating beam is given by

\begin{eqnarray}
		\label{kin_energy_eq1}
		\KEOp & = &
		\frac{1}{2} \iiint_{\domain{B}_b}
		\rho \, \dotprod{\vec{\mbox{v}} }{ \vec{\mbox{v}}} \,dx\,dy\,dz ~ + ~\\ \nonumber
		 &  &
		\frac{1}{2} \iiint_{\domain{B}_b}
		\rho \, \dotprod{\vec{\omega}}{
		\left( \dotprod{\vec{r}}{\vec{r}} \, \tensor{I} - \tensorprod{\vec{r}}{\vec{r}} \right)
		\vec{\omega}} \,dx\,dy\,dz,
\end{eqnarray}

\noindent
where first triple integral corresponds to beam translational kinetic energy,
and the second one is associated to beam rotational kinetic energy.
In this equation, $\tensor{I}$ denotes the identity tensor,
the symbol $\dotprod{}{}$ represents the standard inner product between
two Euclidean vectors, and the symbol $\tensorprod{}{}$ is used to designate
the tensor product.


Developing  the vector operations indicated in Eq.(\ref{kin_energy_eq1}),
using (\ref{def_domain}) and (\ref{def_sec_transv}) to define the limits of
integration, using the definitions of $A$, $I_{yy}$, $I_{zz}$, and $I_{xx}$,
and making the other calculations one can show that Eq.(\ref{kin_energy_eq1}) 
is equivalent to

\begin{eqnarray}
		\label{kin_energy_eq2}
		\KEOp & = &
		\frac{1}{2} \integraldef{x=0}{L}{\rho \, A \left( \dot{u}^2 +  \dot{v}^2 +  \dot{w}^2 \right)} ~ + ~ \\ \nonumber
		&  &
		\frac{1}{2} \integraldef{x=0}{L}{2 \, \rho \, I_4 \left(\dot{\theta}_x + \dot{\theta}_z \theta_y \right)^2 } ~ + ~ \\ \nonumber
		&  &
		\frac{1}{2} \integraldef{x=0}{L}{ \rho \, I_4 \left(\dot{\theta}_y \cos{\theta_x} - \dot{\theta}_z \sin{\theta_x} \right)^2 } ~+~\\ \nonumber
		&  &
		\frac{1}{2} \integraldef{x=0}{L}{\rho \, I_4  \left(\dot{\theta}_y \sin{\theta_x} + \dot{\theta}_z \cos{\theta_x}  \right)^2 }.
\end{eqnarray}


\subsection{Strain energy}

The analysis of the beam assumes that it is subjected to
large displacements, and small deformations. In this way, its strain
energy is given by

\begin{equation}
		\SEOp =
		\frac{1}{2} \iiint_{\domain{B}_b}
		\ddotprod{ \tensor{\epsilon} }{ \tensor{\sigma} } \,dx\,dy\,dz,
		\label{strain_energy_eq1}
\end{equation}

\noindent
where $\tensor{\epsilon}$ denotes the Green-Lagrangian strain tensor,
$\tensor{\sigma}$ is the second Piola-Kirchhoff stress tensor, and the
symbol $\ddotprod{}{}$ represents the double inner product between
two tensors.

It is further considered that the beam is made of an isotropic material,
such that stress and strain are related by the following constitutive equation
(Hooke's law)

\begin{equation}
		\tensor{\sigma} =
		2 \, G \, \tensor{\epsilon} +
		\lambda \, \trace{\tensor{\epsilon}} \, \tensor{I},
		\label{hooks_law_3d}
\end{equation}

\noindent
where $\trace{\cdot}$ represents the trace operator,
$G$ is material shear modulus, and $\lambda$ is used to designate
the material first Lam\'{e} parameter. In terms of the elastic modulus
$E$ and the Poisson's ratio $\nu$, these elastic parameters
can be written as

\begin{equation}
		G =
		\frac{E}{2 \, (1 + \nu)},
		\qquad \mbox{and} \qquad
		\lambda =
		\frac{E \, \nu}{(1 + \nu)(1 - 2 \, \nu)}.
		\label{elastic_param_def}
\end{equation}

According to the beam theory used in this work, there is no tension in
any cross section of the beam that is perpendicular to $x$ axis, i.e.,
$\sigma_{yy} = 0$, $\sigma_{zz} = 0$, $\sigma_{yz} = 0$, and $\sigma_{zy} = 0$.
When this hypothesis is combined with the three-dimensional Hooke's law,
represented by Eq.(\ref{hooks_law_3d}), one can conclude that
$\sigma_{xx} = E \, \epsilon_{xx}$, $\sigma_{xy} = 2 \, G \, \epsilon_{xy}$, and
$\sigma_{xz} = 2 \, G \, \epsilon_{xz}$, which is an one-dimensional
version of Hooke's law.

Combining this one-dimensional Hooke's law with the 
stress tensor symmetry, one can express the double contraction
between strain and stress tensors, within the integral in Eq.(\ref{strain_energy_eq1}),
as a quadratic form

\begin{equation}
		\ddotprod{\tensor{\epsilon}}{ \tensor{\sigma}} =
		       E \, \epsilon_{xx}^2 +
		4 \, G \, \epsilon_{xy}^2 +
		4 \, G \, \epsilon_{xz}^2,
\end{equation}

\noindent
which is modified, by the introduction of shearing
factor $\kappa_{s}$, as

\begin{equation}
		\ddotprod{\tensor{\epsilon}}{ \tensor{\sigma}} =
		                            E \, \epsilon_{xx}^2 +
		4 \, \kappa_{s} \, G \, \epsilon_{xy}^2 +
		4 \, \kappa_{s} \, G \, \epsilon_{xz}^2.
		\label{strain_strees_quad_form}
\end{equation}

\noindent
This modification aims to take into account the effect of
shear deformation in the beam cross section area, which is neglected
when one uses the one-dimensional Hooke's law.

Hence, after replace Eq.(\ref{strain_strees_quad_form})
in Eq.(\ref{strain_energy_eq1}), one finally obtains

\begin{equation}
		\SEOp =
		\frac{1}{2} \iiint_{\domain{B}_b}
		\left(						    E \, \epsilon_{xx}^2 +
				4 \, \kappa_{s} \, G \, \epsilon_{xy}^2 +
				4 \, \kappa_{s} \, G \, \epsilon_{xz}^2 \right) \,dx\,dy\,dz.
		\label{strain_energy_eq2}
\end{equation}

As the analysis is using large displacements, one has

\begin{eqnarray}
		\label{epsilon_xx_eq}
		\epsilon_{xx} & = &
		\frac{1}{2} \left( \dpd{u_{x}}{x} + \dpd{u_{x}}{x} \right) ~ + ~\\ \nonumber
		& &
		\frac{1}{2} \left( \dpd{u_{x}}{x} \dpd{u_{x}}{x} +
								  \dpd{u_{y}}{x} \dpd{u_{y}}{x} +
								  \dpd{u_{z}}{x} \dpd{u_{z}}{x}
						 \right),
\end{eqnarray}

\begin{eqnarray}
		\label{epsilon_xy_eq}
		\epsilon_{xy} & = &
		\frac{1}{2} \left( \dpd{u_{y}}{x} + \dpd{u_{x}}{y} \right) ~ + ~ \\ \nonumber
		& &
		\frac{1}{2} \left( \dpd{u_{x}}{x} \dpd{u_{x}}{y} +
								  \dpd{u_{y}}{x} \dpd{u_{y}}{y} +
								  \dpd{u_{z}}{x} \dpd{u_{z}}{y}
						 \right),
\end{eqnarray}

\noindent
and

\begin{eqnarray}
		\label{epsilon_xz_eq}
		\epsilon_{xz} & = &
		\frac{1}{2} \left( \dpd{u_{z}}{x} + \dpd{u_{x}}{z} \right) ~ + ~\\ \nonumber
		& &
		\frac{1}{2} \left( \dpd{u_{x}}{x} \dpd{u_{x}}{z} +
								  \dpd{u_{y}}{x} \dpd{u_{y}}{z} +
								  \dpd{u_{z}}{x} \dpd{u_{z}}{z}
						 \right),
\end{eqnarray}

\noindent
where the quadratic terms on the right hand side of the above equations
are associated to beam model geometric nonlinearity.

Substituting the kinematic hypothesis of Eq.(\ref{kinematic_hyp}) in
Eqs.(\ref{epsilon_xx_eq}) to (\ref{epsilon_xz_eq}), and then calculating the partial
derivatives, one concludes that the deformations are respectively given by

\begin{eqnarray}
\resizebox{.97\hsize}{!}{$
  \begin{split}
		\epsilon_{xx} & = & u' - y \, \theta_{z}' + z \, \theta_{y}' +
										u' \left( z \, \theta_{y}' - y \, \theta_{z}'\right) - y \, z \, \theta_{y}' \, \theta_{z}' ~ + ~~~~~~~~ \\
		                      &     & \theta_{x}' \left( \left( y\,w' - z\,v' \right) \cos{\theta_{x}} - \left( y\,v' + z\,w' \right) \sin{\theta_{x}} \right) ~ + ~~~~~~\\
		                      &     & \frac{1}{2} \left( u'^{\,2} + v'^{\,2} + w'^{\,2} + y^2 \, \theta_{z}'^{\,2} +
		                                    z^2 \, \theta_{y}'^{\,2} + \left(y^2+z^2 \right) \theta_{x}'^{\,2}
		                      				\right),
  \end{split}$
  }
		\label{epsilon_xx_eq2}
\end{eqnarray}

\begin{eqnarray}
		\epsilon_{xy} & = & \frac{1}{2} \left( v' \, \cos{\theta_{x}} + w' \sin{\theta_{x}} - z \, \theta_{x}' \right) ~ + ~ \\ \nonumber
		                      &     & \frac{1}{2} \, \theta_{z} \left( y \, \theta_{z}' - z \theta_{y}' - u' - 1 \right),
		\label{epsilon_xy_eq2}
\end{eqnarray}

\noindent
and

\begin{eqnarray}
		\epsilon_{xz} & = & \frac{1}{2} \left( w' \, \cos{\theta_{x}} - v' \sin{\theta_{x}} + y \, \theta_{x}' \right) ~ + ~ \\ \nonumber
		                      &    & \frac{1}{2} \, \theta_{y} \left( - y \, \theta_{z}' + z \theta_{y}' + u' + 1 \right),
		\label{epsilon_xz_eq2}
\end{eqnarray}

\noindent
where prime is used as an abbreviation for space derivative, 
i.e., $\square' \equiv \pd{\square}{x}$.


\subsection{Energy dissipation function}

It is assumed that the beam loses energy through
a mechanism of viscous dissipation, with a (dimensionless)
damping constant $c$.
In this way, there is an energy dissipation function (per unit of length)
associated to the system, which is given by

\begin{eqnarray}
		\label{dissip_func_eq1}
		\DampPotOp & = &
		\frac{1}{2} \iint_{\domain{S}_b}
		c \, \rho \, \dotprod{\vec{\mbox{v}}}{ \vec{\mbox{v}}} \,dy\,dz ~ + ~ \\ \nonumber
		&  &
		\frac{1}{2} \iint_{\domain{S}_b}
		c \, \rho \, \dotprod{\vec{\dot{\theta}}}{
		\left( \dotprod{\vec{r}}{\vec{r}} \, \tensor{I} - \tensorprod{\vec{r}}{\vec{r}} \right)
		\vec{\dot{\theta}}} \,dy\,dz,
\end{eqnarray}

\noindent
where the first term is a dissipation potential due to the  translational movement,
and the second term represents a dissipation potential due to the movement of rotation.

Making a development almost similar to the one performed to obtain
Eq.(\ref{kin_energy_eq2}), it can be shown that

\begin{eqnarray}
		\label{dissip_func_eq2}
		\DampPotOp & = &
		\frac{1}{2} \, c \, \rho \, A \left( \dot{u}^2 +  \dot{v}^2 +  \dot{w}^2 \right) ~ + ~ \\ \nonumber
		&  &
		\frac{1}{2} \, c \, \rho \, I_4 \left( 2 \, \dot{\theta}^2_x +  \dot{\theta}^2_y +  \dot{\theta}^2_z \right).
\end{eqnarray}


\subsection{External forces work}

The work done by the external forces acting on the beam is given by

\begin{eqnarray}
		\label{ext_work_eq2}
		\ExtWorkOp =
		- \integraldef{x=0}{L}{\rho\, A \, g \, w\,}  +  \ExtWorkOp_{\FS} + \ExtWorkOp_{\BR},
\end{eqnarray}

\noindent
where the first term is due to gravity, the second one is associated to 
friction and shock effects, and the last term accounts 
the work done by the force/torque that comes from bit-rock interaction.

Note that, due to non-holonomic nature of the forces and torques
that comes from the effects of friction/shock, and bit-rock interaction,
it is not possible to write explicit formulas for $\ExtWorkOp_{\FS}$
and $\ExtWorkOp_{\FS}$ \cite{lanczos1986}.

However, it is known that the virtual work of $\ExtWorkOp_{\FS}$,
denoted by $\varOp{\ExtWorkOp_{\FS}}$, is written as

\begin{eqnarray}
		\label{ext_work_W_FS}
\resizebox{.98\hsize}{!}{$
  \begin{split}
		\varOp{\ExtWorkOp_{\FS}}  & = & \sum_{m=1}^{N_{nodes}}
													\left( F_{\FS}^{a} \, \varOp{u} +
													  F_{\FS}^{n} \left( v \, \varOp{v} +
								                    		  											w\, \varOp{w}
								                    		  									\right)/r +
												    		  T_{\FS} \, \varOp{\theta_{x}}
												    \right) \Big\vert_{x = x_{m}}
  \end{split}$
}
\end{eqnarray}

\noindent
where $x_{m}$ are the global coordinates of finite element nodes,
$N_{nodes}$ is the number of nodes in finite element mesh, and
$\varOp{u}$, $\varOp{v}$, $\varOp{w}$, and $\varOp{\theta_{x}}$
respectively denote the variations of $u$, $v$, $w$, and $\theta_{x}$.

On the other hand, the virtual work of $\ExtWorkOp_{\BR}$,
denoted by $\varOp{\ExtWorkOp_{\BR}}$, reads as

\begin{equation}
		\varOp{\ExtWorkOp_{\BR}}  = F_{\BR} \, \varOp{u} \Big\vert_{x = L} +
												       T_{\BR} \, \varOp{\theta_{x}} \Big\vert_{x = L}.
		\label{ext_work_W_BR}
\end{equation}


\section{Mathematical model for the problem}
\label{math_model}

\subsection{Equation of motion of the nonlinear dynamics}
\label{eq_motion_nonlinear_dyn}

A modified version of Hamilton's extended principle \cite{lanczos1986}
is employed to derive the equations which describe the mechanical system
nonlinear dynamics, so that the first variation is expressed as

\begin{eqnarray}
\resizebox{.9\hsize}{!}{$
  \begin{split}
		\int_{t=t_0}^{t_f} \left( \varOp{\KEOp} - \varOp{\SEOp}	+ \varOp{\ExtWorkOp}  \right) dt
		~ - ~
		\int_{t=t_0}^{t_f}
		\integraldef{x=0}{L}{ \dotprod{\varOp{\vec{U}}}{\dpd{\DampPotOp}{\vec{\dot{U}}}} \,} \, dt = 0,
  \end{split}$
  }
		\label{least_action_var_eq3}
\end{eqnarray}

\noindent
where the first term corresponds to dynamics conservative part,
and the second one is associated to energy dissipation. Also, $\vec{U}$ is a
vector field which lumps the field variables, the initial and final instants of
observation are respectively denoted by $t_0$ and $t_f$, and
the symbol $\varOp{}$ represents the variation operator \cite{sagan1992}.

The development of Eq.(\ref{least_action_var_eq3}) results in the following
weak equation of motion

\begin{eqnarray}
		\label{weak_eq}
\resizebox{.9\hsize}{!}{$
  \begin{split}
		\MassOp \left( \vec{\psi}, \vec{\ddot{U}} \right) +
		\DampOp \left( \vec{\psi}, \vec{\dot{U}} \right) +
		\StiffOp \left( \vec{\psi}, \vec{U} \right)
		= \ForceOp \left( \vec{\psi}, \vec{U}, \vec{\dot{U}}, \vec{\ddot{U}} \right),
  \end{split}$
}
\end{eqnarray}

\noindent
valid for any $\vec{\psi}$ chosen in a ``suitable" space of weight functions,
where the field variables and their corresponding weight functions are represented
by the vector fields $\vec{U} = \left(u, v, w, \theta_{x}, \theta_{y}, \theta_{z} \right)$, and
$\vec{\psi} = \left(\wfunc{u}, \wfunc{v}, \wfunc{w}, \wfunc{\theta_{x}}, \wfunc{\theta_{y}}, \wfunc{\theta_{z}} \right)$.

\pagebreak
Furthermore,

\begin{eqnarray}
		\label{mass_op_beam}
		\MassOp \left( \vec{\psi}, \vec{\ddot{U}} \right) & = &
		\integraldef{x=0}{L}{\rho \, A \left( \wfunc{u} \, \ddot{u} +
															 	\wfunc{v} \, \ddot{v} +
															 	\wfunc{w} \, \ddot{w}
												\right)} ~ + ~\\ \nonumber
		&  &
		\integraldef{x=0}{L}{\rho \, I_4 \left( 2 \, \wfunc{\theta_{x}} \, \ddot{\theta}_{x} +
																  		  \wfunc{\theta_{y}} \, \ddot{\theta}_{y} +
																  		  \wfunc{\theta_{z}} \, \ddot{\theta}_{z}
														\right)},
\end{eqnarray}

\noindent
represents the mass operator,

\begin{eqnarray}
		\label{damp_op_beam}
		\DampOp \left( \vec{\psi}, \vec{\dot{U}} \right) & = &
		\integraldef{x=0}{L}{c \, \rho \, A \left(  \wfunc{u} \, \dot{u} +
															 			\wfunc{v} \, \dot{v} +
															 			\wfunc{w} \, \dot{w}
												\right)} ~ + ~\\ \nonumber
		& &
		\integraldef{x=0}{L}{c \, \rho \, I_4 \left( 2 \, \wfunc{\theta_{x}} \, \dot{\theta}_{x} +
																		  	    \wfunc{\theta_{y}} \, \dot{\theta}_{y} +
																  				\wfunc{\theta_{z}} \, \dot{\theta}_{z}
													\right)},
\end{eqnarray}

\noindent
is the damping operator,

\begin{eqnarray}
		\label{stiff_op_beam}
		\StiffOp \left(\vec{\psi}, \vec{U}\right)
		& = &
		\integraldef{x=0}{L}{E \, A \, \wfunc{u}' \, u' \,} ~ + ~\\ \nonumber
		&  &
		\integraldef{x=0}{L}{E \, I_4 \left(\wfunc{\theta_{y}}' \, \theta'_{y} + \wfunc{\theta_{z}}' \, \theta'_{z}\right)} ~ + ~\\ \nonumber
		&  &
		\integraldef{x=0}{L}{2 \, \kappa_{s} \, G \, I_4 \, \wfunc{\theta_{x}}' \, \theta'_{x} \,} ~ + ~\\ \nonumber
		&  &
		\integraldef{x=0}{L}{\kappa_{s} \, G \, A  \left(\wfunc{\theta_{y}}+\wfunc{w}'\right)\left( \theta_{y} + w'\right)} ~+~ \\ \nonumber
		&  &
		\integraldef{x=0}{L}{\kappa_{s} \, G \, A \left(\wfunc{\theta_{z}}-\wfunc{v}'\right)\left( \theta_{z} - v'\right)},
\end{eqnarray}

\noindent
is the stiffness operator, and

\begin{eqnarray}
		\label{NLforce_op}
		\ForceOp \left( \vec{\psi}, \vec{U}, \vec{\dot{U}}, \vec{\ddot{U}} \right) & = &
		 \ForceOp_{\KE} \left( \vec{\psi}, \vec{U}, \vec{\dot{U}}, \vec{\ddot{U}} \right) + \\ \nonumber
		& &
		\ForceOp_{\SE} \left( \vec{\psi}, \vec{U} \right)  ~ + ~
		\ForceOp_{\FS} \left( \vec{\psi}, \vec{U} \right) ~ + ~ \\ \nonumber
		& &
		\ForceOp_{\BR} \left( \vec{\psi}, \vec{\dot{U}} \right) ~ + ~
		\ForceOp_{\G} \left( \vec{\psi} \right),
\end{eqnarray}

\noindent
is the force operator, which is divided into five parts.
A nonlinear force due to inertial effects

\begin{eqnarray}
		\label{NLforce_op_ke}
		\ForceOp_{\KE}
		& = &
		~ - ~ \integraldef{x=0}{L}{2 \, \rho \, I_4 \, \wfunc{\theta_{x}} \, \left(\theta_{y} \, \ddot{\theta}_{z} +
																														\dot{\theta}_{y} \, \dot{\theta}_{z}
																											   \right)} ~~~~~~~~~ \\ \nonumber
		& &
		~ + ~ \integraldef{x=0}{L}{2 \, \rho \, I_4 \, \wfunc{\theta_{y}} \left(\theta_{y} \, \dot{\theta}^{2}_{z} +
																													\dot{\theta}_{x} \, \dot{\theta}_{z}
																											\right)} \\ \nonumber
		& &
		~ - ~ \integraldef{x=0}{L}{2 \, \rho \, I_4 \, \wfunc{\theta_{z}} \left(\theta_{y} \, \ddot{\theta}_{x} +
																												   \theta^{2}_{y} \, \ddot{\theta}_{z}
																 								   			\right)} \\ \nonumber
		& &
		~ - ~ \integraldef{x=0}{L}{2 \, \rho \, I_4 \, \wfunc{\theta_{z}} \left(\dot{\theta}_{x} \, \dot{\theta}_{y} +
																 								   					2 \, \theta_{y} \, \dot{\theta}_{y} \, \dot{\theta}_{z}
																 								   			\right)},
\end{eqnarray}

\noindent
a nonlinear force due to geometric nonlinearity

\begin{eqnarray}
		\label{NLforce_op_se}
		\ForceOp_{\SE}
		& = &
		\integraldef{x=0}{L}{\left(\wfunc{\theta_{x}} \, \Gamma_{1} +
												\wfunc{\theta_{y}} \, \Gamma_{2} +
												\wfunc{\theta_{z}}  \, \Gamma_{3}
										\right)} ~ + ~\\ \nonumber
		& &
		\integraldef{x=0}{L}{ \left( \wfunc{u}' \, \Gamma_{4} +
												  \wfunc{v}' \,  \Gamma_{5} +
												  \wfunc{w}' \, \Gamma_{6}
										\right)} ~+~ \\ \nonumber
		& &
		\integraldef{x=0}{L}{ \left(  \wfunc{\theta_{x}}' \Gamma_{7} +
												   \wfunc{\theta_{y}}' \, \Gamma_{8} +
												   \wfunc{\theta_{z}}' \, \Gamma_{9}
										\right)},
\end{eqnarray}

\noindent
a nonlinear force due to friction and shock effects

\begin{eqnarray}
		\label{NLforce_op_shock}
\resizebox{.98\hsize}{!}{$
  \begin{split}
		\ForceOp_{\FS}
		& = &
		\sum_{m=1}^{N_{nodes}}
													 \left( F_{\FS}^{a} \, \wfunc{u} +
													          F_{\FS}^{n} \left( v \, \wfunc{v} +
								                    		  					  w \, \wfunc{w} \right) / r +
												    		  T_{\FS} \, \wfunc{\theta_{x}}
												    \right) \Big\vert_{x = x_{m}}
  \end{split}$
}
\end{eqnarray}

\noindent
a nonlinear force due to bit-rock interaction

\begin{equation}
		\ForceOp_{\BR}  =
								   F_{\BR} \, \wfunc{u} \Big\vert_{x = L} +
								   T_{\BR} \, \wfunc{\theta_{x}} \Big\vert_{x = L},
		\label{NLforce_op_bitrock}
\end{equation}

\noindent
and a linear force due to gravity

\begin{equation}
		\ForceOp_{\G} =
		- \integraldef{x=0}{L}{ \rho\, A \, g \, \wfunc{w} \,}.
		\label{NLforce_op_gravity}
\end{equation}

The nonlinear functions $\Gamma_n$, with $n=1, \cdots, 9$,
in Eq.(\ref{NLforce_op_se}) are very complex and,
for sake of space limitation, are not presented in this section.
But they can be seen in Appendix~\ref{nonlinear_gammas}.

The model presented above is an adaptation, 
for horizontal drillstrings, of the model proposed by 
\cite{ritto2009p865,ritto2010} to describe the nonlinear 
dynamics of vertical drillstrings. To be more precise, in
the reference problem gravity is parallel to drillstring main axis, 
while in this work, it is perpendicular to the structure primal direction. 
Therefore, the former problem primarily addresses the dynamics of a column, 
while the new problem deals with the dynamics of a beam. Also, the original
problem treated the nonlinear dynamics around a pre-stressed equilibrium 
configuration, while the new problem does not consider the dynamics around 
any particular configuration. It is worth mentioning that changes made in 
the modeling of friction and shock effects are significant. For instance,
a nonlinear shock model that also takes into account the dissipation of energy 
during an impact is introduced, in contrast to the reference work, that only 
consider the linear elastic deformation effects. In addition, the boundary conditions 
are different, as well as the bit-rock interaction model. On the other hand, 
for the sake of simplicity, the fluid structure interaction effects, considered in reference 
work are neglected in this study.


\subsection{Initial conditions}

With regard to mechanical system initial state, it is
assumed that the beam presents neither displacement nor rotations, i.e.,
$u(x,0) = 0$, $v(x,0) = 0$, $w(x,0) = 0$, $\theta_{x} (x,0) = 0$,
$\theta_{y} (x,0) = 0$, and $\theta_{z} (x,0) = 0$. These field variables, except for
$u$ and $\theta_{x}$, also have initial velocities and rate of rotations equal to zero, i.e.
$\dot{v}(x,0) = 0$, $\dot{w}(x,0) = 0$, $\dot{\theta}_{y} (x,0) = 0$, and
$\dot{\theta}_{z} (x,0) = 0$.

It is also assumed that, initially, the beam moves horizontally
with a constant axial velocity $V_0$, and rotates around the $x$ axis
with a constant angular velocity $\Omega$. Thereby, one has that
$\dot{u} (x,0) = V_0$, and \mbox{$\dot{\theta}_{x} (x,0) = \Omega$.}

Projecting the initial conditions in 
a ``suitable" space of weight functions, weak forms 
for them are obtained, respectively, given by

\begin{equation}
		\MassOp \left( \vec{\psi}, \vec{U} (0) \right)  =
		\MassOp \left( \vec{\psi}, \vec{U}_{0}\right) ,
		\label{weak_ic_eq1}
\end{equation}

\noindent
and

\begin{equation}
		\MassOp \left( \vec{\psi}, \vec{\dot{U}} (0) \right)  =
		\MassOp \left( \vec{\psi}, \vec{\dot{U}}_{0}\right),
		\label{weak_ic_eq2}
\end{equation}

\noindent
where $\vec{U}_{0} =\left(0,0,0,0,0,0\right)$ and
$\vec{\dot{U}}_{0} =\left(V_{0},0,0,\Omega,0,0\right)$.

In formal terms, the weak formulation of the initial--boundary
value problem that describes the mechanical system 
nonlinear dynamics consists in find a vector field $\vec{U}$,
``sufficiently regular", which satisfies the weak equation of
motion given by Eq.(\ref{weak_eq}) for all ``suitable" $\vec{\psi}$,
as well as the weak form of initial conditions, given by
Eqs.(\ref{weak_ic_eq1}), and (\ref{weak_ic_eq2}) \cite{hughes2000}.


\subsection{Associated linear conservative dynamics}
\label{conserv_dyn_assoc}

Consider the linear homogeneous equation given by

\begin{equation}
		\MassOp \left( \vec{\psi}, \vec{\ddot{U}} \right) +
		\StiffOp \left( \vec{\psi}, \vec{U} \right) = 0,
    \label{homog_weak_eq}
\end{equation}

\noindent
obtained from Eq.(\ref{weak_eq}) when one discards the damping,
and force operators, and which is valid for all $\vec{\psi}$
in the space of weight functions.

Suppose that Eq.(\ref{homog_weak_eq}) has a solution
of the form $\vec{U} = e^{i \omega t} \vec{\phi}$,
where $\omega$ is a natural frequency (in rad/s),
$\vec{\phi}$ is the associated normal mode,
and $i = \sqrt{-1}$ is the imaginary unit.
Replacing this expression of $\vec{U}$ in
Eq.(\ref{homog_weak_eq}) and using the linearity of
operators $\MassOp$, and $\StiffOp$, one gets

\begin{equation}
		\left(  -\omega^2 \MassOp \left( \vec{\psi}, \vec{\phi} \right) +
			   	    \StiffOp \left( \vec{\psi}, \vec{\phi} \right) \right)  e^{i \omega t} = 0,
\end{equation}

\noindent
which is equivalent to

\begin{equation}
		-\omega^2 \MassOp \left( \vec{\psi}, \vec{\phi} \right) +
		\StiffOp \left( \vec{\psi}, \vec{\phi} \right) = 0,
		\label{gen_eigen_eq1}
\end{equation}

\noindent
a generalized eigenvalue problem.

Since operator $\MassOp$ is positive-definite, and
operator $\StiffOp$ is positive semi-definite,
the generalized eigenvalue problem
above has a denumerable number of solutions. The solutions of
this eigenproblem have the form $(\omega_{n}^{2},\vec{\phi}_n)$,
where $\omega_{n}$ is the $n$-th natural frequency and $\vec{\phi}_n$ is the
$n$-th normal mode \cite{hagedorn2007}.

Also, the symmetry of operators $\MassOp$, and $\StiffOp$ 
implies the following orthogonality relations

\begin{equation}
		\MassOp \left( \vec{\phi}_n, \vec{\phi}_m \right) = \delta_{nm},
		~~ \mbox{and} ~~
		\StiffOp \left( \vec{\phi}_n, \vec{\phi}_m \right) = \omega_{n}^{2} \, \delta_{nm},
		\label{orthogonality_relations}
\end{equation}

\noindent
where $\delta_{nm}$ represents the Kronecker delta symbol.
See \cite{hagedorn2007} for more details.

The generalized eigenvalue problem of Eq.(\ref{gen_eigen_eq1}),
as well as the properties of (\ref{orthogonality_relations}),
will be useful to construct a reduced order model
for discretized dynamical system which approximates the solution 
of the weak initial--boundary value problem of
Eqs.(\ref{weak_eq}), (\ref{weak_ic_eq1}), and (\ref{weak_ic_eq2}).



\section{Computational model for the problem}
\label{comp_model}

\subsection{Discretization of the nonlinear dynamics}
\label{discretization_dynamics}

To proceed with the discretization of the weak initial--boundary 
value problem of Eqs.(\ref{weak_eq}), (\ref{weak_ic_eq1}), and (\ref{weak_ic_eq2}),
which describes the rotating beam nonlinear dynamics,
it is used the standard finite element method (FEM)
\cite{hughes2000}, where the spaces of basis and weight
functions are constructed by the same (finite dimensional) class of functions.

\begin{figure}[h]
	\centering
	\includegraphics[scale=0.6]{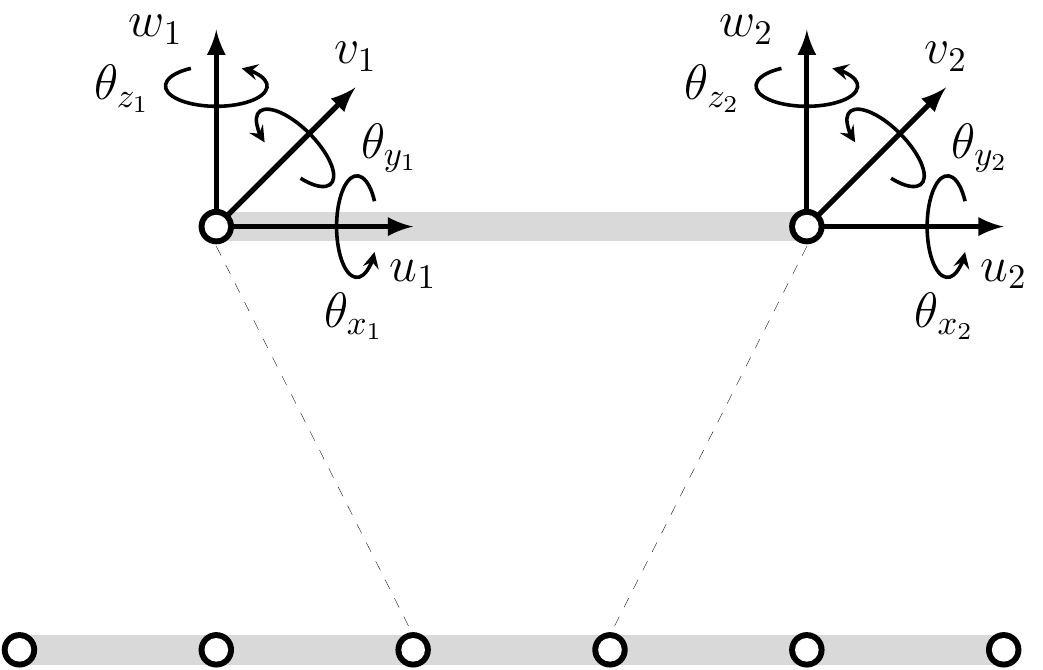}
	\caption{Illustration of FEM mesh/element used to discretize the beam geometry.}
	\label{fem_mesh_fig}
\end{figure}

In this procedure, the beam geometry is discretized by a FEM mesh with $N_{elem}$
finite elements. Each one of these elements is composed by two nodes, and each
one of these nodes has six degrees of freedom associated, one for each field variable
in the beam model described in section~\ref{eq_motion_nonlinear_dyn}.
Thus, the number of degrees of freedom associated with FEM model is
$N_{dofs} = 6 (N_{elem} + 1)$. An illustration of FEM mesh/element can be
seen in Figure~\ref{fem_mesh_fig}.

Concerning the shape functions, it is adopted an interdependent
interpolation scheme which avoids shear-locking effect
\cite{reddy1997p113}. This scheme uses, for
transverse displacements/rotations, Hermite cubic polynomials, and,
for the fields of axial displacement/torsional rotation, affine functions
\cite{bazoune2003p3473}.

Thus, each field variable of the physical model is approximated
by a linear combination of basis functions, in such way that

\begin{eqnarray}
		\label{fem_approx}
		u(x,t) & \approx & \sum_{m = 1}^{N_{dofs}} Q_{m}(t) \, \intfunc{N}{m}{} (x), \\ \nonumber
		\theta_{x}(x,t) & \approx & \sum_{m = 1}^{N_{dofs}} Q_{m}(t) \, \intfunc{N}{m}{} (x),\\ \nonumber
		v(x,t) & \approx & \sum_{m = 1}^{N_{dofs}} Q_{m}(t) \, \intfunc{H}{m}{(1)} (x),\\ \nonumber
		w(x,t) & \approx & \sum_{m = 1}^{N_{dofs}} Q_{m}(t) \, \intfunc{H}{m}{(1)} (x), \\ \nonumber
		\theta_{y}(x,t) & \approx & \sum_{m = 1}^{N_{dofs}} Q_{m}(t) \, \intfunc{H}{m}{(2)} (x),\\ \nonumber
		\theta_{z}(x,t) & \approx & \sum_{m = 1}^{N_{dofs}} Q_{m}(t) \, \intfunc{H}{m}{(2)} (x),
\end{eqnarray}

\noindent
where $\intfunc{N}{m}{} (x)$, $\intfunc{H}{m}{(1)} (x)$, and $\intfunc{H}{m}{(2)} (x)$
are the (position dependent) shape functions, and $Q_{m}(t)$ are the (time dependent)
coefficients of the linear combination. In physical terms, each one of these temporal 
coefficients represents a degree of freedom of the FEM model.

The discretization results is the $N_{dofs} \times N_{dofs}$
nonlinear system of ordinary differential equations given by

\begin{eqnarray}
		\label{galerkin_eq}
		\mat{\MassOp} \vec{\ddot{Q}}(t) +
		\mat{\DampOp} \vec{\dot{Q}}(t) +
		\mat{\StiffOp} \vec{Q}(t)
		= \vec{\ForceOp} \left( \vec{Q}, \vec{\dot{Q}}, \vec{\ddot{Q}} \right),
\end{eqnarray}

\noindent
where $\vec{Q}(t)$ is the nodal displacement vector (translations and rotations),
$\vec{\dot{Q}}(t)$ is the nodal velocity vector, and
$\vec{\ddot{Q}}(t)$ is the nodal acceleration vector.
The other objects in Eq.(\ref{galerkin_eq}) are
the mass matrix $\mat{\MassOp}$,
the damping matrix $\mat{\DampOp}$,
the stiffness matrix $\mat{\StiffOp}$, and
the force vector $\vec{\ForceOp}$.

A discretization procedure similar to one presented above is applied
to the initial conditions of Eqs.(\ref{weak_ic_eq1}) and (\ref{weak_ic_eq2}),
which results in linear systems of algebraic equations given by

\begin{equation}
		\label{galerkin_ic_eq}
    \mat{\MassOp} \vec{Q}(0) =  \vec{Q}_{0},
    \qquad \mbox{and} \qquad
    \mat{\MassOp} \vec{\dot{Q}}(0) =  \vec{\dot{Q}}_{0}.
\end{equation}


\subsection{Reduction of finite element model}

In order to reduce the dimension of finite element model developed
in section~\ref{discretization_dynamics}, it is considered a finite dimensional
version of the generalized eigenvalue problem presented in
section~\ref{conserv_dyn_assoc}, which is defined by

\begin{equation}
		\mat{\StiffOp} \vec{\phi}_n =
		\omega_{n}^{2} \mat{\MassOp} \vec{\phi}_n.
		\label{gen_eigen_eq2}
\end{equation}

Due to the properties of $\MassOp$, and $\StiffOp$ operators,
discussed in section~\ref{conserv_dyn_assoc}, and inherited by finite
dimensional operators $\mat{\MassOp}$ and $\mat{\StiffOp}$, 
the above eigenvalue problem has $N_{dofs}$ solutions.
But Eq.(\ref{gen_eigen_eq2}) is solved only
for $n=1,2,\cdots,N_{red}$, where the reduced model dimension $N_{red}$
is an integer chosen such that $N_{red} \ll N_{dofs}$.

The procedure that follows consists in project the nonlinear dynamics, 
defined by the initial value problem of Eqs.(\ref{galerkin_eq}) and (\ref{galerkin_ic_eq}), into
the vector space spanned by $\{\vec{\phi}_{1},\vec{\phi}_{2},\cdots,\vec{\phi}_{N_{red}}\}$.

For this purpose, define the $N_{dofs} \times N_{red}$ projection matrix by

\begin{equation}
		\mat{\Phi} =
		\left[ \begin{array}{cccc}
			 |                     & |                      &            & |\\
			 \vec{\phi}_{1} & \vec{\phi}_{2} & \cdots & \vec{\phi}_{N_{red}}\\
			 |                     & |                     &             & |
		\end{array} \right],
		\label{projection_matrix_eq}
\end{equation}

\noindent
make in Eqs.(\ref{galerkin_eq}) and (\ref{galerkin_ic_eq})
the change of basis defined by

\begin{equation}
		\vec{Q}(t) = \mat{\Phi} \vec{q}(t),
		\label{modal_reduction_eq}
\end{equation}

\noindent
and then pre-multiply the resulting equations by matrix
$\transp{\mat{\Phi}}$, where superscript $\transp{}$
represents the transposition operation.

This development results in the reduced initial value problem given by

\begin{equation}
		\mat{M} \vec{\ddot{q}}(t) +
		\mat{C} \vec{\dot{q}}(t) +
		\mat{K} \vec{q}(t) =
		\vec{f} \left( \vec{q}(t), \vec{\dot{q}}(t), \vec{\ddot{q}}(t) \right),
		\label{reduced_galerkin_eq}
\end{equation}

\noindent
and

\begin{equation}
    \vec{q}(0) =  \vec{q}_0,
    \qquad \mbox{and} \qquad
    \vec{\dot{q}}(0) =  \vec{\dot{q}}_0,
    \label{reduced_galerkin_ic_eq}
\end{equation}

\noindent
where $\vec{q}(t)$ is the reduced displacement vector,
$\vec{\dot{q}}(t)$ is the reduced velocity vector,
$\vec{\ddot{q}}(t)$ is the reduced acceleration vector.
The reduced matrices of mass, damping, and stiffness,
as well as the reduced vectors of force,
initial displacement, and
initial velocity are,
respectively, defined by
$\mat{M} = \transp{\mat{\Phi}} \mat{\MassOp} \mat{\Phi}$,
$\mat{C} = \transp{\mat{\Phi}} \mat{\DampOp} \mat{\Phi}$,
$\mat{K} = \transp{\mat{\Phi}} \mat{\StiffOp} \mat{\Phi}$,
$\vec{f} = \transp{\mat{\Phi}} \vec{\ForceOp}\left( \mat{\Phi} \vec{q}(t), \mat{\Phi} \vec{\dot{q}}(t), \mat{\Phi} \vec{\ddot{q}}(t) \right)$,
$\vec{q}_{0} = \transp{\mat{\Phi}} \vec{Q}_{0}$,
$\vec{\dot{q}}_{0} = \transp{\mat{\Phi}} \vec{\dot{Q}}_{0}$.
These matrices are $N_{red} \times N_{red}$, while these vectors
are $N_{red} \times 1$. Furthermore, due to the orthogonality
properties defined by Eq.(\ref{orthogonality_relations}), that are
inherited by the operators in finite dimension, these matrices are
diagonal.

Thus, although the initial value problem of
Eqs.(\ref{reduced_galerkin_eq}) and (\ref{reduced_galerkin_ic_eq})
is apparently similar to the one defined by Eqs.(\ref{galerkin_eq})
and (\ref{galerkin_ic_eq}), the former has a structure that makes
it much more efficient in terms of computational cost, and so,
it will be used to analyze the nonlinear dynamics under study.


\subsection{Integration of discretized nonlinear dynamics}

In order to solve the initial value problem of
Eqs.(\ref{reduced_galerkin_eq}) and (\ref{reduced_galerkin_ic_eq}),
it is employed the Newmark method \cite{newmark1959p67},
which defines the following implicit integration scheme

\begin{equation}
    \vec{\dot{q}}_{n+1} =  \vec{\dot{q}}_n +
    (1 - \gamma) \Delta t \,\vec{\ddot{q}}_n + \gamma \Delta t \,\vec{\ddot{q}}_{n+1},
    \label{newmark_vel}
\end{equation}

\begin{equation}
    \vec{q}_{n+1} =  \vec{q}_n +
    \Delta t \,\vec{\dot{q}}_n +
    \left(\frac{1}{2} - \beta \right) \Delta t^2 \,\vec{\ddot{q}}_{n} +
    \beta \, \Delta t^2 \,\vec{\ddot{q}}_{n+1},
    \label{newmark_disp}
\end{equation}

\noindent
where $\vec{q}_n$, $\vec{\dot{q}}_n$ and $\vec{\ddot{q}}_n$ are approximations
to $\vec{q}(t_n)$, $\vec{\dot{q}}(t_n)$ and $\vec{\ddot{q}}(t_n)$,
respectively, and $t_n = n \Delta t$ is an instant in a temporal mesh
defined over the interval $[t_0,t_f]$, with an uniform time step $\Delta t$.
The parameters $\gamma$ and $\beta$ are associated with
accuracy and stability of the numerical scheme \cite{hughes2000}, and
for the simulations reported in this work they are assumed as
$\gamma = 1/2 + \alpha$, and $\beta = 1/4\left(1/2 + \gamma \right)^2$,
with $\alpha = 15/1000$.

Handling up properly Eqs.(\ref{newmark_vel}) and (\ref{newmark_disp}), and
the discrete version of Eq.(\ref{reduced_galerkin_eq}), one arrives in a nonlinear
system of algebraic equations, with unknown vector $\vec{q}_{n+1}$,
which is represented by

\begin{equation}
		\hat{\mat{K}} \vec{q}_{n+1} =  \hat{\vec{f}}_{n+1} \left( \vec{q}_{n+1} \right),
		\label{newmark_nonlinear_system}
\end{equation}

\noindent
where $\hat{\mat{K}}$ is the effective stiffness matrix, and
$\hat{\vec{f}}_{n+1}$ is the (nonlinear) effective force vector.


\subsection{Incorporation of boundary conditions}

As can be seen in Figure~\ref{drillstring_geometry_fig},
the mechanical system has the following boundary conditions:
(i) left extreme with no transversal displacement, nor
transversal rotation; (ii) right extreme with no transversal
displacement. It is also assumed that the left end has: 
(iii) constant axial and rotational velocities in $x$,
respectively equal to $V_{0}$ and $\Omega$.

Hence, for $x = 0$, it is true that
$u(0,t) = V_{0} \, t$,
$v(0,t) = 0$,
$w(0,t) = 0$,
$\theta_{x} (0,t) = \Omega \, t$,
$\theta_{y} (0,t) = 0$, and
$\theta_{z} (0,t) = 0$.
On the other hand,  for $x = L$, one has
$v(L,t) = 0$, and $w(L,t) = 0$.

The variational formulation presented in section~\ref{eq_motion_nonlinear_dyn},
was made for a free-free beam, i.e. the above geometric boundary 
conditions were not considered. For this reason, they are included 
in the formulation as constraints, using the Lagrange multipliers method 
\cite{hughes2000}. The details of this procedure are presented below.

Observe that the boundary conditions can be rewritten in matrix form as

\begin{equation}
	    \label{constraint_eq1}
		\mat{\BoundOp} \vec{Q}(t) = \vec{h}(t),
\end{equation}

\noindent
where the constraint matrix $\mat{\BoundOp}$ is $8 \times N_{dofs}$
and has almost all entries equal to zero. The exceptions are
$\mat{\BoundOp}_{ii} = 1$ for $i=\{1,\cdots,6\}$,
$\mat{\BoundOp}_{7(N_{dofs}-5)} = 1$, and
$\mat{\BoundOp}_{8(N_{dofs}-4)} = 1$.
The constraint vector is given by

\begin{equation}
	    \label{constraint_vector}
		\vec{h}(t) =
		\left( \begin{array}{ccccccccc}
			 u(0,t) \\
			 v(0,t) \\
			 w(0,t) \\
			 \theta_{x} (0,t) \\
			 \theta_{y} (0,t) \\
			 \theta_{z} (0,t)  \\
			 v(L,t) \\
			 w(L,t) \\
		\end{array} \right).
\end{equation}

Making the change of basis defined by Eq.(\ref{modal_reduction_eq}),
one can rewrite Eq.(\ref{constraint_eq1}) as

\begin{equation}
	    \label{constraint_eq2}
		\mat{B} \vec{q}(t) = \vec{h}(t),
\end{equation}

\noindent
where the $8 \times N_{red}$ reduced constraint matrix is defined by
$\mat{B} = \mat{\BoundOp} \mat{\Phi}$.

The discretization of Eq.(\ref{constraint_eq2}) results in

\begin{equation}
	    \label{constraint_eq4}
		\mat{B} \vec{q}_{n+1} = \vec{h}_{n+1},
\end{equation}

\noindent
where $\vec{h}_{n+1}$ is an approximation to $\vec{h}(t_{n+1})$.
This equation defines the constraint that must be satisfied by 
the variational problem ``approximate solution".

In what follows it is helpful to think that Eq.(\ref{newmark_nonlinear_system})
comes from the minimization of an energy functional
$\vec{q}_{n+1} \mapsto \mathscr{F} \left( \vec{q}_{n+1} \right)$,
which is the weak form of this nonlinear system of algebraic equations.

Then, one defines the Lagrangian as

\begin{eqnarray}
	    \label{lagrange_eq}
\resizebox{.9\hsize}{!}{$
  \begin{split}
		\mathscr{L} \left( \vec{q}_{n+1}, \vec{\lambda}_{n+1} \right) =
		\mathscr{F} \left( \vec{q}_{n+1} \right)  +
		\transp{\vec{\lambda}}_{n+1} \left( \mat{B} \vec{q}_{n+1} - \vec{h}_{n+1} \right),
  \end{split}$
}
\end{eqnarray}

\noindent
being the (time-dependent) Lagrange multipliers vector of the form

\begin{equation}
		\vec{\lambda}_{n+1} =
		\left( \begin{array}{c}
			 \lambda_{1}(t_{n+1}) \\
			 \lambda_{2}(t_{n+1}) \\
			 \lambda_{3}(t_{n+1}) \\
			 \lambda_{4}(t_{n+1}) \\
			 \lambda_{5}(t_{n+1}) \\
			 \lambda_{6}(t_{n+1}) \\
			 \lambda_{7}(t_{n+1}) \\
			 \lambda_{8}(t_{n+1})
		\end{array} \right).		
\end{equation}

Invoking the Lagrangian stationarity condition one arrives 
in the following $(N_{red}+8) \times (N_{red}+8)$ system of nonlinear 
algebraic equations

\begin{equation}
		\left[ \begin{array}{ll}
			 \hat{\mat{K}} & \transp{\mat{B}} \\
			 \mat{B}          & \mat{0}
		\end{array} \right]
		\left( \begin{array}{l}
			 \vec{q}_{n+1} \\
			 \vec{\lambda}_{n+1}
		\end{array} \right) =
		\left( \begin{array}{l}
			 \hat{\vec{f}}_{n+1} \\
			 \vec{h}_{n+1}
		\end{array} \right),
		\label{extended_nonlinear_system1}
\end{equation}

\noindent
where $\mat{0}$ is a $8 \times 8$ null matrix. The unknowns are
$\vec{q}_{n+1}$ and $\vec{\lambda}_{n+1}$,
and must be solved for each instant of time in the temporal mesh,
in order to construct an approximation to the mechanical system 
dynamic response.

The solution of the nonlinear system of algebraic equations,
defined by Eq.(\ref{extended_nonlinear_system1}), is carried out
first obtaining and solving a discrete Poisson equation for
$\vec{\lambda}_{n+1}$ \cite{golub2013}, and then using the
first line of (\ref{extended_nonlinear_system1})
to obtain $\vec{q}_{n+1}$. To solve these equations,
a procedure of fixed point iteration is used in combination
with a process of successive over relaxation \cite{young2003}.



\section{Probabilistic modeling of system-parameter uncertainties}
\label{prob_mod_data_uncert}

The mathematical model used to describe the physical behavior of the
mechanical system is an abstraction of reality, and its use does not
consider some aspects of the problem physics. Regarding the system
modeling, either the beam theory used to describe the structure
dynamics \cite{ritto2009p865}, as the friction and shock model used
\cite{hunt1975p440} are fairly established physical models, who have gone
through several experimental tests to prove their validity, and have been
used for many years in similar situations. On the other hand, the bit-rock
interaction model adopted in this work, until now was used only in a purely
numeric context \cite{ritto2013p145}, without any experimental validation.
Thus, it is natural to conclude that bit-rock interaction law is the weakness
of the model proposed in this work.

In this sense, this work will focus on modeling and quantifying the uncertainties
that are introduced in the mechanical system by bit-rock interaction model.
For convenience, it was chosen to use a parametric probabilistic approach
\cite{soize2012}, where only the uncertainties of system parameters are considered, 
and the maximum entropy principle is employed to construct the probability distributions.

\subsection{Probabilistic framework}

Let $\randvar{X}$ be a real-valued random variable, 
defined on a probability space $(\SS, \SA, \PM)$, 
for which the probability distribution $P_{\randvar{X}}(dx)$ on 
$\R$ admits a density $x\mapsto p_{\randvar{X}}(x)$ with respect to $dx$. 
The support of the probability density function (PDF) $p_{\randvar{X}}$ 
will be denoted by $\supp {\randvar{X}} \subset \R$. 
The mathematical expectation of $\randvar{X}$ is defined by

\begin{equation}
	\expval{\randvar{X}} = \int_{\supp {\randvar{X}}} x\, p_{\randvar{X}}(x) \, dx \, ,	
	\label{def_expval_op}
\end{equation}

\noindent
and any realization of random variable $\randvar{X}$ 
will be denoted by $\randvar{X}(\SSpt)$ for $\SSpt \in \SS$.
Let $\mean{\randvar{X}} = \expval{\randvar{X}}$ be the mean value,
$\var{\randvar{X}} = \expval{\left( \randvar{X} - \mean{\randvar{X}} \right)^2}$ 
be the variance, and $\stddev{\randvar{X}} = \sqrt{\var{\randvar{X}}}$ 
be the standard deviation of $\randvar{X}$. The Shannon entropy of
$p_{\randvar{X}}$ is defined by
$\shannon{\pdf{\randvar{X}}} = - \expval{ \ln \pdf{\randvar{X}}(\randvar{X})}$.

\subsection{Probabilistic model for bit-rock interface law}

Recalling that bit-rock interaction force and torque are,
respectively, given by Eqs.(\ref{br_force_eq}) and (\ref{br_torque_eq}),
the reader can see that this bit-rock interface law is characterized
by three parameters, namely, $\alpha_{\BR}$, $\Gamma_{\BR}$, and $\mu_{\BR}$.
The construction of the probabilistic model for each one parameter of these
parameters, which are respectively modeled by random variables $\randvar{\bbalpha}_{\BR}$,
$\randvar{\Gamma}_{\BR}$, and $\randvar{\bbmu}_{\BR}$, is presented below.

\subsection{Distribution of force rate of change}
\label{distr_rate_change}

As the rate of change $\alpha_{\BR}$ is positive, it is reasonable
to assume $\supp{\randvar{\bbalpha}_{\BR}} = ]0,\infty[$. Therefore,
the PDF of $\randvar{\bbalpha}_{\BR}$ is a nonnegative function
$\pdf{\randvar{\bbalpha}_{\BR}}$, such that

\begin{equation}
	\int_{\alpha = 0}^{+\infty} \pdf{\randvar{\bbalpha}_{\BR}} (\alpha) \, d\alpha = 1.
	\label{norm_cond_alpha_br}
\end{equation}

It is also convenient to assume that the mean value of $\randvar{\bbalpha}_{\BR}$
is a known positive number, denoted by $\mean{\randvar{\bbalpha}_{\BR}}$, i.e.,

\begin{equation}
	\expval{\randvar{\bbalpha}_{\BR}} = \mean{\randvar{\bbalpha}_{\BR}} > 0.
	\label{meanval_alpha_br}
\end{equation}

One also need to require that

\begin{equation}
	\expval{\ln \left( \randvar{\bbalpha}_{\BR} \right) } = q_{\randvar{\bbalpha}_{\BR}}, ~~ |q_{\randvar{\bbalpha}_{\BR}}|< + \infty,
	\label{finite_log_alpha_br}
\end{equation}

\noindent
which ensures, as can be seen in \cite{soize2000p277,soize2001p1979,soize2005p623},
that the inverse of $\randvar{\bbalpha}_{\BR}$ is second order random variable.
This condition is necessary to guarantee that the stochastic dynamical system
associated to this random variable is of second order, i.e., it has finite variance.
Employing the principle of maximum entropy 
one need to maximize the entropy function $\shannon{\pdf{\randvar{\bbalpha}_{\BR}}}$,
respecting the constraints imposed by (\ref{norm_cond_alpha_br}),
(\ref{meanval_alpha_br}) and (\ref{finite_log_alpha_br}).

The desired PDF corresponds to the gamma distribution and is given by

\begin{equation}
  \begin{split}
	\pdf{\randvar{\bbalpha}_{\BR}}(\alpha) =
	\indfunc{]0,\infty[} (\alpha) \,
	\frac{1}{\mean{\randvar{\bbalpha}_{\BR}}}
	\left( \frac{1}{\delta_{\randvar{\bbalpha}_{\BR}}^2} \right)^{1/\delta_{\randvar{\bbalpha}_{\BR}}^2} ~~~~~~~~~~~~~~~\\
	\times
	\frac{1}{\Gamma(1/\delta_{\randvar{\bbalpha}_{\BR}}^2)}
	\left( \frac{\alpha}{ \mean{\randvar{\bbalpha}_{\BR}} } \right)^{1/\delta_{\randvar{\bbalpha}_{\BR}}^2 -1}
	\exp\left( \frac{- \alpha}{\delta_{\randvar{\bbalpha}_{\BR}}^2 \mean{\randvar{\bbalpha}_{\BR}} } \right),
	\label{prob_distr_alpha_br}
  \end{split}
\end{equation}

\noindent
where the symbol $\indfunc{]0,\infty[} (\alpha)$ denotes the indicator function
of the interval $]0,\infty[$,
$0 \leq  \delta_{\randvar{\bbalpha}_{\BR}} = \stddev{\randvar{\bbalpha}_{\BR}} / \mean{\randvar{\bbalpha}_{\BR}} < 1/\sqrt{2}$
is a type of dispersion parameter, and

\begin{equation}
	\Gamma(z) =
	\int_{y = 0}^{+\infty}  y^{z-1} \, e^{- y} \, dy,
\label{gamma_func}
\end{equation}

\noindent
is the gamma function.


\subsection{Distribution of limit force}

The parameter $\Gamma_{\BR}$ is also positive, in a way that
\linebreak$\supp{\randvar{\Gamma}_{\BR}} = ]0,\infty[$,~and~consequently

\begin{equation}
	\int_{\gamma = 0}^{+\infty} \pdf{\randvar{\Gamma}_{\BR}} (\gamma) \, d\gamma = 1.
	\label{norm_cond_Gamma_br}
\end{equation}

The hypothesis that the mean is a known positive number $\mean{\randvar{\Gamma}_{\BR}}$
is also done, i.e.,

\begin{equation}
	\expval{\randvar{\Gamma}_{\BR}} = \mean{\randvar{\Gamma}_{\BR}} > 0,
	\label{meanval_Gamma_br}
\end{equation}

\noindent
as well as that the technical condition, required for the stochastic dynamical system
associated be of second order, is fulfilled, i.e.

\begin{equation}
	\expval{\ln \left( \randvar{\Gamma}_{\BR} \right) } = q_{\randvar{\Gamma}_{\BR}}, ~~ |q_{\randvar{\Gamma}_{\BR}}|< + \infty.
	\label{finite_log_Gamma_br}
\end{equation}

In a similar way to the procedure presented in section~\ref{distr_rate_change},
it can be shown that PDF of maximum entropy is also gamma distributed,
and given by

\begin{equation}
  \begin{split}
	\pdf{\randvar{\Gamma}_{\BR}}(\gamma) =
	\indfunc{]0,\infty[} (\gamma) \,
	\frac{1}{\mean{\randvar{\Gamma}_{\BR}}}
	\left( \frac{1}{\delta_{\randvar{\Gamma}_{\BR}}^2} \right)^{1/\delta_{\randvar{\Gamma}_{\BR}}^2} ~~~~~~~~~~~~~~~\\
	\times
	\frac{1}{\Gamma(1/\delta_{\randvar{\Gamma}_{\BR}}^2)}
	\left( \frac{\gamma}{ \mean{\randvar{\Gamma}_{\BR}} } \right)^{1/\delta_{\randvar{\Gamma}_{\BR}}^2 -1}
	\exp\left( \frac{- \gamma}{\delta_{\randvar{\Gamma}_{\BR}}^2 \mean{\randvar{\Gamma}_{\BR}} } \right).
	\label{prob_distr_Gamma_br}
  \end{split}
\end{equation}


\subsection{Distribution of friction coefficient}

With respect to the parameter $\mu_{\BR}$, one know it is nonnegative
and bounded above by the unity. Thus, one can safely assume that
$\supp{\bbmu_{\BR}} = [0,1]$, so that the normalization condition
read as

\begin{equation}
	\int_{\mu = 0}^{1} \pdf{\randvar{\bbmu}_{\BR}} (\mu) \, d\mu = 1.
	\label{norm_cond_mu_br}
\end{equation}

The following two conditions are also imposed

\begin{equation}
	\expval{\ln \left( \randvar{\bbmu}_{\BR} \right) } = q^1_{\randvar{\bbmu}_{\BR}}, ~~ |q^1_{\randvar{\bbmu}_{\BR}}|< + \infty,
	\label{finite_log_mu_br}
\end{equation}

\begin{equation}
	\expval{\ln \left( 1 - \randvar{\bbmu}_{\BR} \right) } = q^2_{\randvar{\bbmu}_{\BR}}, ~~ |q^2_{\randvar{\bbmu}_{\BR}}|< + \infty,
	\label{finite_log_1_mu_br}
\end{equation}

\noindent
representing a weak decay of the PDF of $\randvar{\bbmu}_{\BR}$ in $0^+$ and
$1^-$ respectively \cite{soize2000p277,soize2001p1979,soize2005p623}.
Evoking again the principle of maximum entropy 
considering now as known information the constraints defined by
(\ref{norm_cond_mu_br}), (\ref{finite_log_mu_br}), and (\ref{finite_log_1_mu_br})
one has that the desired PDF is given by

\begin{equation}
	\pdf{\randvar{\bbmu}_{\BR}}(\mu) =
	\indfunc{[0,1]} (\mu) \,
	\frac{\Gamma(a+b)}{\Gamma(a)\,\Gamma(b)} \,
	\mu^{a-1} \, \left( 1 - \mu \right)^{b-1},
	\label{prob_distr_mu_br}
\end{equation}

\noindent
which corresponds to the beta distribution

The parameters $a$ and $b$
are associated with the shape of the probability distribution, and can be
related with $\mean{\randvar{\bbmu}_{\BR}}$ and $\delta_{\randvar{\bbmu}_{\BR}}$ by

\begin{equation}
	a = \frac{ \mean{\randvar{\bbmu}_{\BR}} }{ \delta_{\randvar{\bbmu}_{\BR}}^2 }
	      \left( \frac{1}{\mean{\randvar{\bbmu}_{\BR}}} - \delta_{\randvar{\bbmu}_{\BR}}^2 -1\right),
\end{equation}

\noindent
and

\begin{equation}
	b = \frac{ \mean{\randvar{\bbmu}_{\BR}} }{ \delta_{\randvar{\bbmu}_{\BR}}^2 } \,
	      \left( \frac{1}{\mean{\randvar{\bbmu}_{\BR}}} - \delta_{\randvar{\bbmu}_{\BR}}^2 -1\right)
		  \left( \frac{1}{ \mean{\randvar{\bbmu}_{\BR}} } -1\right).
\end{equation}


\subsection{Stochastic nonlinear dynamical system}

Due to the randomness of parameters $\randvar{\bbalpha}_{\BR}$,
$\randvar{\Gamma}_{\BR}$, and $\randvar{\bbmu}_{\BR}$, the 
mechanical system physical behavior is now described, 
for all $\theta$ in $\Theta$, by the stochastic nonlinear dynamical 
system defined by

\begin{equation}
		\mat{M} \randproc{\ddot{q}} (t,\SSpt) +
		\mat{C} \randproc{\dot{q}} (t,\SSpt) +
		\mat{K} \randproc{q} (t,\SSpt) =
		\randproc{f} \left( \randproc{q}, \randproc{\dot{q}}, \randproc{\ddot{q}} \right),
		\label{stoch_galerkin_eq}
\end{equation}

\begin{equation}
    \randproc{q} (0,\SSpt) =  \vec{q}_{0},
    \qquad \mbox{and} \qquad
    \randproc{\dot{q}} (0,\SSpt) =  \vec{\dot{q}}_{0}, \quad a.s.
    \label{stoch_galerkin_ic_eq}
\end{equation}

\noindent
where $\randproc{q}(t)$ is the random reduced displacement vector,
$\randproc{\dot{q}}(t)$ is the random reduced velocity vector, and
$\randproc{\ddot{q}}(t)$ is the random reduced acceleration vector,
and $\randproc{f}$ is the random reduced nonlinear force vector.

The methodology used to calculate the propagation of uncertainties through
this stochastic dynamical system is Monte Carlo (MC) method \cite{kroese2011},
employing a strategy of parallelization described in \cite{cunhajr2014p1355}.


\section{Numerical experiments and discussions}
\label{num_results}

In order to simulate the mechanical system nonlinear dynamics,
the physical parameters presented in Table~\ref{physical_param_tab}
are adopted, as well as $L = 100~m$, the rotational and axial
velocities in x, respectively given by $\Omega = 2 \pi ~ rad/s$, and
$V_0 = 1/180~m/s$. The values of these parameters do not correspond
exactly to actual values used in a real drillstring, but are of the
same order of magnitude. For this configuration, the beam geometry is
discretized by 500 finite elements, and the interval of integration
$[t_0,t_f] = [0,10]~s$ is considered.

\begin{table}[ht!]
	\centering
	\caption{Physical parameters of the mechanical system that are used in the simulation.}
	\begin{tabular}{ccl}
		\toprule
		parameter & value & unit  \\
		\midrule
		$\rho$           & $7900$                              & $kg/m^3$ \\
		$g$                & $9.81$                               & $m/s^2$ \\
		$\nu$            & $0.3$                                  & --- \\
		$\kappa_{s}$ & $6/7$                                 & ---\\
		$c$                & $0.01$                                & ---\\
		$E$                & $203 \times 10^{9}$          & $Pa$ \\
		$R_{bh}$        & $~95 \times 10^{-3}$        & $m$ \\
		$R_{int}$        & $~50 \times 10^{-3}$        & $m$ \\
		$R_{ext}$       & $~80 \times 10^{-3}$        & $m$ \\
		\bottomrule
	\end{tabular}
	\label{physical_param_tab}
\end{table}

For the friction and shock model constants,
are considered the values shown in Table~\ref{shock_param_tab},
which have order of magnitude typical of a borehole wall made of steel
\cite{zhang2009p051002}. The low value for $\mu_{\FS}$ is justified 
by the fact that, in a real system, there is a fluid between the column
and borehole wall, which carries a substantial reduction in the
torsional friction.

\begin{table}[ht!]
	\centering
	\caption{Parameters of the friction and shock model that are used in the simulation.}
	\begin{tabular}{cll}
		\toprule
		parameter & value & unit  \\
		\midrule
		$k_{\FS_1}$  & $1 \times 10^{10}$ & $N/m$ \\
		$k_{\FS_2}$  & $1 \times 10^{16}$ & $N/m^3$ \\
		$c_{\FS}$      & $1 \times 10^{6}$ & $(N/m^3)/(m/s)$ \\
		$\mu_{\FS}$ & $0.25$ & ---\\
		\bottomrule
	\end{tabular}
	\label{shock_param_tab}
\end{table}

The bit-rock interaction model constants can be
seen in Table~\ref{bitrock_param_tab}, and were estimated
following a similar strategy as that shown in \cite{ritto2013p145}.

\begin{table}[ht!]
	\centering
	\caption{Parameters of the bit-rock interaction model that are used in the simulation.}
	\begin{tabular}{cll}
		\toprule
		parameter & value & unit  \\
		\midrule
		$\Gamma_{\BR}$ & $30 \times 10^{3}$   & $N$\\
		$\alpha_{\BR}$    & $400$                          & $1/(m/s)$\\
		$\mu_{\BR}$        & $0.4$ & ---\\
		\bottomrule
	\end{tabular}
	\label{bitrock_param_tab}
\end{table}

\subsection{``Validation'' of the computational model}

In order to validate to computational model developed in
this work, the most widely accepted approach would be 
through comparison with experimental data
\cite{oberkampf2010}. Unfortunately, it is very dificult
to obtain such experimental measurements for the type of 
mechanical system analyzed in this work. Accordingly, 
as in many analyzes of dynamic systems with complex behavior
(e.g. aerospace, automotive, etc), this work used numerical simulation 
without experimental analysis. However, it is clear that such numerical 
simulations do have meaning, and are only useful, if prior validation 
of models, formulations, and numerical simulation software were obtained.
In the cases treated, ``validations'' of the model, formulation, and software 
were obtained, step by step, by comparing the results obtained with the software
with the following reference results: (i) linear static; (ii) nonlinear static with contact; 
(iii) linear dynamic nonrotating without contact; (iv) linear dynamic rotating without contact; 
(v) nonlinear nonrotating dynamic with contacts (shocks); and
(vi) nonlinear rotating dynamic with contacts (shocks). All tests were performed 
and the analyzes were used to ``validate'' the model, the formulation, and the 
results of numerical simulations.


\subsection{Modal analysis of the mechanical system}

In this section, the modal content of the mechanical system
is investigated. This investigation aims to identify the natural
frequencies of the system, and, especially, to check the influence
of \emph{slenderness ratio}, defined as the ratio between beam
length and external diameter, in the natural frequencies distribution.

Therefore, the dimensionless frequency band for the problem
is assumed as being $\textsl{B} = [0,4]$, with the dimensionless
frequency defined by

\begin{equation}
	\dimless{f} = \frac{f \, L}{c_{L}},
\end{equation}

\noindent
where $f$ is the dimensional frequency (Hz), and
$c_{L} = \sqrt{E/\rho}$ is the longitudinal wave velocity.
Once it was defined in terms of a dimensionless frequency,
the band of analysis does not change when the beam length
is varied. Also, the reader can check that this band is representative
for the mechanical system dynamics, once the beam rotates at $2\pi~rad/s$,
which means that it is excited at $1~Hz$.

\begin{figure*}
	\centering
	\subfigure[slenderness = 312.5]{\includegraphics[scale=0.32]{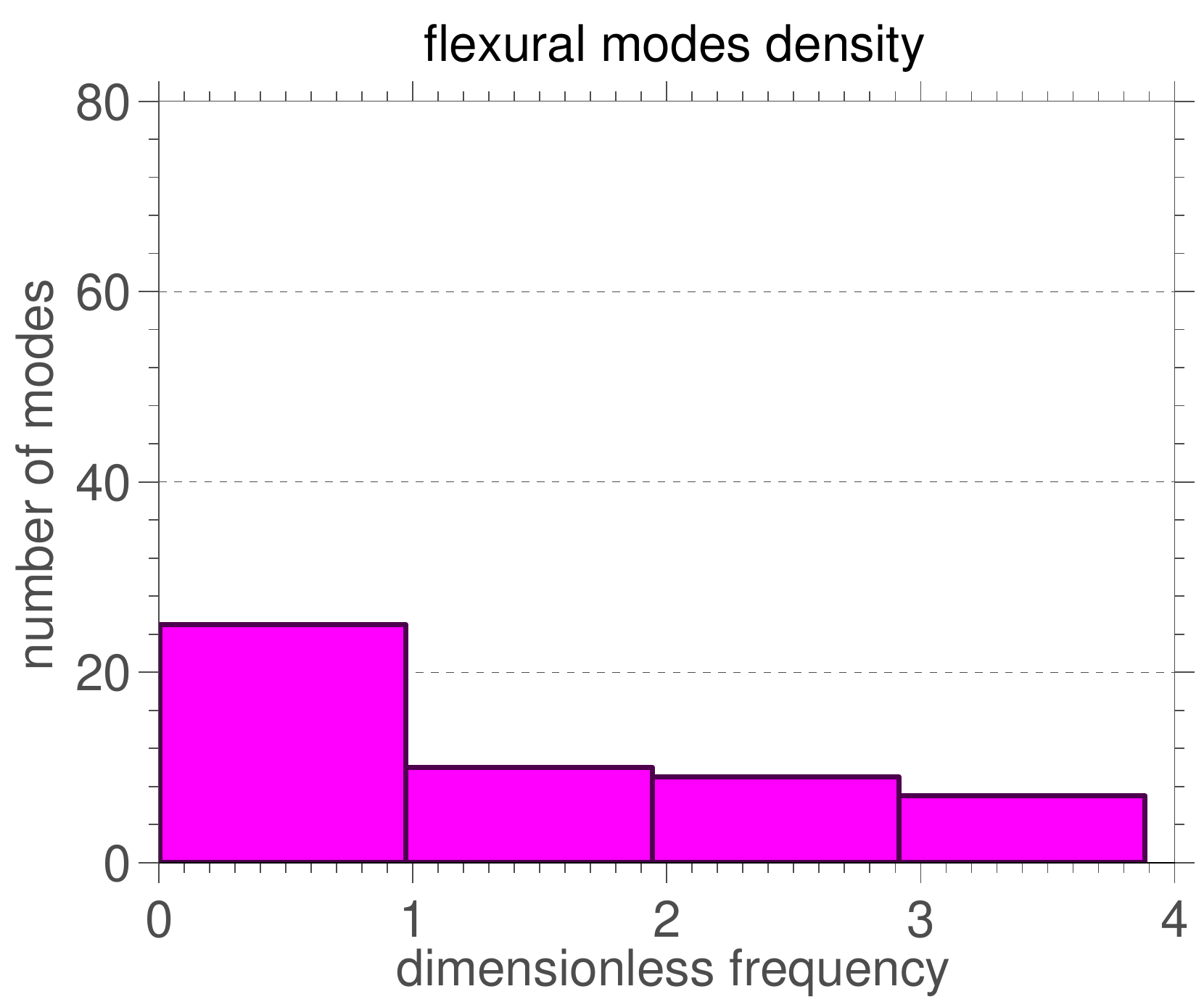}}~~
	\subfigure[slenderness = 625.0]{\includegraphics[scale=0.32]{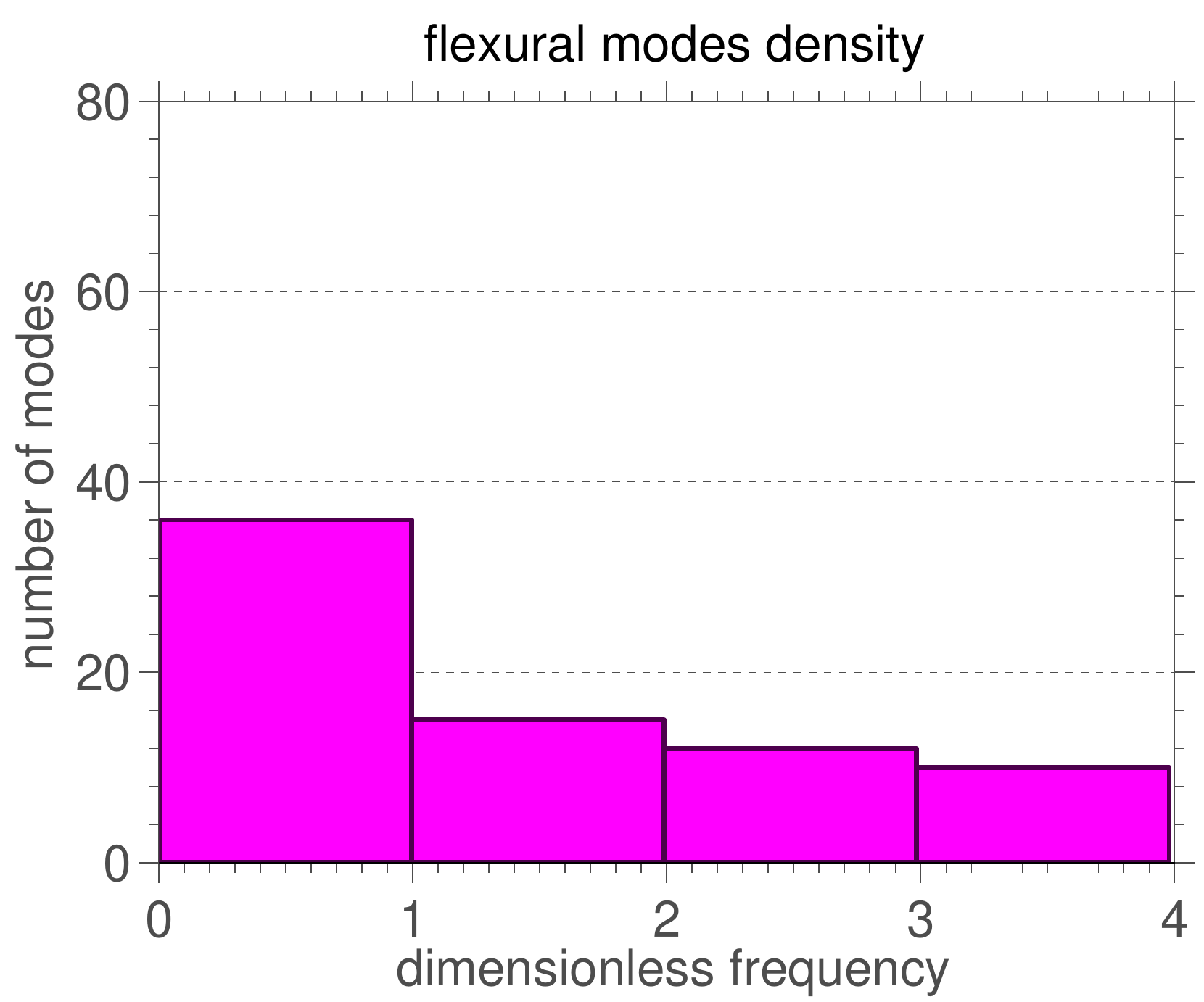}}~~
	\subfigure[slenderness = 937.5]{\includegraphics[scale=0.32]{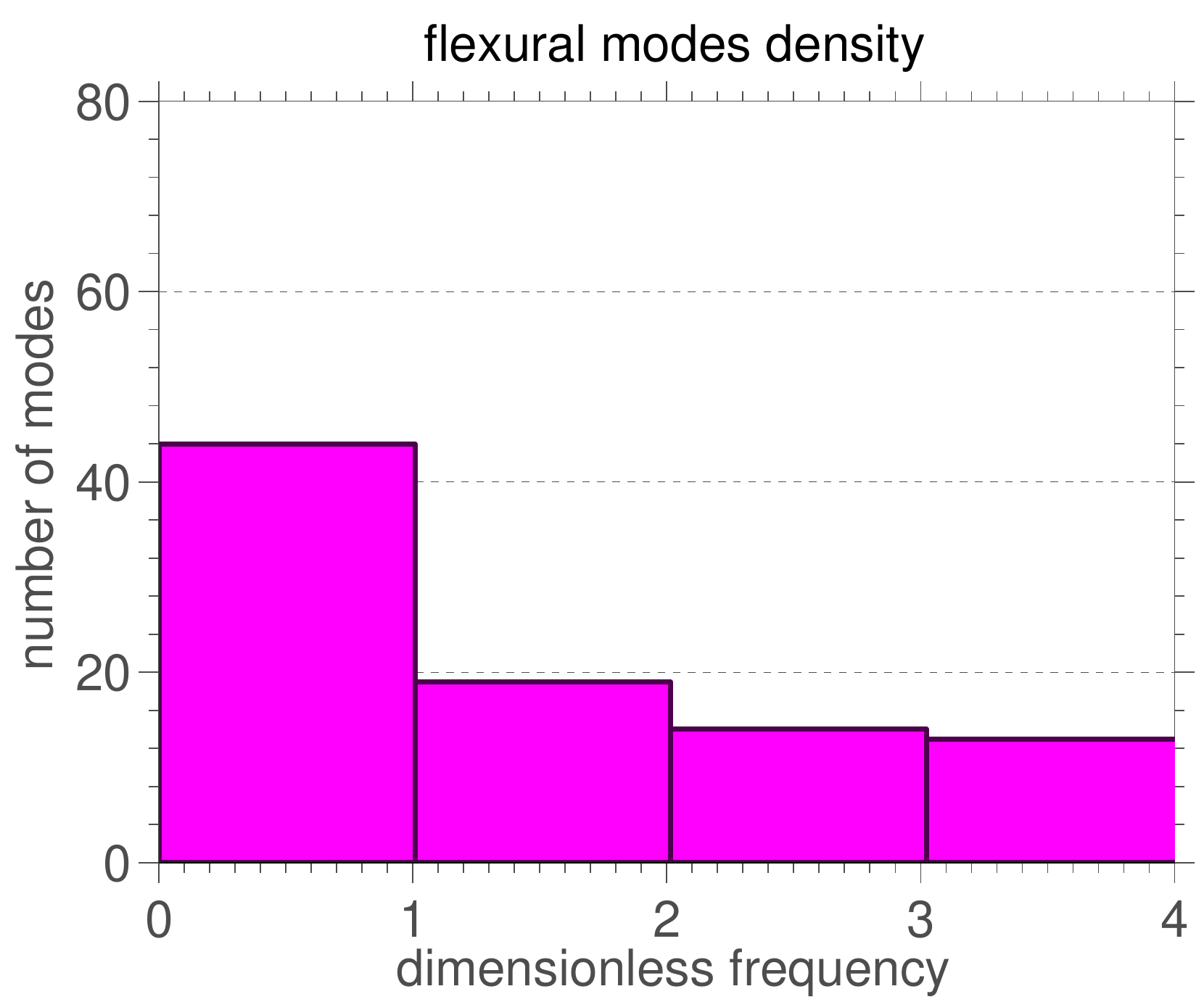}}~~
	\caption{Distribution of the flexural modes as function of dimensionless frequency, for several values of slenderness ratio.}
	\label{mdensity_flex_y_fig}
\end{figure*}

In Figure~\ref{mdensity_flex_y_fig} one can see the distribution of the flexural modes
as a function of dimensionless frequency, for several values of slenderness ratio.
Clearly it is observed that the flexural modes are denser in the low frequency range.
Further, when the slenderness ratio increases, the modal density in the low frequencies
range tend to increase.

\begin{figure*}
	\centering
	\subfigure[slenderness = 312.5]{\includegraphics[scale=0.32]{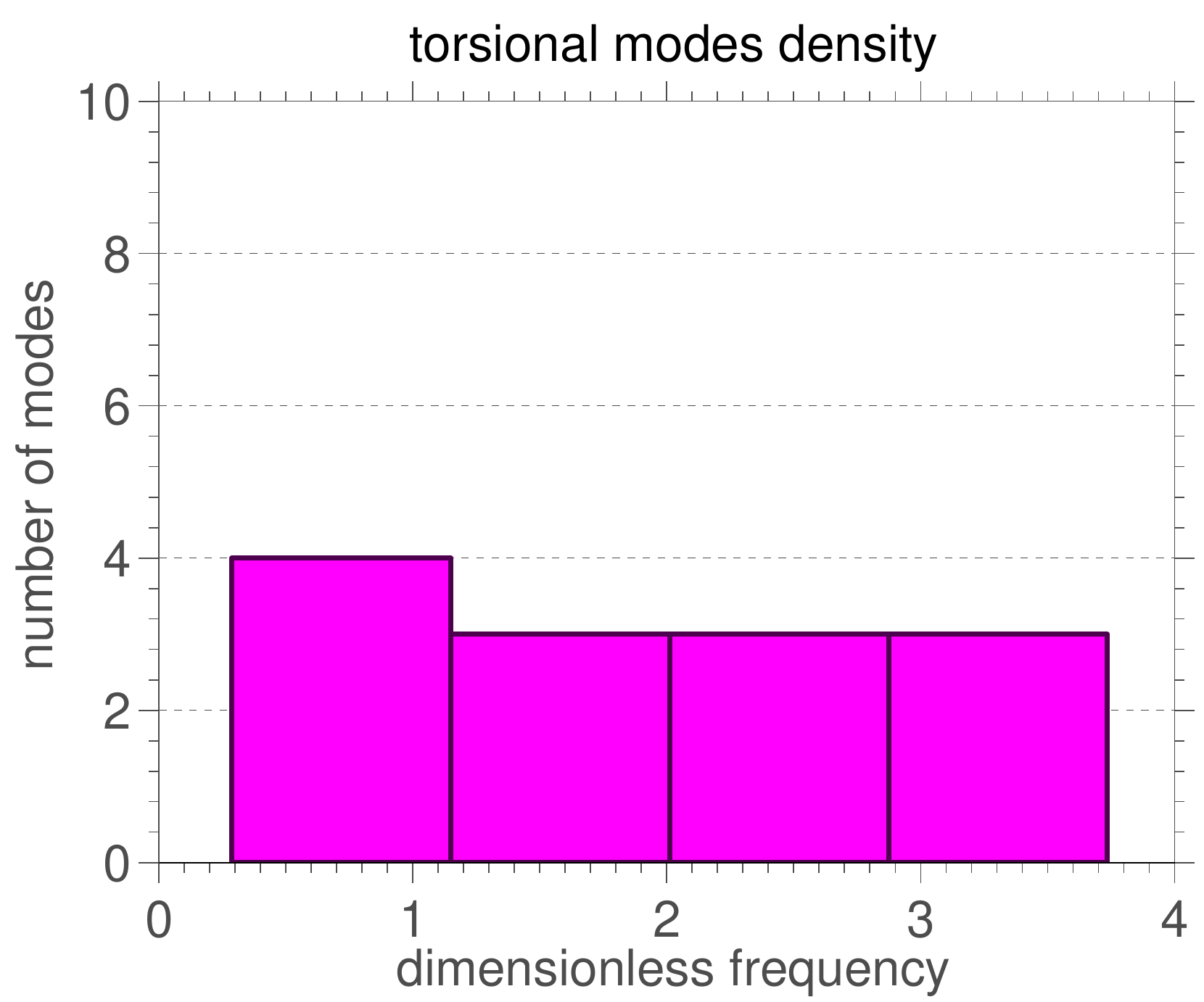}}~~
	\subfigure[slenderness = 625.0]{\includegraphics[scale=0.32]{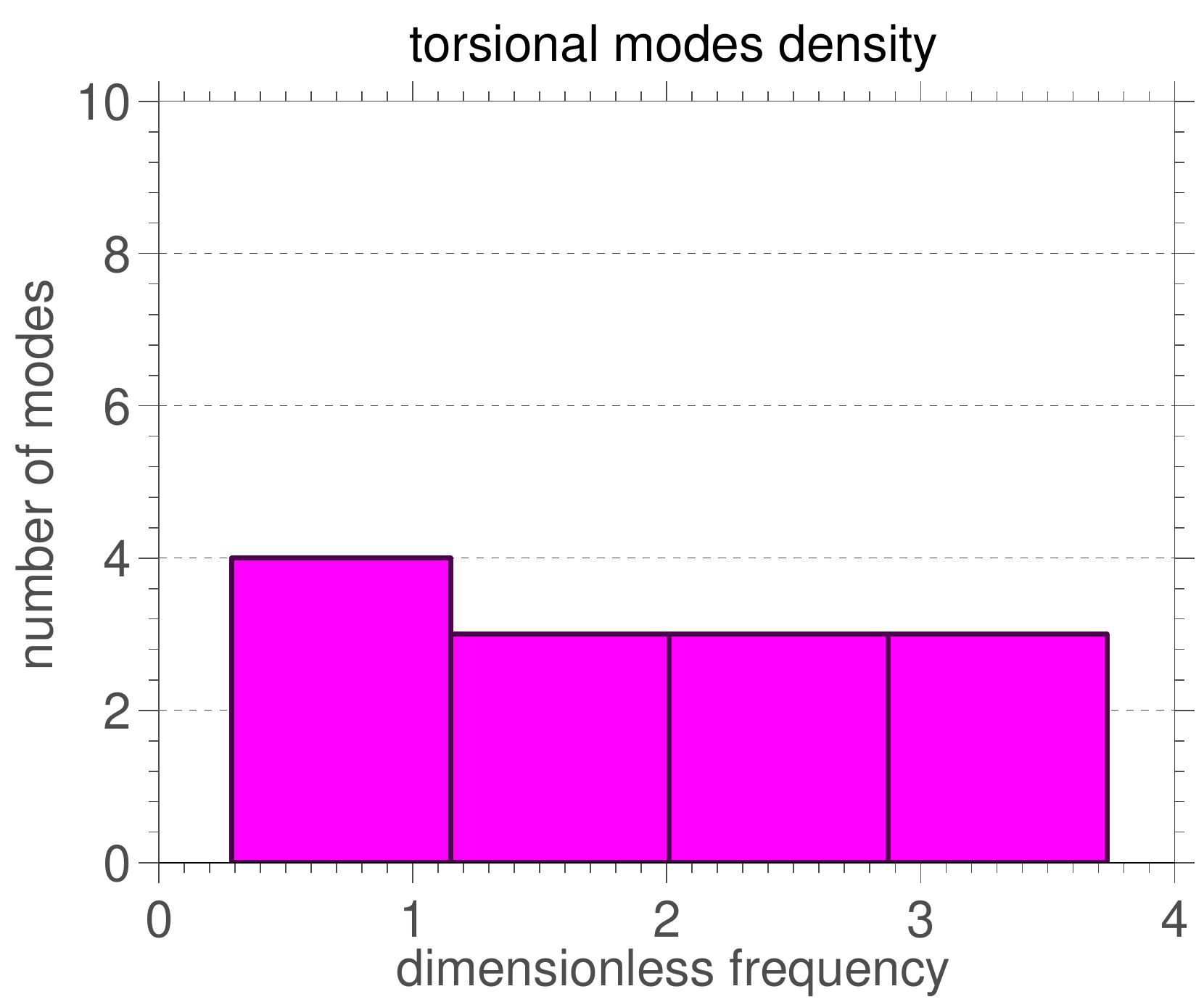}}~~
	\subfigure[slenderness = 937.5]{\includegraphics[scale=0.32]{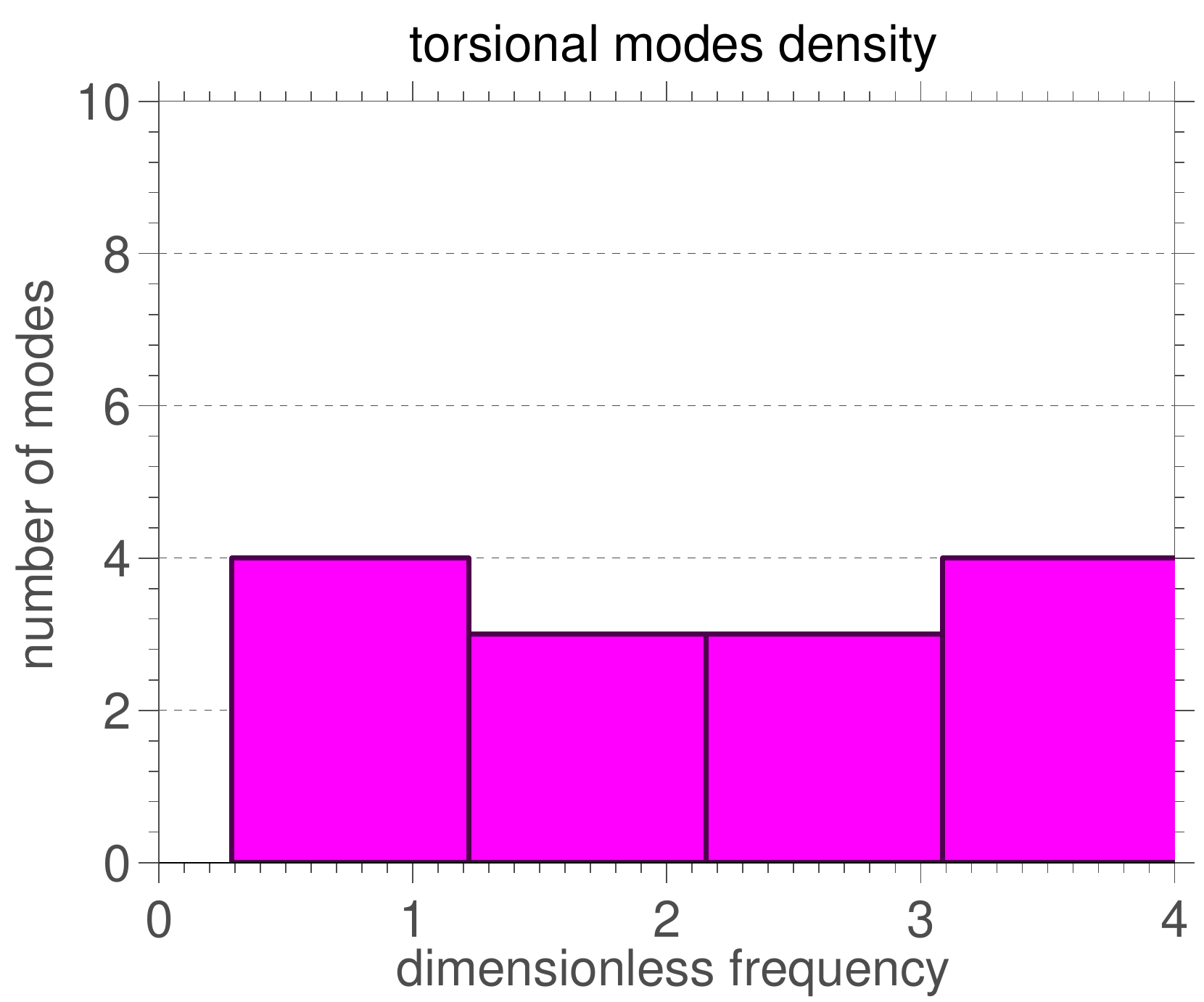}}~~
	\caption{Distribution of the torsional modes as function of dimensionless frequency, for several values of slenderness ratio.}
	\label{mdensity_tors_x_fig}
\end{figure*}

\begin{figure*}
	\centering
	\subfigure[slenderness = 312.5]{\includegraphics[scale=0.32]{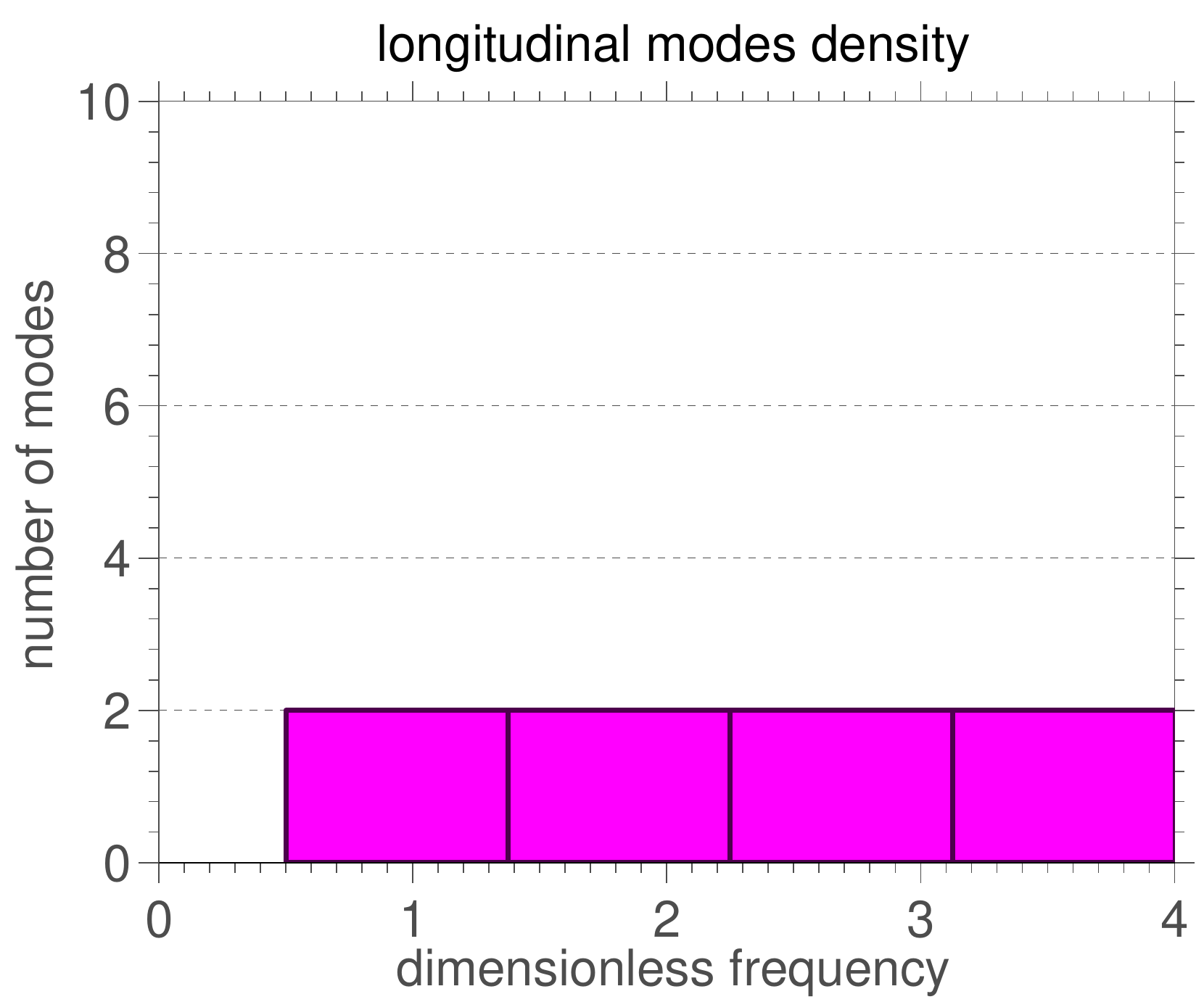}}~~
	\subfigure[slenderness = 625.0]{\includegraphics[scale=0.32]{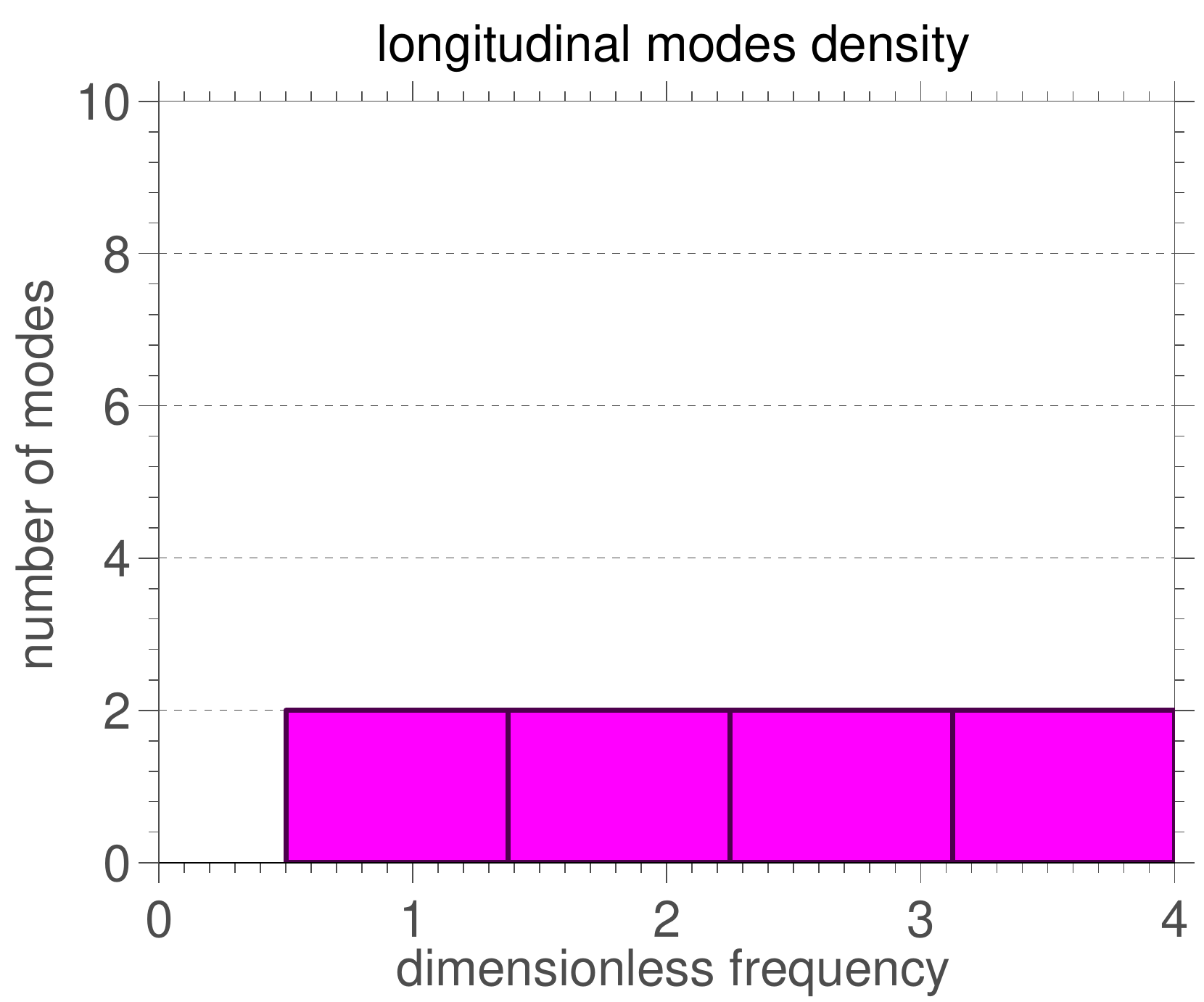}}~~
	\subfigure[slenderness = 937.5]{\includegraphics[scale=0.32]{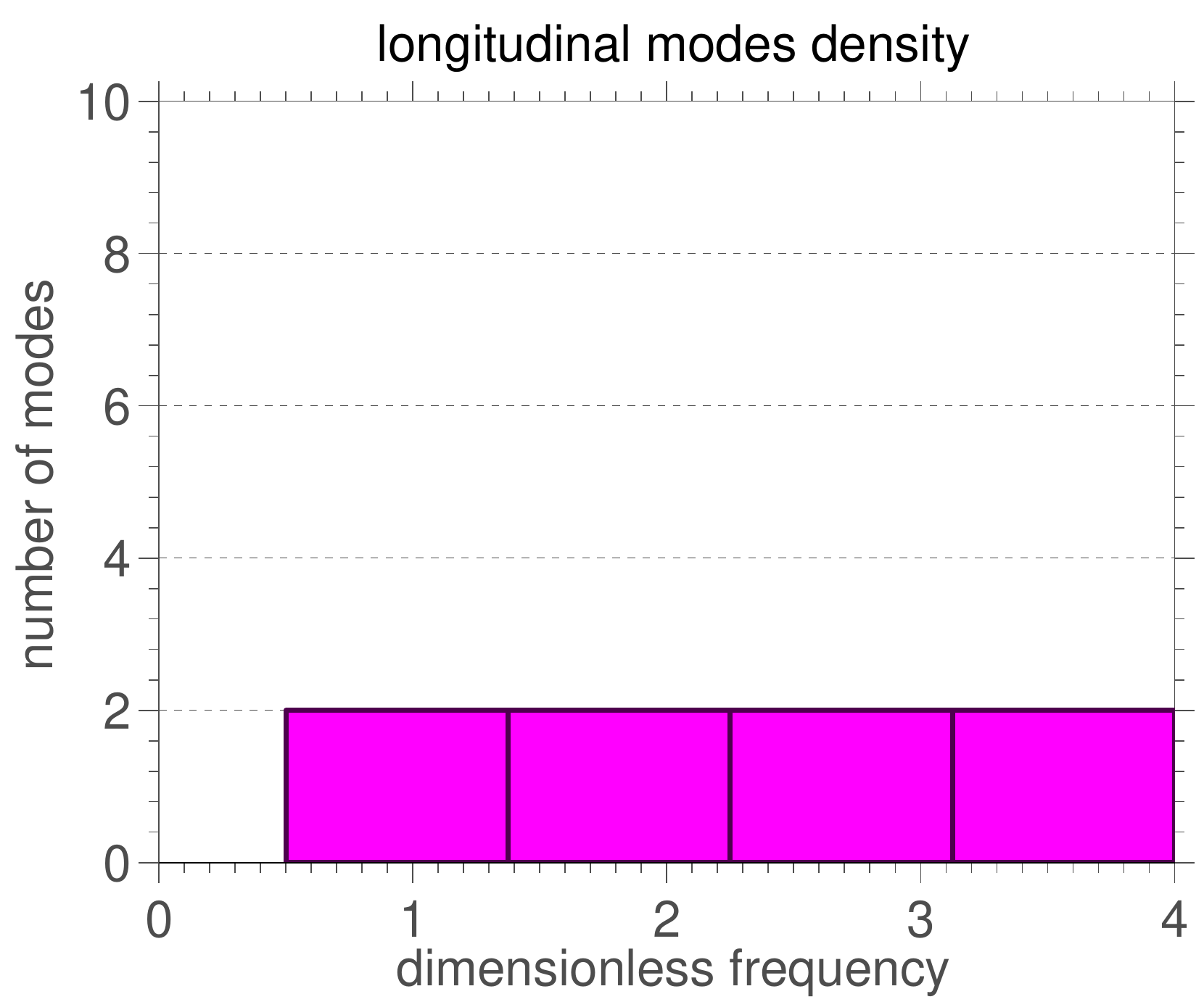}}~~
	\caption{Distribution of the longitudinal modes as function of dimensionless frequency, for several values of slenderness ratio.}
	\label{mdensity_long_x_fig}
\end{figure*}

A completely different behavior is observed for the torsional and longitudinal (traction-compression)
modes of vibration, as can be seen in Figures~\ref{mdensity_tors_x_fig} and \ref{mdensity_long_x_fig},
respectively. One can note that, with respect to these two modes of vibration, the modal distribution
is almost uniform with respect to dimensionless frequency, and invariant to changes in the slenderness ratio.

It may also be noted from Figures~\ref{mdensity_flex_y_fig} to \ref{mdensity_long_x_fig}
that lowest natural frequencies are associated with flexural mechanism.
This is because beam flexural stiffness is much smaller than torsional stiffness,
which, in turn, is less than axial stiffness. In other words, it is much easier to bend
the beam than twisting it. However, twists the beam is easier than buckling it.

The dimensionless frequency band adopted in the analysis corresponds
to a maximum dimensional frequency of $f_{max} = 4 \, c_{L}/L$. In this way,
a nominal time step of $\Delta t = (2 \, f_{max})^{-1}$ is adopted for time integration.
This time step is automatically refined by the algorithm of integration,
whenever necessary, to capture the shock effects.


\subsection{Construction of the reduced model}

In the reduced model construction, are taken into account
the mechanical system rigid body modes, as well as the
modes of bending, torsion and traction-compression. The construction
strategy consists of including:
(i) the two rigid body modes (translation and rotation);
(ii) all flexural modes such that $0 < \dimless{f} \leq 5 \, L/c_{L}$;
(iii) all torsional modes such that $0 < \dimless{f} \leq 4$;
(iv) all longitudinal modes such that $0 < \dimless{f} \leq 4$.

In this way, the total number of modes used in FEM model is a
function of beam length. In Table~\ref{reduced_model_tab}
the reader can see a comparison, for different values of $L$,
of the full FEM model dimension and the corresponding 
reduced order model dimension. Note that using 
the above strategy, the reduced model dimension is always much smaller
than the full model dimension.

\begin{table}[ht!]
	\centering
	\caption{Dimension of the FEM model as a function of beam length.}
	\begin{tabular}{ccc}
		\toprule
		beam length & full model & reduced model \\
		              (m) &         DoFs & DoFs\\
		\midrule
		  50 & 306 & 37\\
		100 & 3006 & 49\\
		150 & 4506 & 60\\
		\bottomrule
	\end{tabular}
	\label{reduced_model_tab}
\end{table}


\begin{figure*}
	\centering
	\includegraphics[scale=0.35]{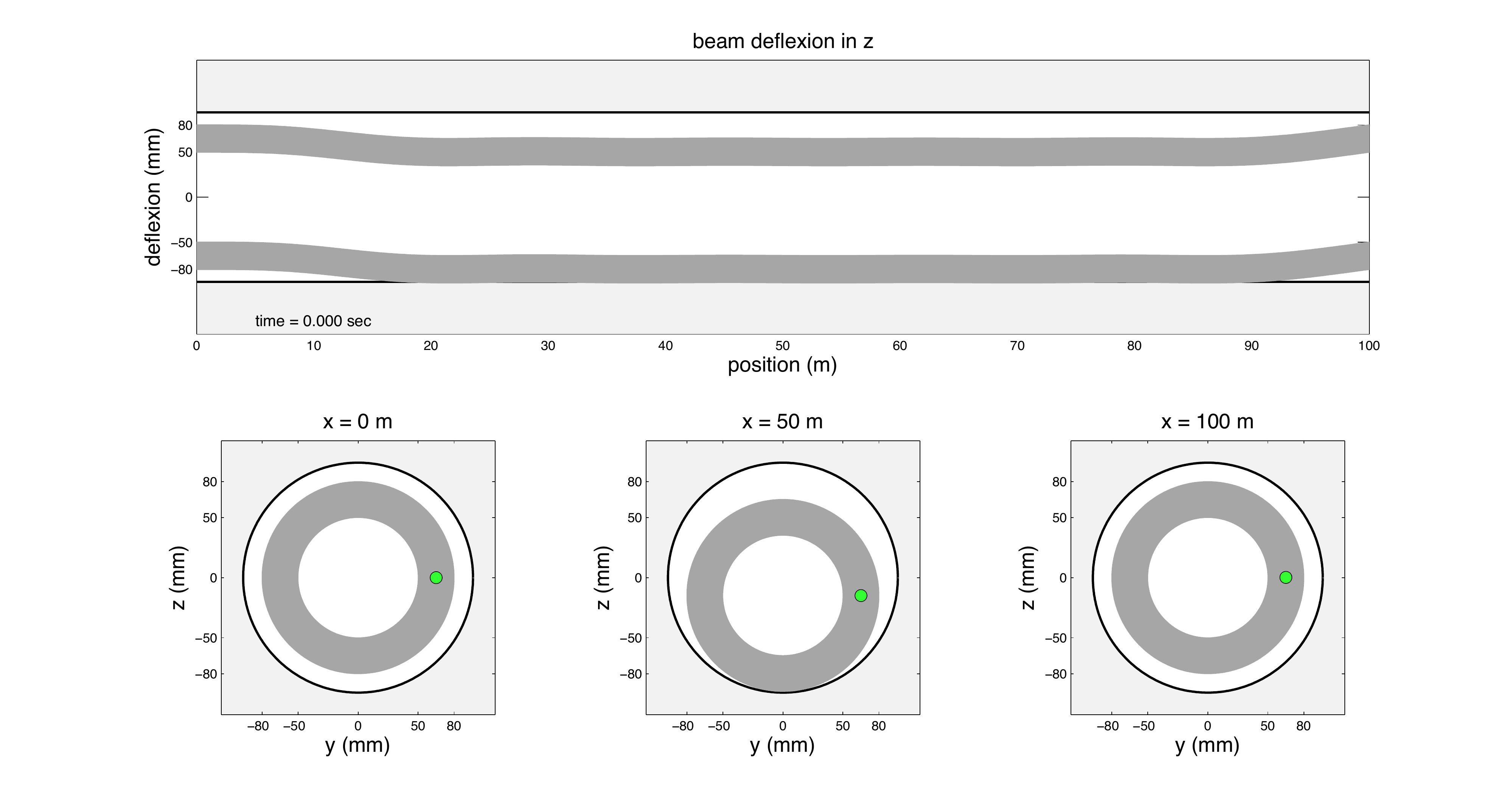}
	\caption{Illustration of static equilibrium configuration of a horizontal drillstring with $100~m$ length.}
	\label{static_equil_100m_fig}
\end{figure*}

\subsection{Calculation of static equilibrium configuration}

Before the beginning of drilling operation, the drillstring is
inserted into the borehole, without axial velocity and rotation imposed.
Due to gravitational effects, the column deflects until it reaches
a static equilibrium configuration. This configuration can be calculated
by temporal integration of the dynamical system defined by
Eqs.(\ref{reduced_galerkin_eq}) and (\ref{reduced_galerkin_ic_eq}),
assuming zero initial conditions, i.e., $\Omega = 0~ rad/s$, and
$V_0 = 0~m/s$. In this way, after a short transient, the system
reaches static equilibrium and remains in this configuration indefinitely.

An illustration of this equilibrium configuration,
for a $100~m$ long column is presented
in Figure~\ref{static_equil_100m_fig}. In this illustration,
one can see the mechanical system sectioned by the plane $y=0~m$,
as well as by the planes $x=\{0, 50, 100\}~m$. A visual
inspection clearly indicates that this equilibrium is stable.
Moreover, as this equilibrium configuration is the initial state of
a real system, it will be used as initial condition 
in all other simulations reported bellow.

An animation which illustrates the beam static
equilibrium calculation can be seen in Online Resource~1.


\subsection{Drill-bit nonlinear dynamic behavior}

\begin{figure}
	\centering
	\includegraphics[scale=0.45]{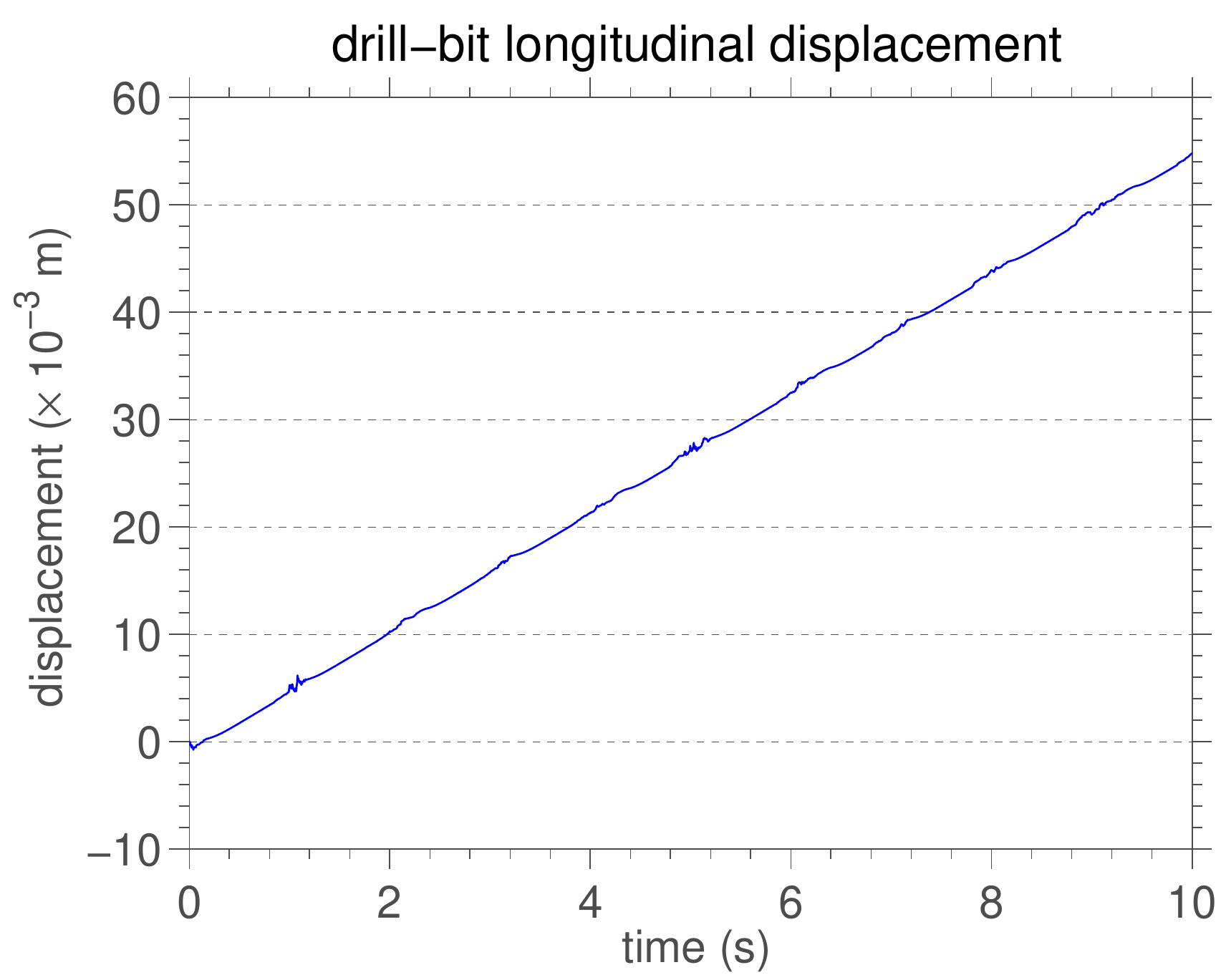}\\
	\vspace{2mm}
	\includegraphics[scale=0.45]{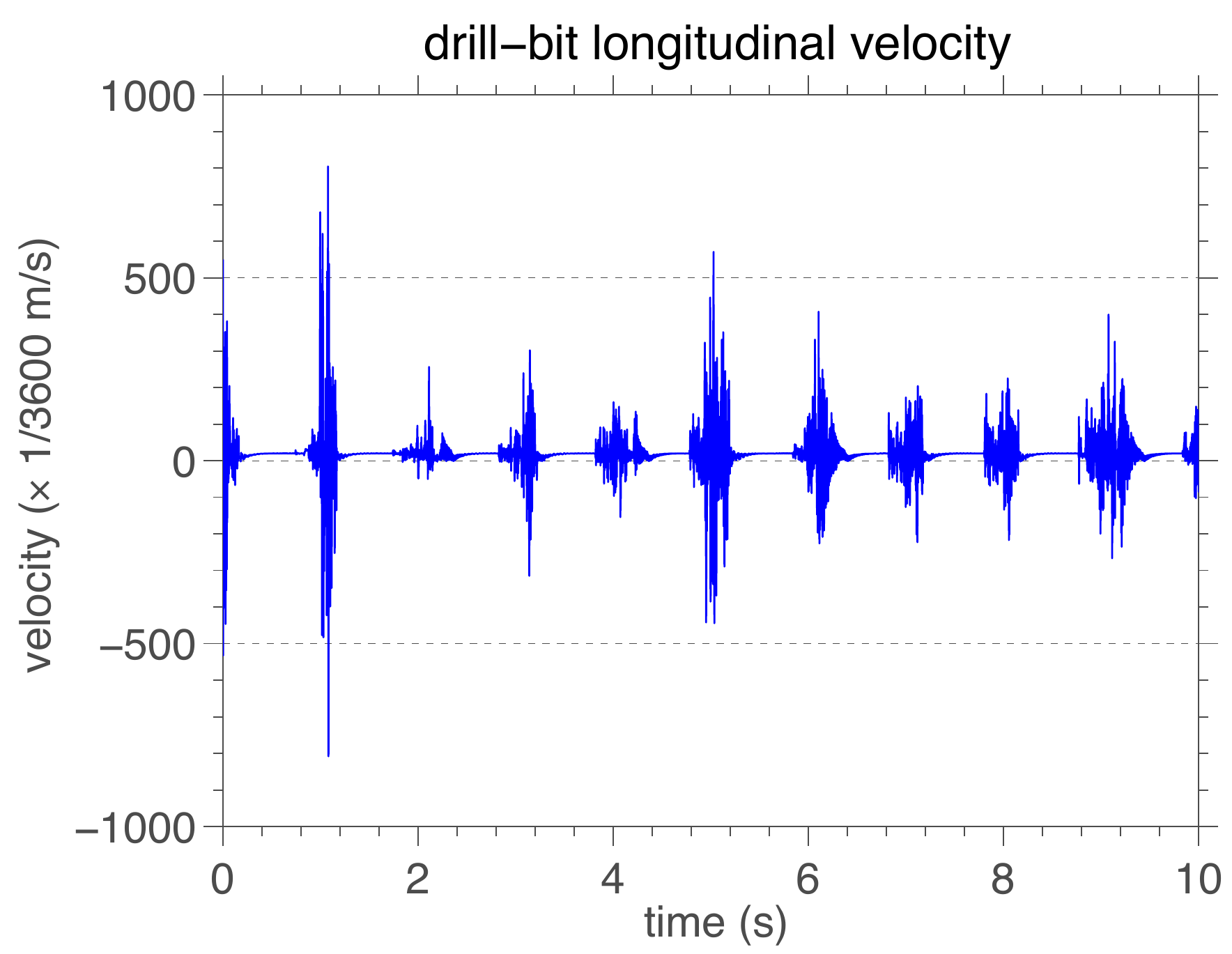}
	\caption{Illustration of drill-bit displacement (top) and drill-bit velocity (bottom).}
	\label{drill-bit_disp_velo}
\end{figure}

The drill-bit longitudinal displacement and velocity, can be seen in
Figure~\ref{drill-bit_disp_velo}. For practical reasons, some scaling
factors were introduced in the units of measure of these quantities.
They allow one to read the displacement in ``millimeter", and the
velocity in ``meters per hour". Accordingly, it is noted that,
during the interval of analysis, the column presents an advance
in forward direction with very small axial oscillations
for displacement. The axial oscillations for velocity curve are
more pronounced, and correspond to the vibration mechanism known as
\emph{bit-bounce}, where the drill-bit loses contact with the soil and then
hits the rock abruptly. This phenomenon, which is widely observed in real systems
\citep{spanos2003p85}, presents itself discreetly in the case analyzed.
Note that the velocity exhibits a mean value of
$19.36$ ``meters per hour", close to the velocity $V_0= 20$ ``meters per hour",
which is imposed on the beam left end. Also, throughout the 
``temporal window" analyzed, one can observe packages where the drill-bit velocity
presents large fluctuations, which can reach up to 40 times the mean value.

\begin{figure}
	\centering
	\includegraphics[scale=0.45]{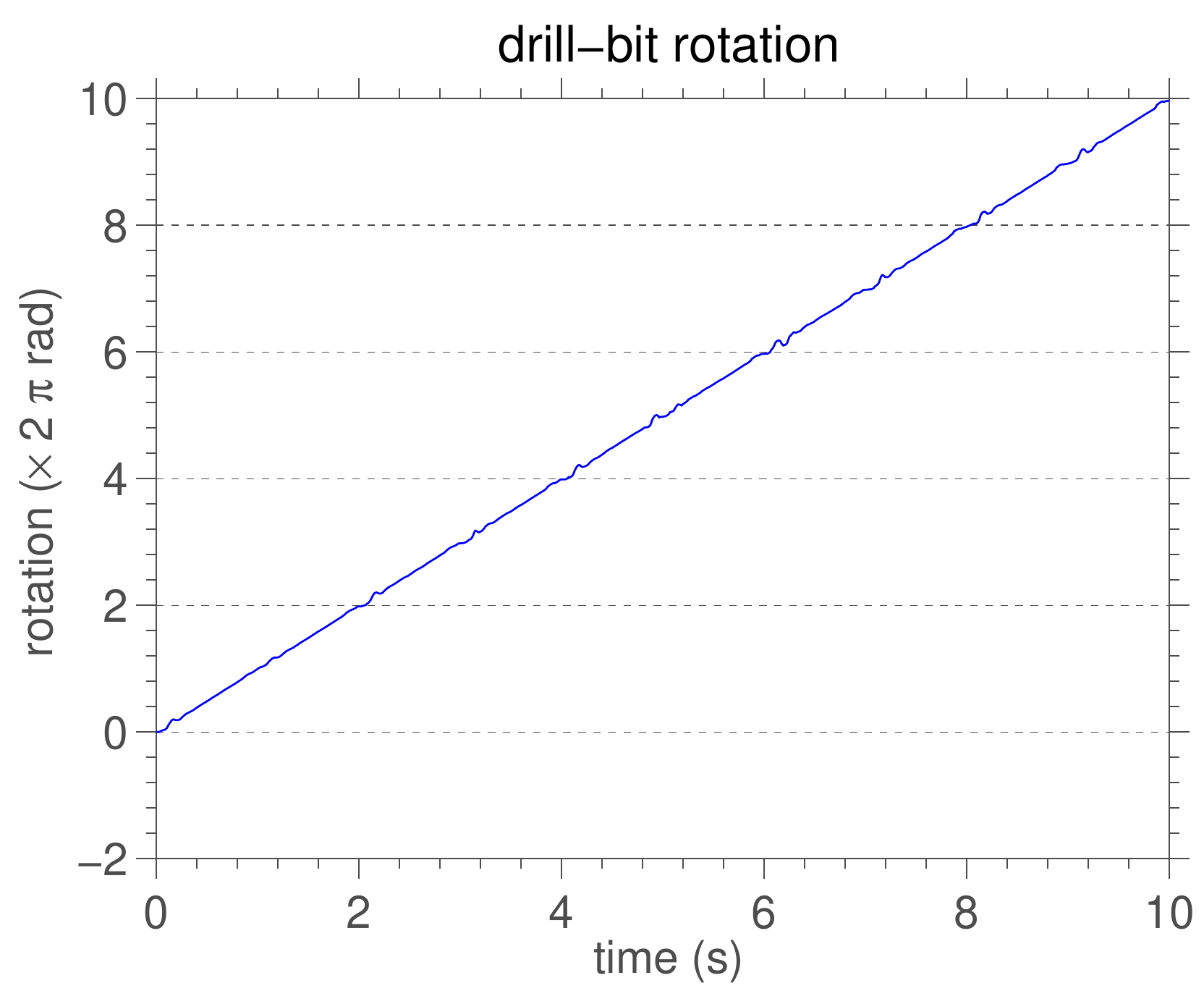}\\
	\vspace{2mm}
	\includegraphics[scale=0.45]{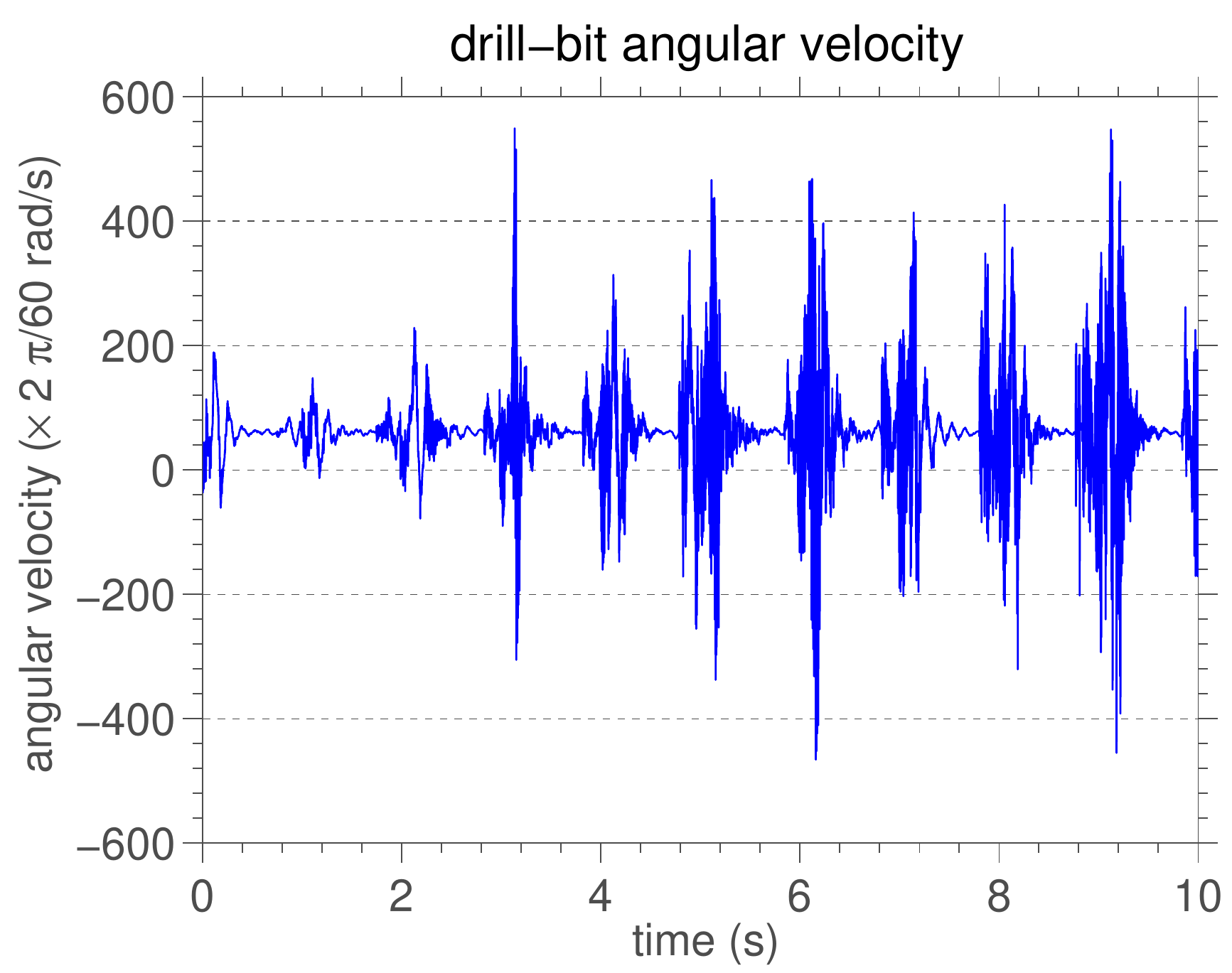}
	\caption{Illustration of drill-bit rotation (top) and drill-bit angular velocity (bottom).}
	\label{drill-bit_rot_angvelo}
\end{figure}

The drill-bit rotation and angular velocity, can be seen in
Figure~\ref{drill-bit_rot_angvelo}. Now the scale factors allow one to
read rotation in ``revolution", and angular velocity in
``revolution per minute". Thus, what is observed is a almost monotonic
rotation. However, when one looks to the angular velocity, it is possible to see packages
of fluctuations with amplitude variations that can reach up to an order of magnitude.
This indicates that drill-bit undergoes a blockage due to torsional friction, and then
it is released subtly, so that its velocity is sharply increased, in a \emph{stick-slip}
phenomenon type. This is also seen experimentally \citep{spanos2003p85} in real
drilling systems, and a serious consequence of this blockage is the reduction of
drilling process efficiency.

\subsection{Transverse nonlinear dynamics of the beam}

\begin{figure}[h!]
	\centering
	\includegraphics[scale=0.45]{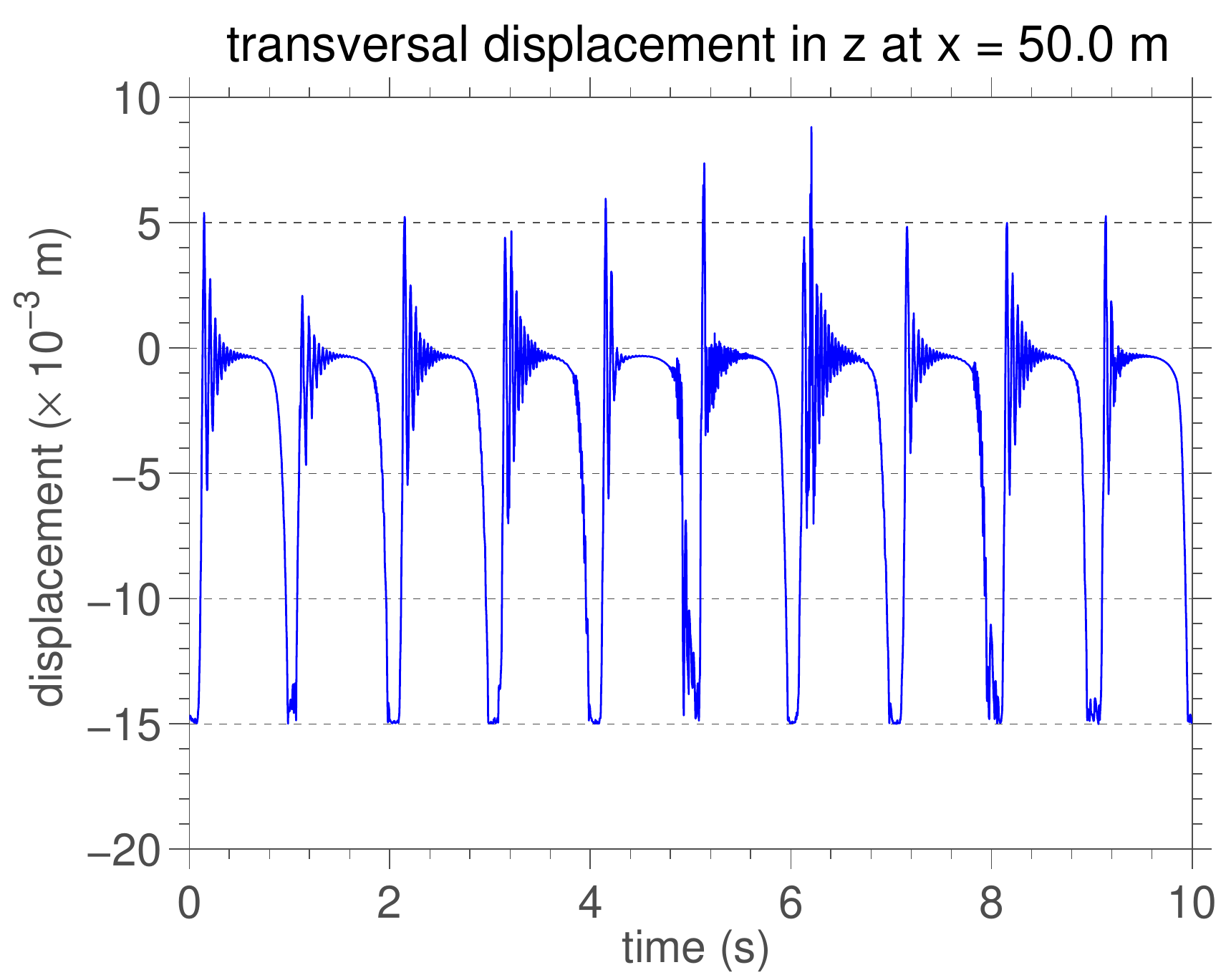}\\
	\vspace{2mm}
	\includegraphics[scale=0.45]{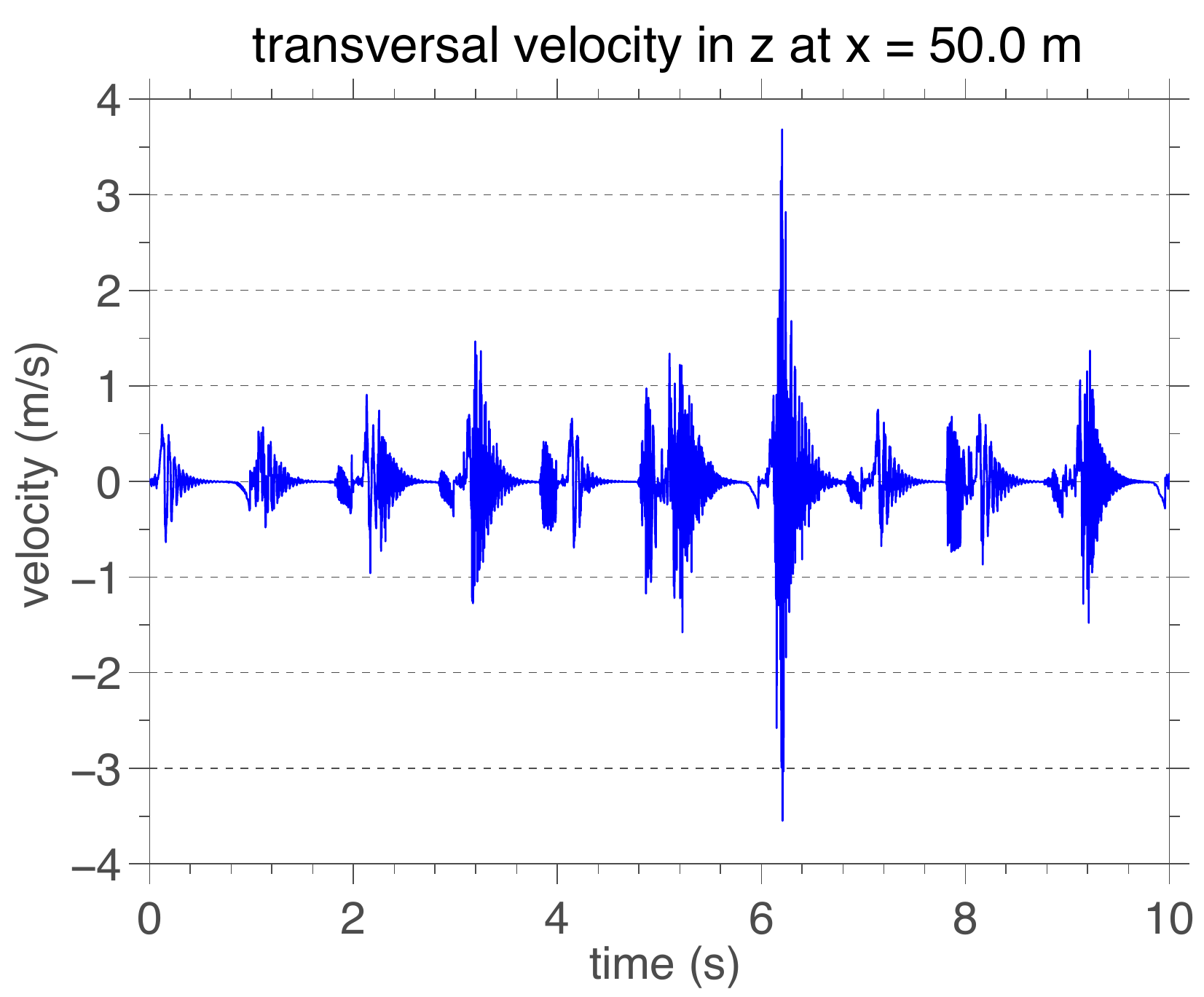}
	\caption{Illustration of transversal displacement (top) and velocity in z (bottom) when $x=50~m$.}
	\label{beam_half_disp_velo}
\end{figure}

Observing the beam cross section at $x=50~m$,
for which transversal displacement (top) and velocity (bottom)
are shown in Figure~\ref{beam_half_disp_velo}, one can see
an asymmetry of the displacement, with respect to the plane $z=0~m$.
This is due to gravity, which favors the beam to move below this plane.
Furthermore, one can note that this signal is composed of ``packages",
which has a recurring oscillatory pattern. As will be seen in section~\ref{infl_trans_impac},
these packages present a strong correlation with the number of impacts which
the mechanical system is subjected.

\begin{figure}
	\centering
	\includegraphics[scale=0.45]{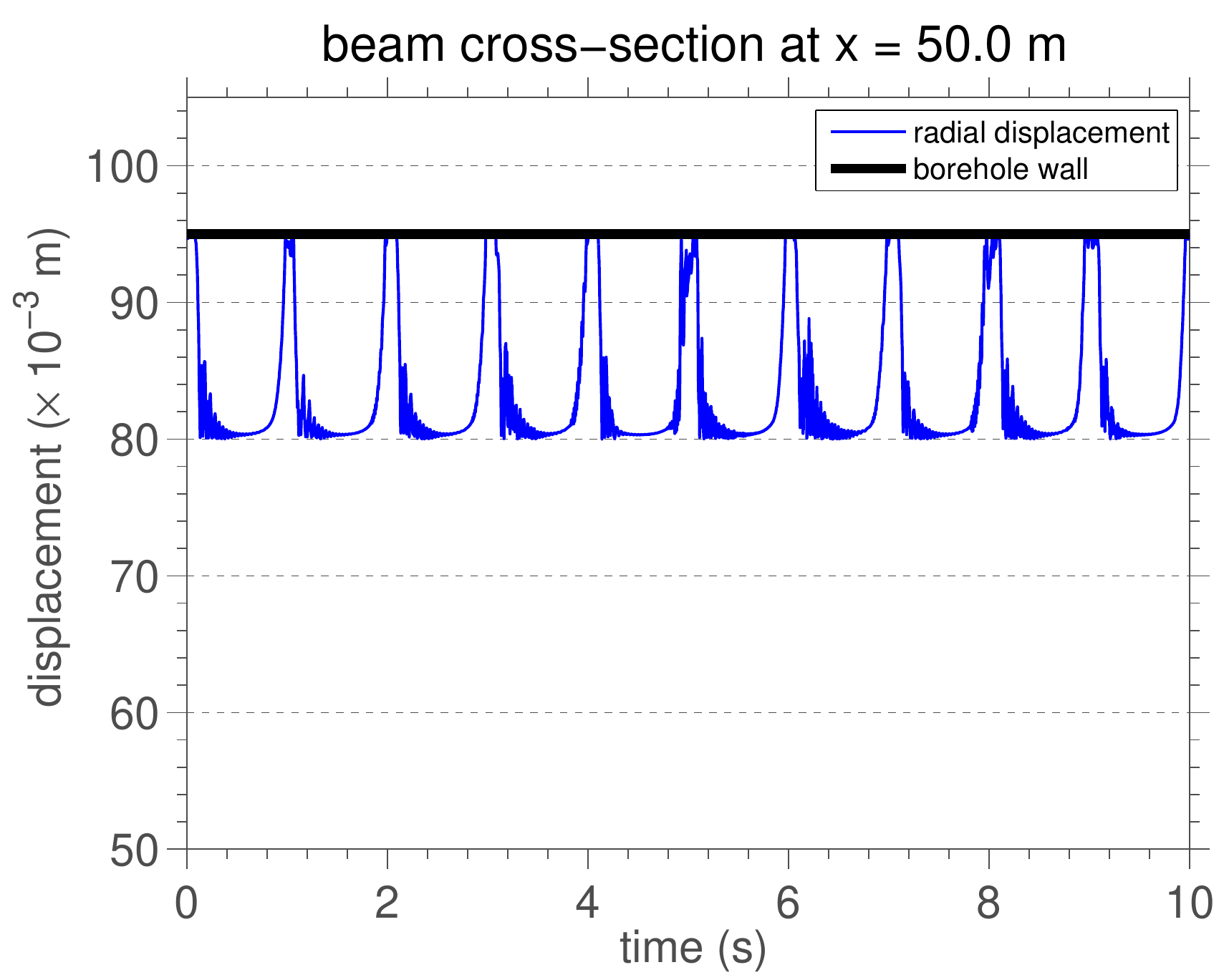}
	\caption{Illustration of beam radial displacement for $x = 50~m$.}
	\label{beam_radial_disp_x_100}
\end{figure}

The evolution of the beam cross-section radial displacement, for $x = 50~m$, 
can be seen in Figure~\ref{beam_radial_disp_x_100}, which shows that several transverse impacts
occur between drillstring and borehole wall during the drilling process.
This fact is also reported experimentally \citep{spanos2003p85}, and is an important cause
of damage for both, well and drillstring.

Note that, after an impact, the oscillations amplitudes decreases until subtly increase sharply,
giving rise to a new impact, and then the entire process repeats again.

\subsection{Influence of transverse impacts on the nonlinear dynamics}
\label{infl_trans_impac}

\begin{figure}
	\centering
	\includegraphics[scale=0.45]{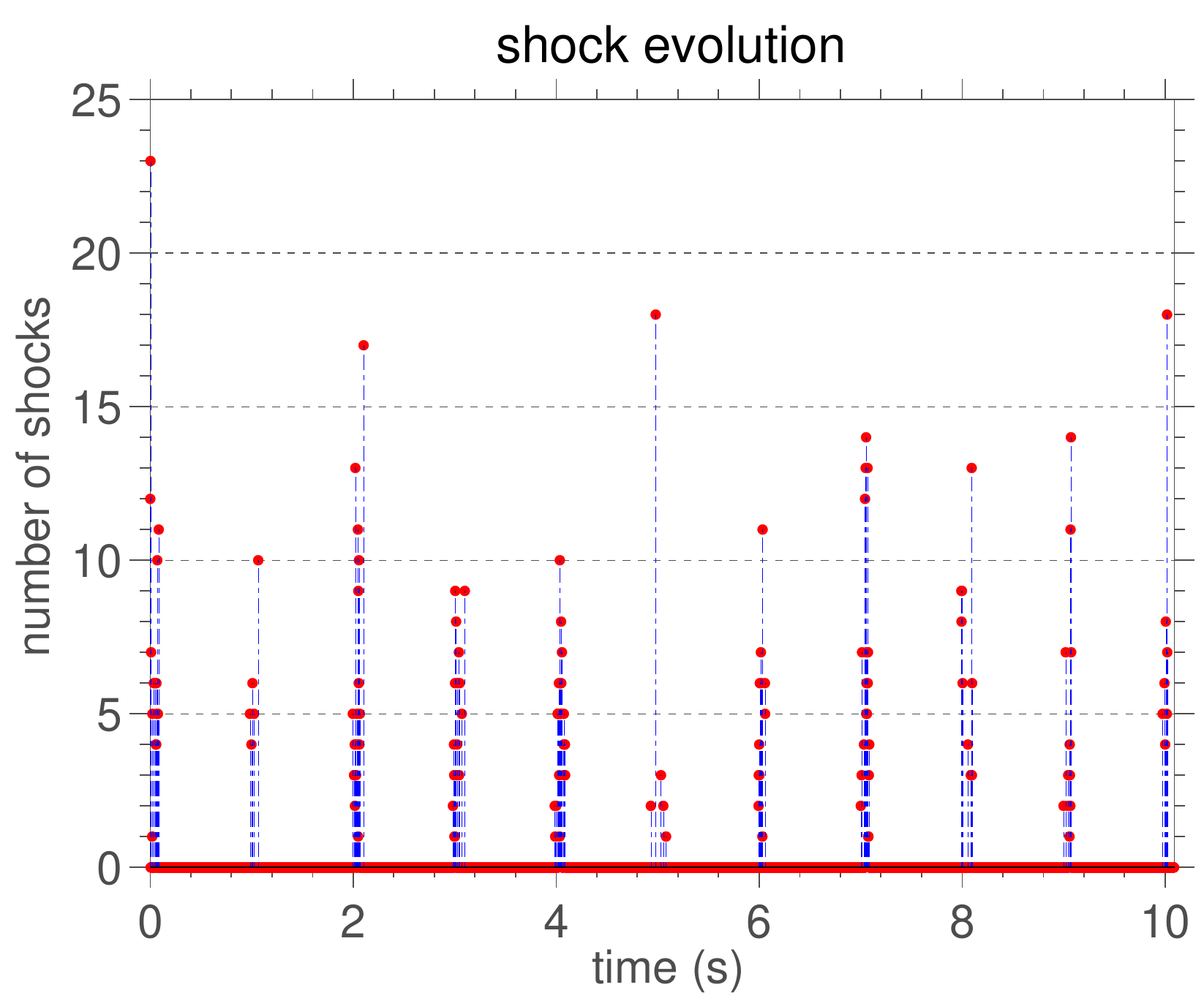}
	\caption{Illustration of the number of impacts suffered by the mechanical system as function of time.}
	\label{shock_number_per_time_fig}
\end{figure}

In Figure~\ref{shock_number_per_time_fig} it is shown the map 
$t \in \R \mapsto \texttt{number of shocks} \in \N$,
which associates for any instant $t$ the number of impacts suffered
by the mechanical system.

The ``packages of fluctuation" observed in
Figures~\ref{drill-bit_disp_velo} to \ref{beam_half_disp_velo}
correspond to transitory periods of the dynamical system,
and are highly correlated with the process of collision between
beam and borehole wall. This assertion can be verified if the reader
compares the graphs of Figures~\ref{drill-bit_disp_velo} to
\ref{beam_half_disp_velo} with the graph of Figure~\ref{shock_number_per_time_fig},
which shows the existence of ``shock packages".
The existence of a correlation is clearly evident.

Whenever there is a shock, the system ``loses it memory" about previous
dynamic behavior, and undergoes a new transient period until reach a steady
state again. This behavior is repeated 11 times in the ``temporal window"
analyzed.

\begin{figure}
	\centering
	\includegraphics[scale=0.45]{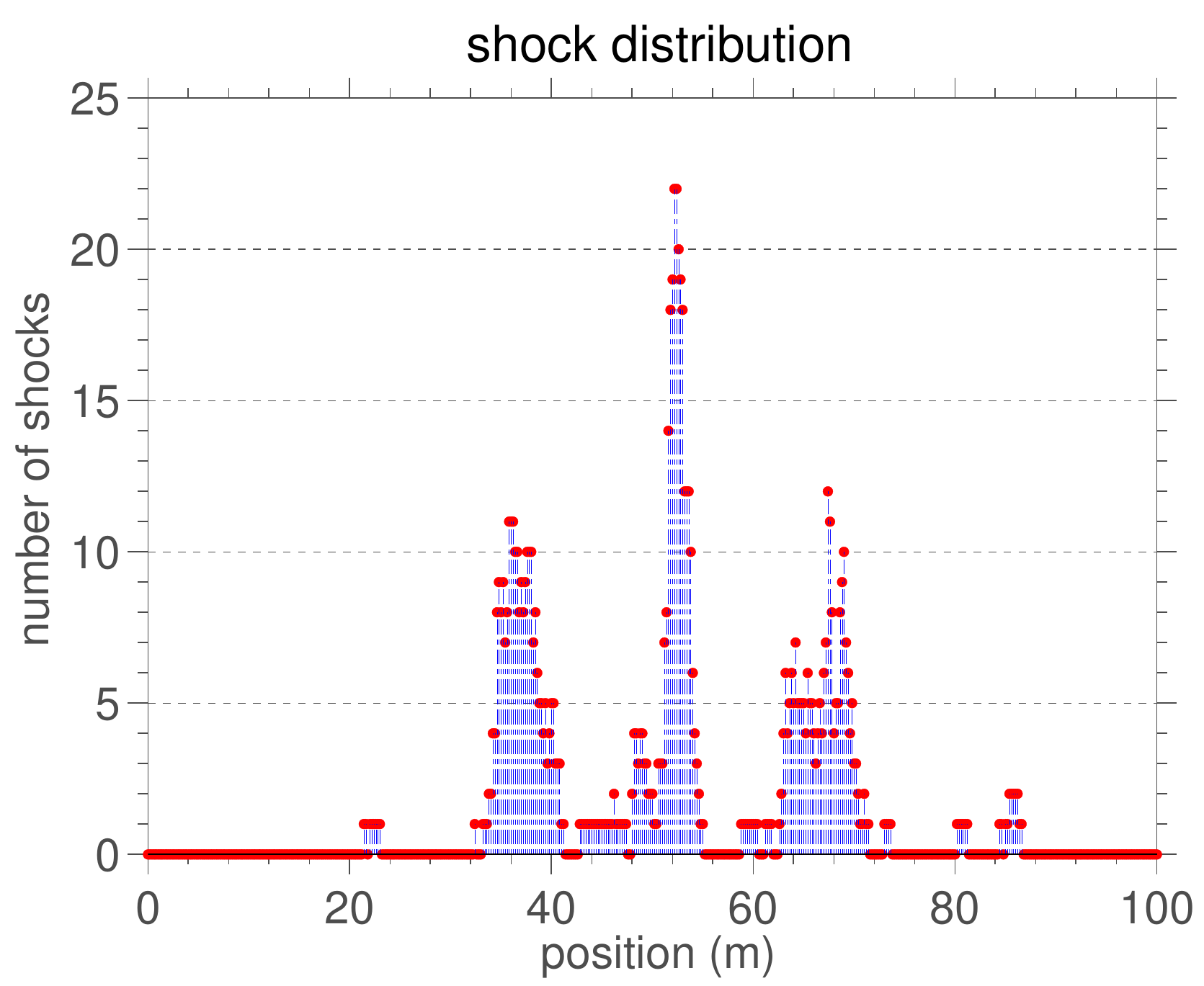}
	\caption{Illustration of the number of impacts suffered by the mechanical system as function of position.}
	\label{shock_number_per_position_fig}
\end{figure}

Regarding the distribution of impacts along the beam,
the map $x \in [0,L] \mapsto \texttt{number of shocks} \in \N$,
which associates for any position $x$ the number of impacts suffered
by the mechanical system, is shown in Figure~\ref{shock_number_per_position_fig}.
It is clear that impacts do not occur near the beam ends. This is natural due to the
restrictions of movement imposed by the boundary conditions.

\begin{figure*}
	\centering
	\subfigure[$t=2.145~s$]{\includegraphics[scale=0.35]{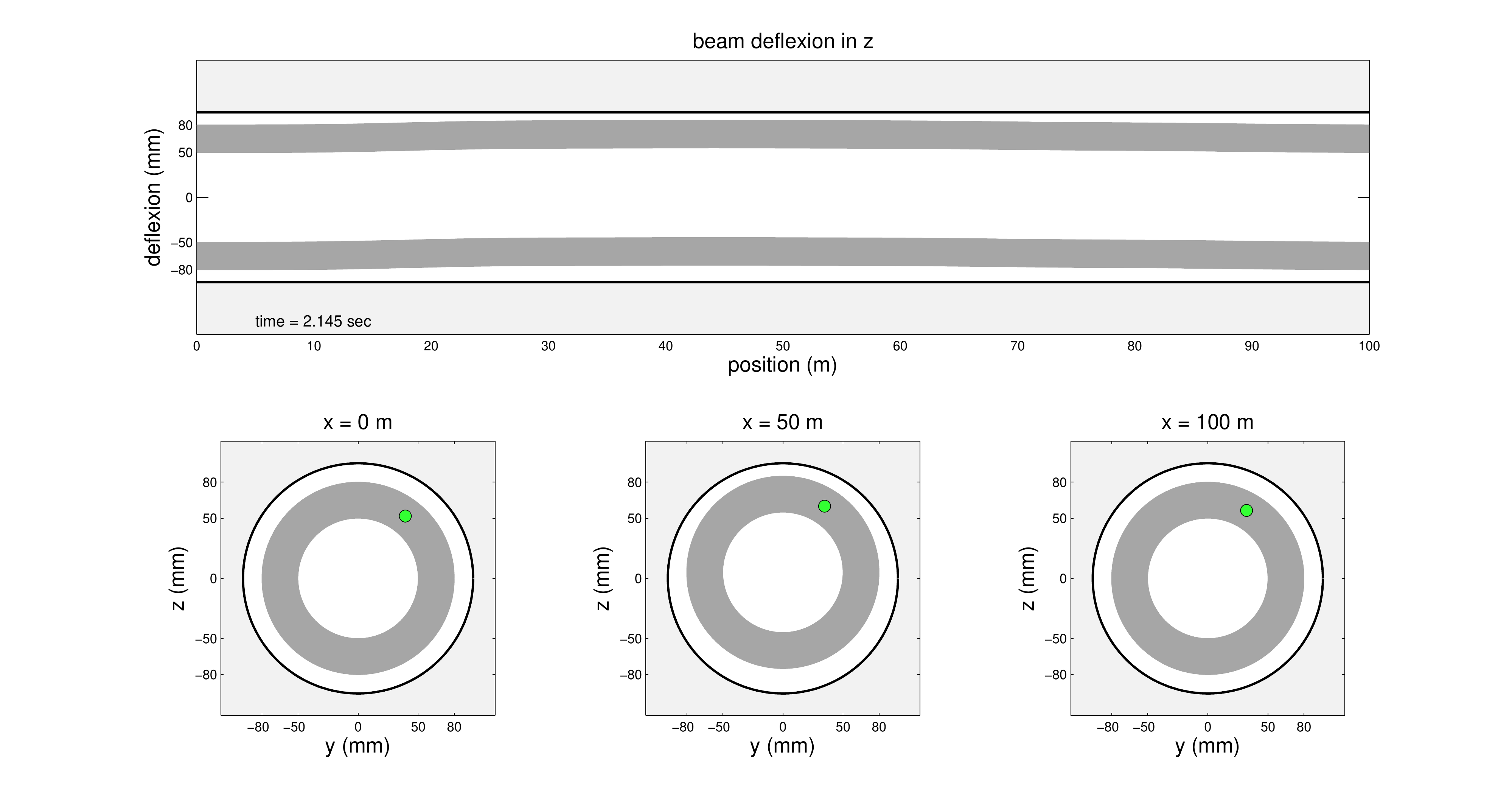}}
	\subfigure[$t=4.932~s$]{\includegraphics[scale=0.35]{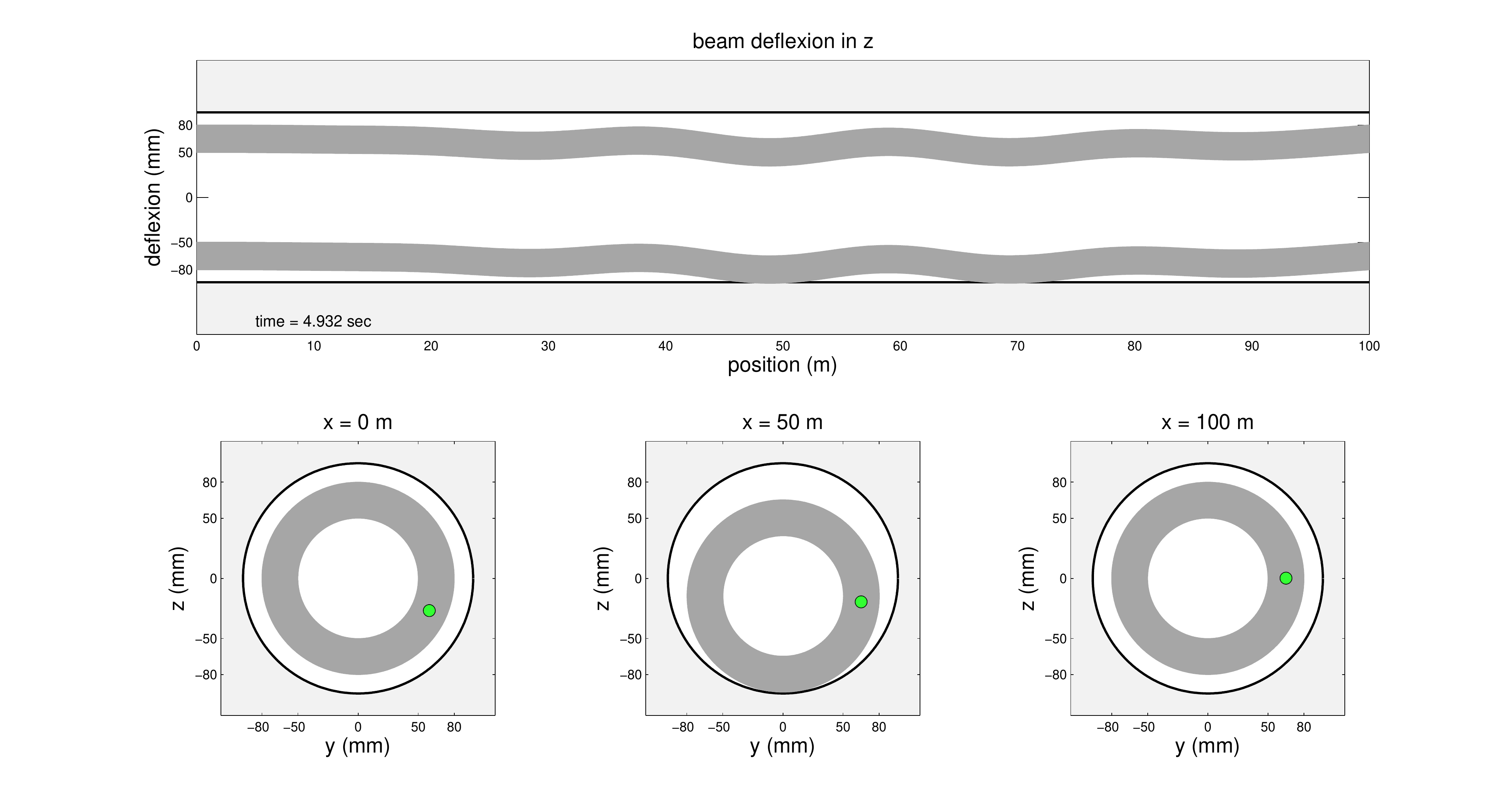}}
	\subfigure[$t=6.214~s$]{\includegraphics[scale=0.35]{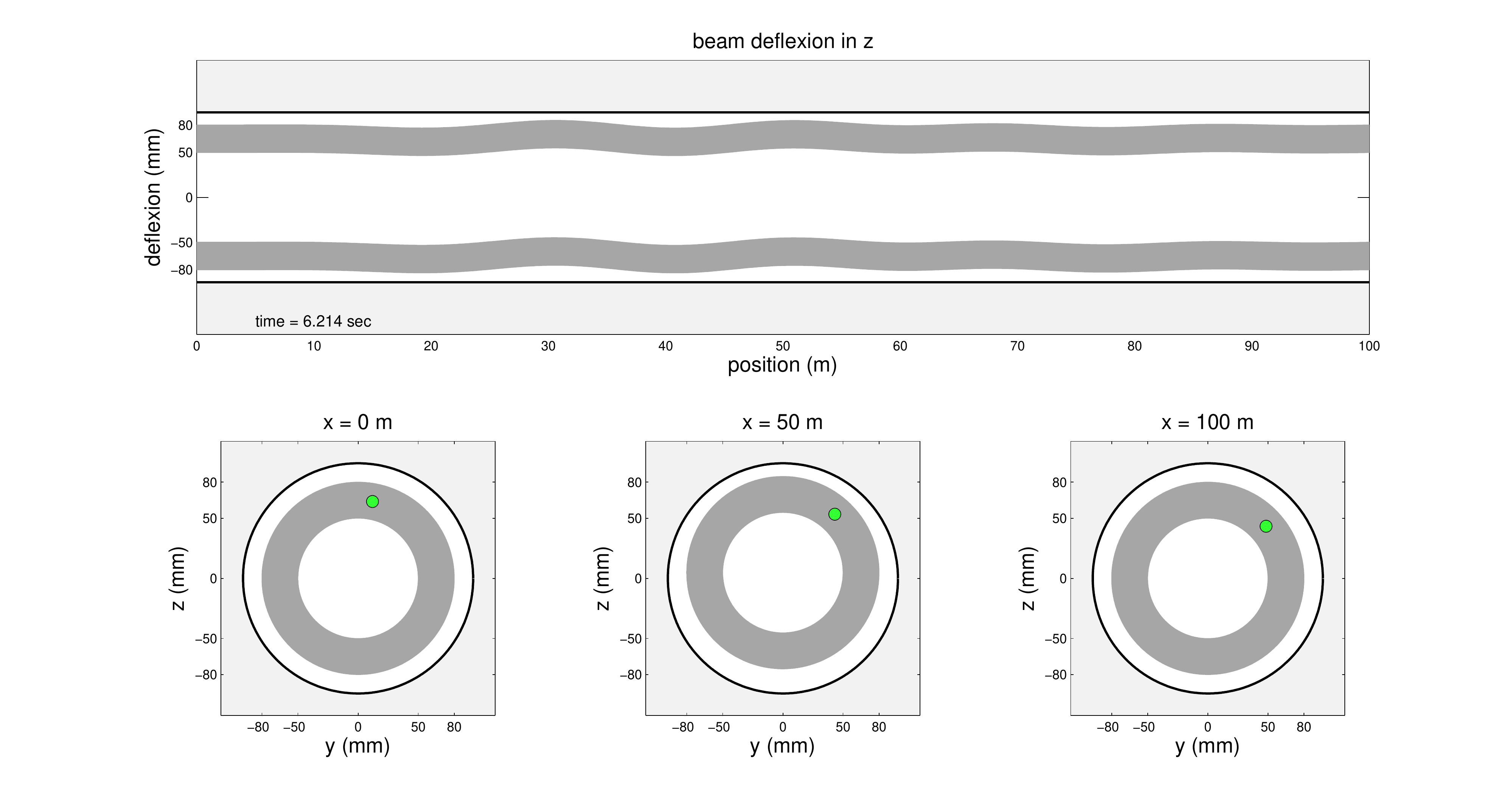}}
	\caption{Illustration of the mechanical system, for several instants, sectioned by the planes $y=0~m$,
	and $x=\{0,50,100\}~m$.}
	\label{mech_system_section_y_fig}
\end{figure*}

The impacts between drillstring and borehole wall generate
nonlinear elastic deformations in the beam, but without residual deformation effects.
In this contact also occurs energy dissipation, due to the normal shock, and the
torsional friction induced by beam rotation. These mechanical contacts
also activate flexural modes of vibration associated to high natural frequencies,
so that the mechanical system assumes complex spatial configurations, as can be seen,
for several instants, in Figure~\ref{mech_system_section_y_fig}.

It is also very clear from Figure~\ref{mech_system_section_y_fig} that,
the mechanical contacts between beam and borehole wall,
do not occur all the time among discrete points, they can also be seen
along continuous line segments.

For a qualitative illustration of the nonlinear dynamics, the reader
can see Online Resource~2.



\subsection{Spectral analysis of the nonlinear dynamics}

All signals presented above, that are associated with the
mechanical system response, have stochastic characteristics.
Thereby, for a good understanding of them, it is necessary to
analyze their spectral content through the power spectral density
(PSD) function \cite{oppenheim2009}.

The PSDs that are presented in this section (magenta line) were estimated using
the periodogram method \cite{oppenheim2009}, and the smooth curves (blue line)
appearing were obtained by a filtering process, using a Savitzky-Golay filter
\cite{savitzky1964p1627}. The PSDs are measured in $\mbox{dB/Hz}$,
where the intensity of reference is adopted as being equal to one.

\begin{figure}
	\centering
	\includegraphics[scale=0.45]{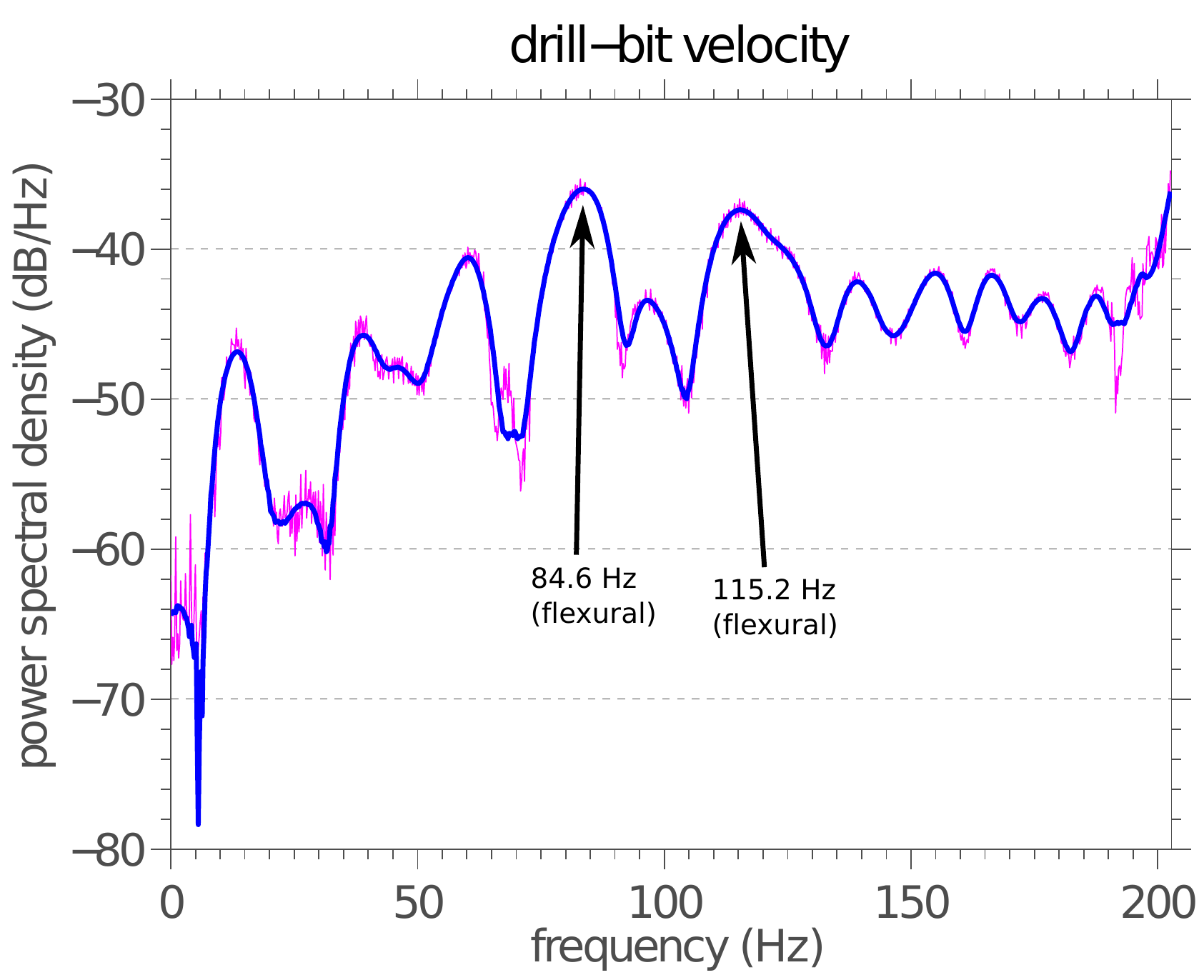}\\
	\vspace{2mm}
	\includegraphics[scale=0.45]{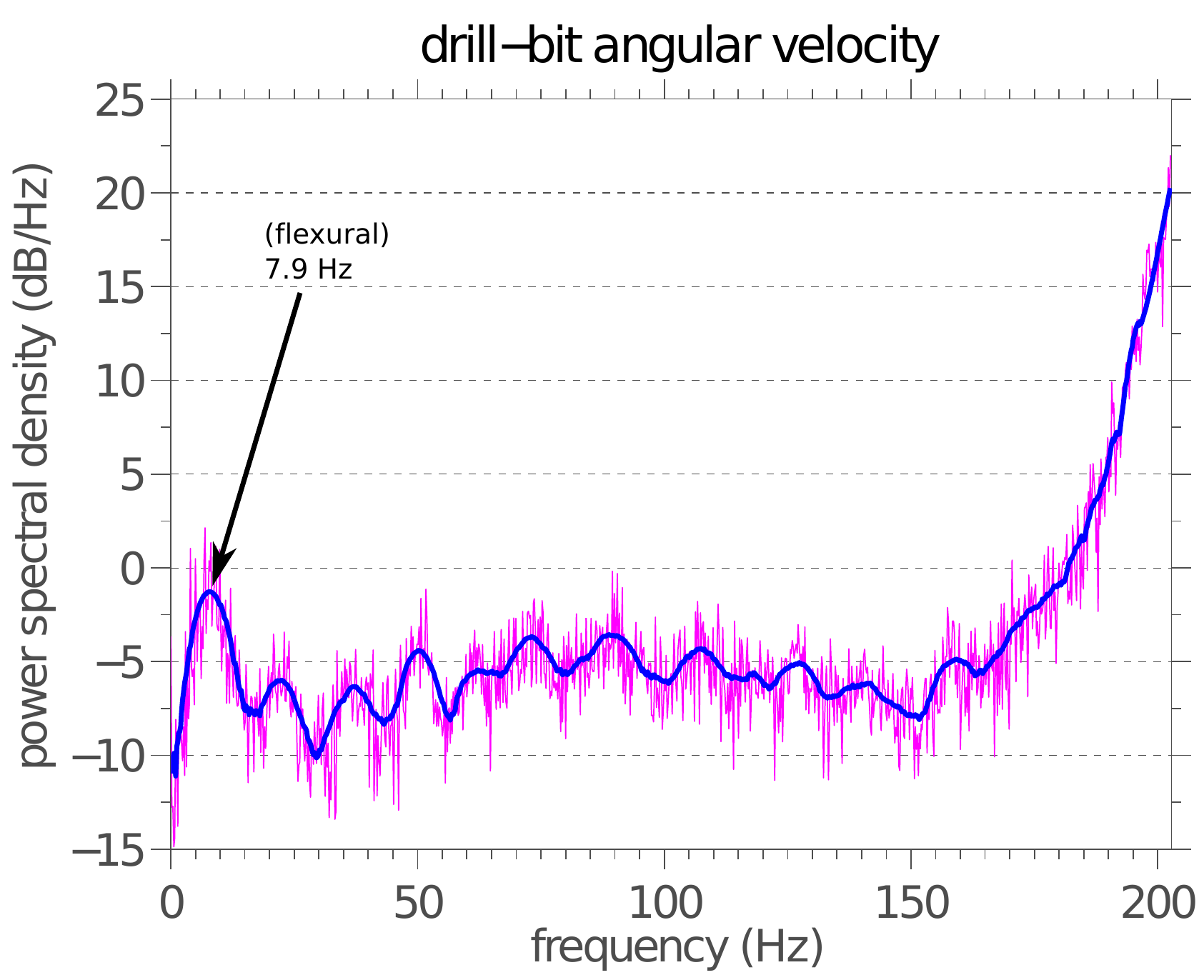}
	\caption{Illustration of power spectral density functions of drill-bit velocity (top) and angular velocity (bottom).}
	\label{psd_velo_angvelo_bit_fig}
\end{figure}

An illustration of PSD functions of drill-bit velocity and angular velocity
is show in Figure~\ref{psd_velo_angvelo_bit_fig}. One can note that,
for velocity, the two peaks of highest amplitude correspond
to the frequencies $84.55~Hz$, and $115.20~Hz$, respectively.
These frequencies are very close to the flexural frequencies $83.65~Hz$,
and $114.27~Hz$, so that drill-bit axial dynamics is controlled by the
transversal mechanisms of vibration. Furthermore, with respect to angular
velocity, it is noted a peak standing out in relation to the others.
This peak is associated with $7.92~Hz$ frequency,
which is very close to the flexural frequency $7.63~Hz$.

\begin{figure}
	\centering
	\includegraphics[scale=0.45]{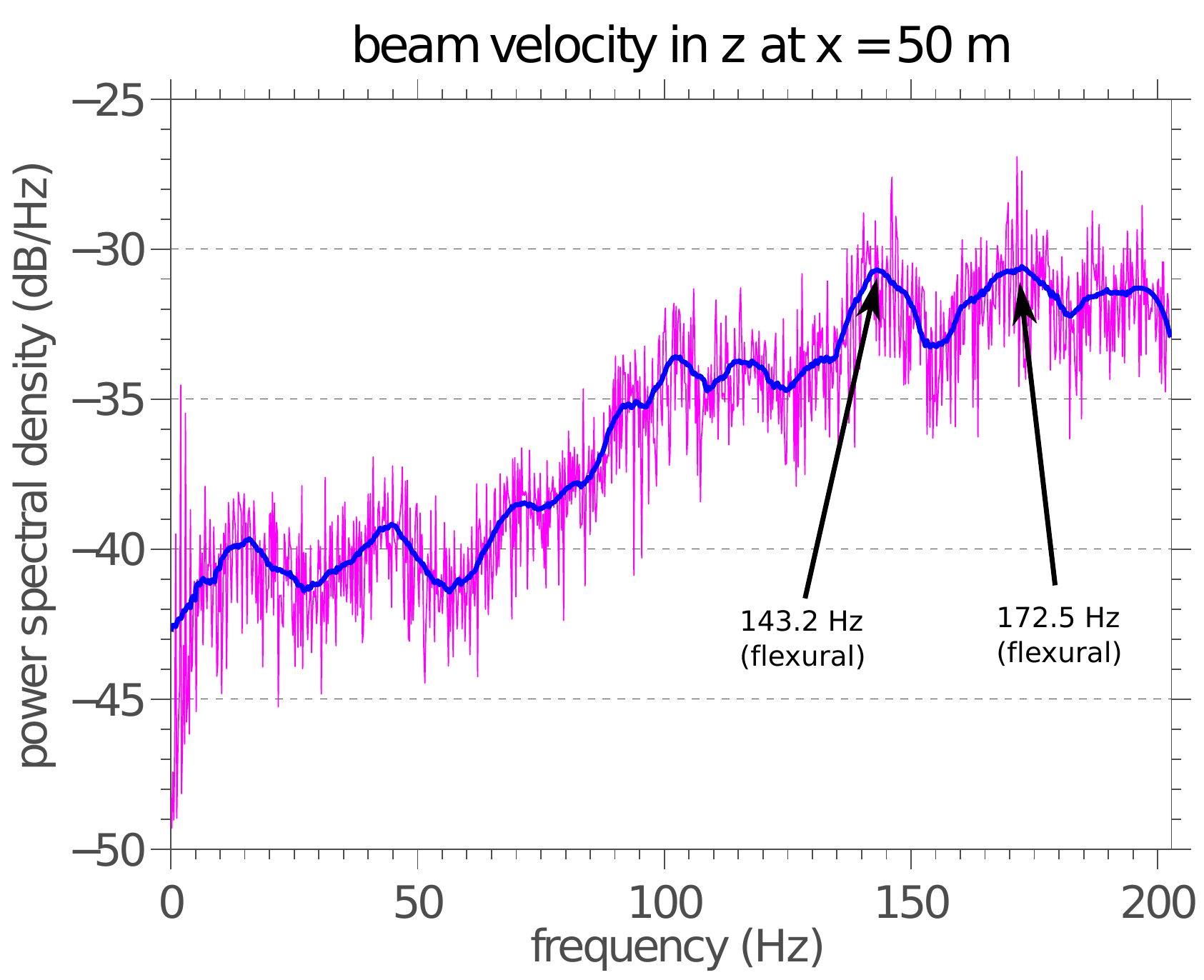}\\
	\vspace{2mm}
	\includegraphics[scale=0.45]{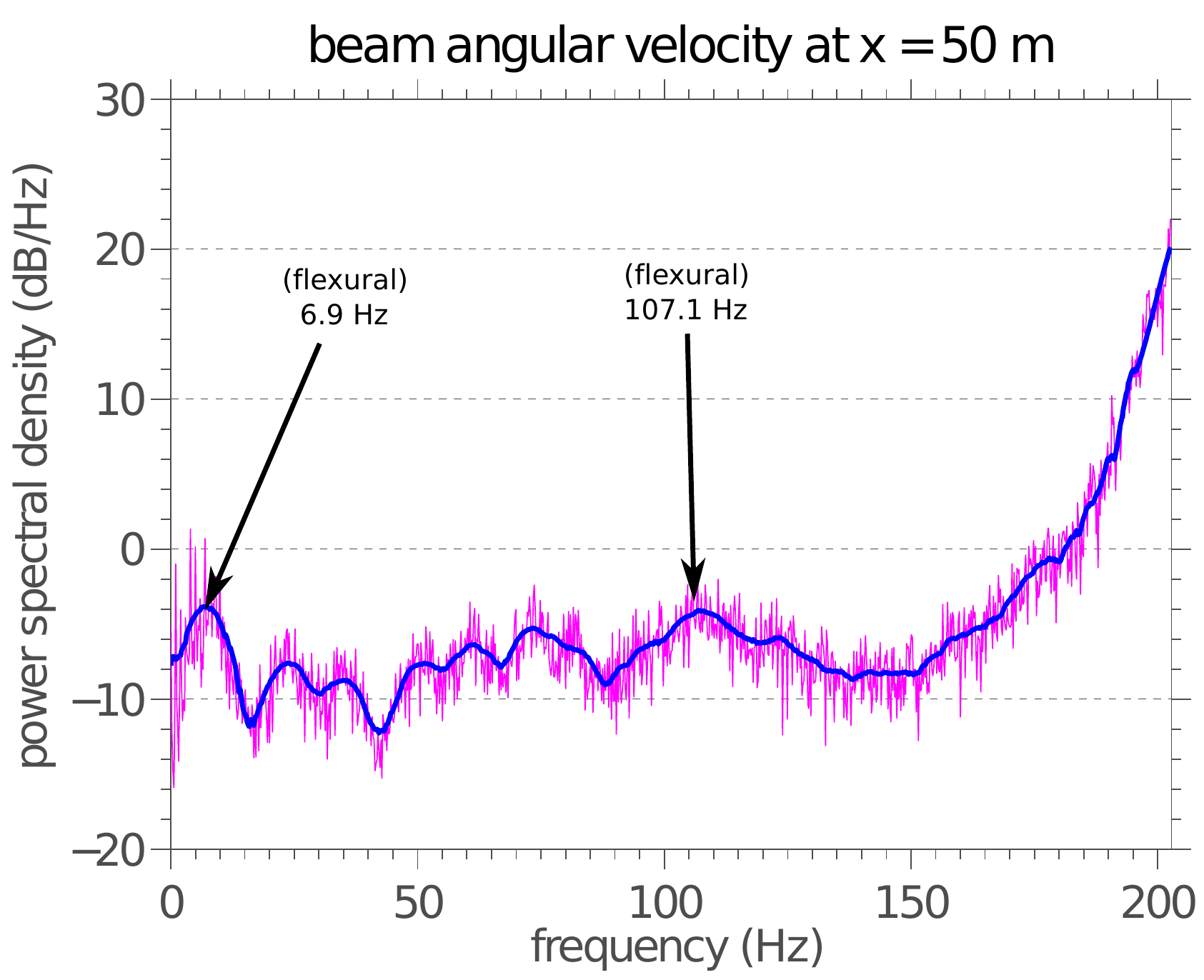}
	\caption{Illustration of power spectral density functions of beam transversal velocity in z (top) and
	angular velocity around x (bottom) when $x=50~m$.}
	\label{psd_w_dot_angvelo_x_half_fig}
\end{figure}

In Figure~\ref{psd_w_dot_angvelo_x_half_fig} the reader can see
an illustration of PSD functions of beam transversal velocity in z and
angular velocity around x when $x=50~m$. The two peaks of highest
amplitude, for velocity in z, correspond to the frequencies
$143.20~Hz$, and $172.50~Hz$, respectively. These frequencies are
close to the flexural frequencies $144.94~Hz$, and $174.07~Hz$,
which indicates that lateral vibrations in z, when $x = 50~m$,
are induced by the transversal vibration mechanism. In addition,
in what concerns angular velocity around x, the two peaks of largest
amplitude are associated to the frequencies $6.93~Hz$, and $107.10~Hz$,
respectively close to the flexural frequencies $6.59~Hz$, and $106.18~Hz$.

\begin{figure}
	\centering
	\includegraphics[scale=0.45]{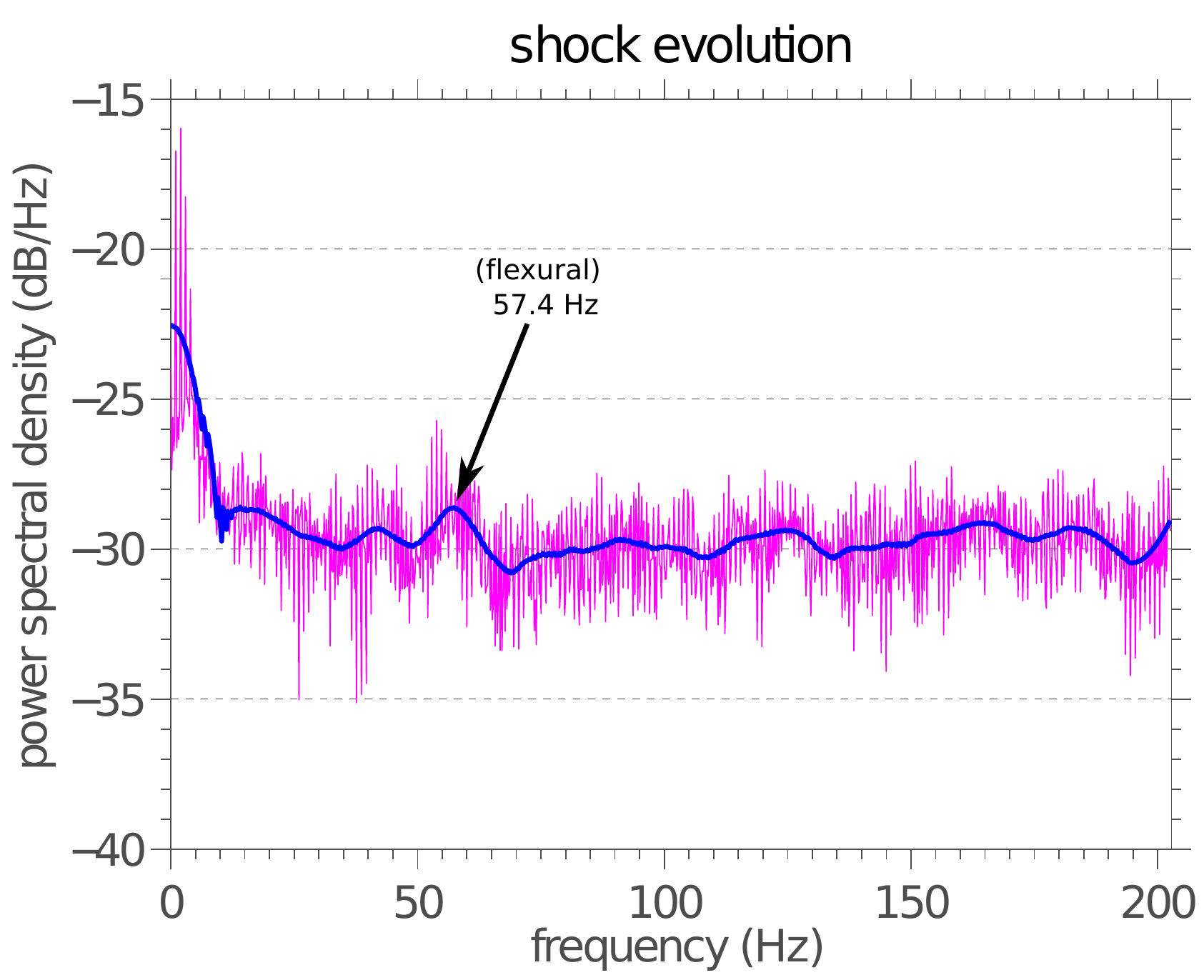}
	\caption{Illustration of power spectral density function of number of shocks per unit of time.}
	\label{psd_shock_per_time_fig}
\end{figure}

According to Figure~\ref{psd_shock_per_time_fig}, flexion is the primary mechanism
of vibration that causes beam and borehole wall impacts, since the
highest peak of the PSD shown in this figure is associated with frequency $57.42~Hz$,
which is close to the flexural frequency $57.77~Hz$. This result is consistent with 
the intuition and with what is reported in the literature of drillstrings in vertical 
configuration \cite{spanos2003p85}.

\subsection{Analysis of drilling process efficiency}

The drilling process efficiency is defined as

\begin{equation}
		\EffOp= \frac{\int_{t_0}^{t_f} \PowerOp_{out} \, dt }{\int_{t_0}^{t_f} \PowerOp_{in} \, dt},
\end{equation}

\noindent
where $\PowerOp_{out}$ is the useful (output) power used in the drilling process,
and $\PowerOp_{in}$ is the total (input) power injected in the system.
The output power is due to drill-bit movements of translation
and rotation so that

\begin{equation}
		\PowerOp_{out} = \dot{u}_{bit}^{+} \left(-F_{\BR} \right)^{+} ~ + ~
							   		 \omega_{bit}^{+} \left( -T_{\BR} \right)^{+},
\end{equation}

\noindent
where the upper script $^{+}$ means the function positive part.
The input power is defined as

\begin{equation}
		\PowerOp_{in} = \dot{u}(0,t)^{+} \, (- \lambda_{1})^{+} ~+~
									\dot{\theta}_x(0,t)^{+} \, (-\lambda_{4})^{+},
\end{equation}

\noindent
where the first and the fourth Lagrange multipliers, respectively, represent
the drilling force and torque on the beam origin. The reason for considering,
in the above definitions, only functions positive part is that negative
powers do not contribute to the drilling process.

\begin{figure}
	\centering
	\includegraphics[scale=0.45]{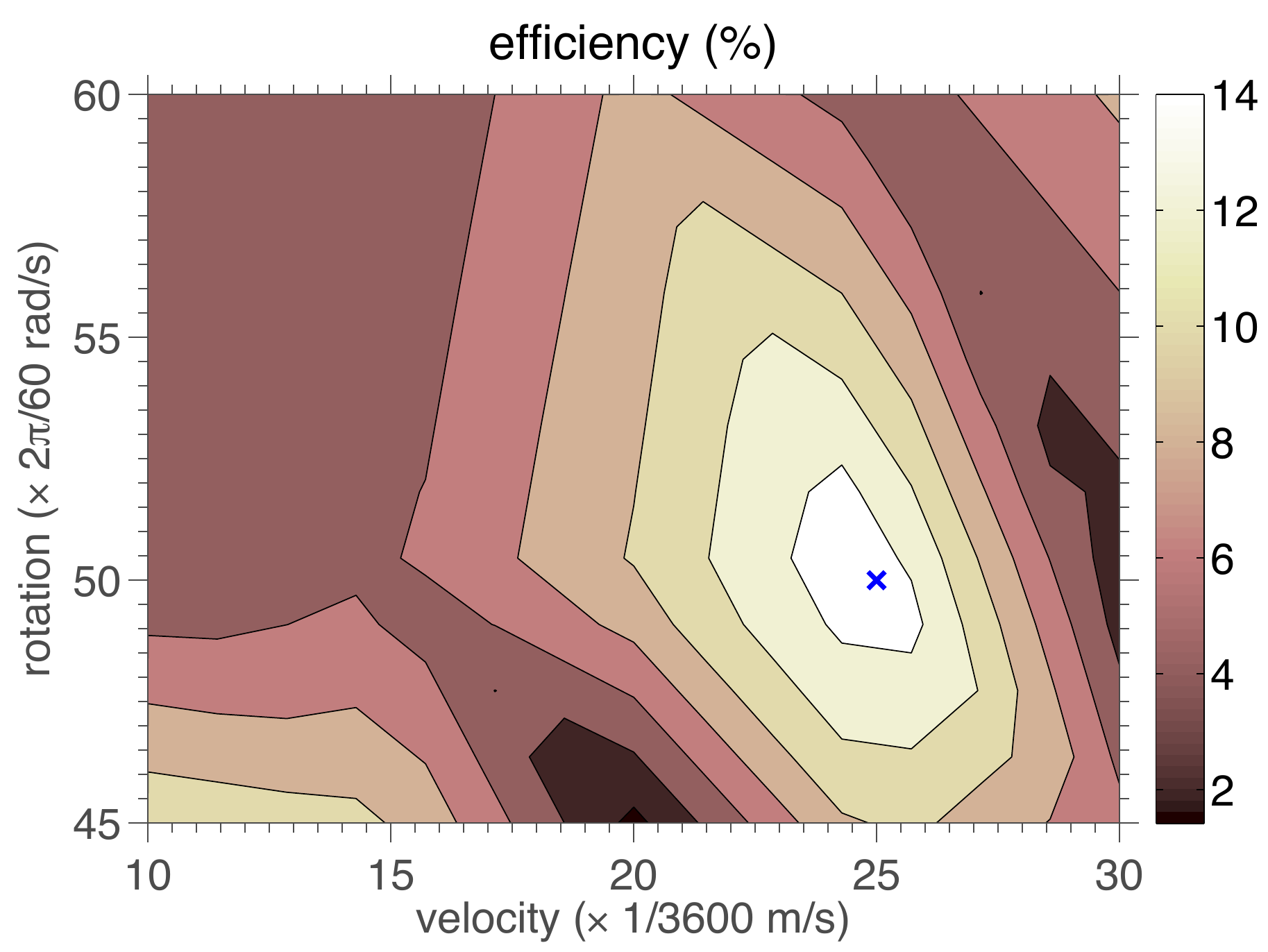}
	\caption{Illustration of efficiency function contour plot, for an
	``operating window" defined by $1/360~m/s \leq V_0 \leq 1/120~m/s$
	and $3\pi/2~rad/s \leq \Omega \leq 2\pi~rad/s$. The maximum is indicated
with a blue cross.}
	\label{eff_func_fig}
\end{figure}

One can observe the contour map of $\EffOp$,
for an ``operating window" defined by $1/360~m/s \leq V_0 \leq 1/120~m/s$
and $3\pi/2~rad/s \leq \Omega \leq 2\pi~rad/s$,
in Figure~\ref{eff_func_fig}. Note that, by operating window of a drillstring,
one means the subset of $\R^{2}$ that provides acceptable values for the pair $(\Omega, \, V_0)$.
In order to facilitate the results interpretation,
some scaling factors were introduced in the units of measure. They allow one to read
velocity in ``meters per hour" and rotation in ``rotation per minute".

Accordingly, it can be noted in Figure~\ref{eff_func_fig} that the
optimum operating condition is obtained at the point
$(V_0,\Omega) = (1/144~m/s,5\pi/3~rad/s)$, which is indicated
with a blue cross in the graph. This point corresponds to
an efficiency of approximately $16\%$. Suboptimal operation conditions occur
in the vicinity of this point, and some points near the ``operating window"
boundary show lower efficiency.

\subsection{Optimization of drillstring rate of penetration}

In order to optimize the drilling process of an oil well in horizontal
configuration, it is necessary to maximize the drillstring ROP into the soil.

The instantaneous rate of penetration is given by the function
$\dot{u}_{bit}(t)$, defined for all instants of analysis. Meanwhile,
only contributes to the column advance, the positive part of this
function $\dot{u}_{bit}^{+}(t)$. In addition, as objective function,
it is more convenient to consider a scalar function.
Thus, the temporal mean of $\dot{u}_{bit}^{+}(t)$ is adopted as rate of
penetration, and, consequently, objective function of the optimization
problem

\begin{equation}
		\texttt{rop} (\Omega,\, V_0) =
		\frac{1}{t_f - t_0} \int_{t=t_0}^{t_f} \dot{u}_{bit}^{+}(t) \, dt.
		\label{rop_def}
\end{equation}

Furthermore, respect the material structural limits 
is indispensable to avoid failures in drillstring during the drilling process. 
For this reason, von Mises criterion of failure is considered, where 
it is established that, for all pairs $(\Omega,\, V_0)$ 
in the ``operating window", one has

\begin{equation}
		\texttt{UTS} - \underset{t_0 \leq t \leq t_f}{\underset{0 \leq x \leq L}{\max}} \left\{\sigma_{VM} (V_0,\, \Omega,\, x,\, t) \right\} \geq 0,
		\label{von_mises_stress_determ}
\end{equation}

\noindent
where $\texttt{UTS}$ is the material ultimate tensile strength, and
$\sigma_{VM}$ is the von Mises equivalent stress.

Regarding the rate of penetration analysis, ``operating window"
is defined by the inequations $1/360~m/s \leq V_0 \leq 1/90~m/s$
and $3\pi/2~rad/s \leq \Omega \leq 7\pi/3~rad/s$, and
$\texttt{UTS} = 650 \times 10^{6}~Pa$.

\begin{figure}
	\centering
	\includegraphics[scale=0.45]{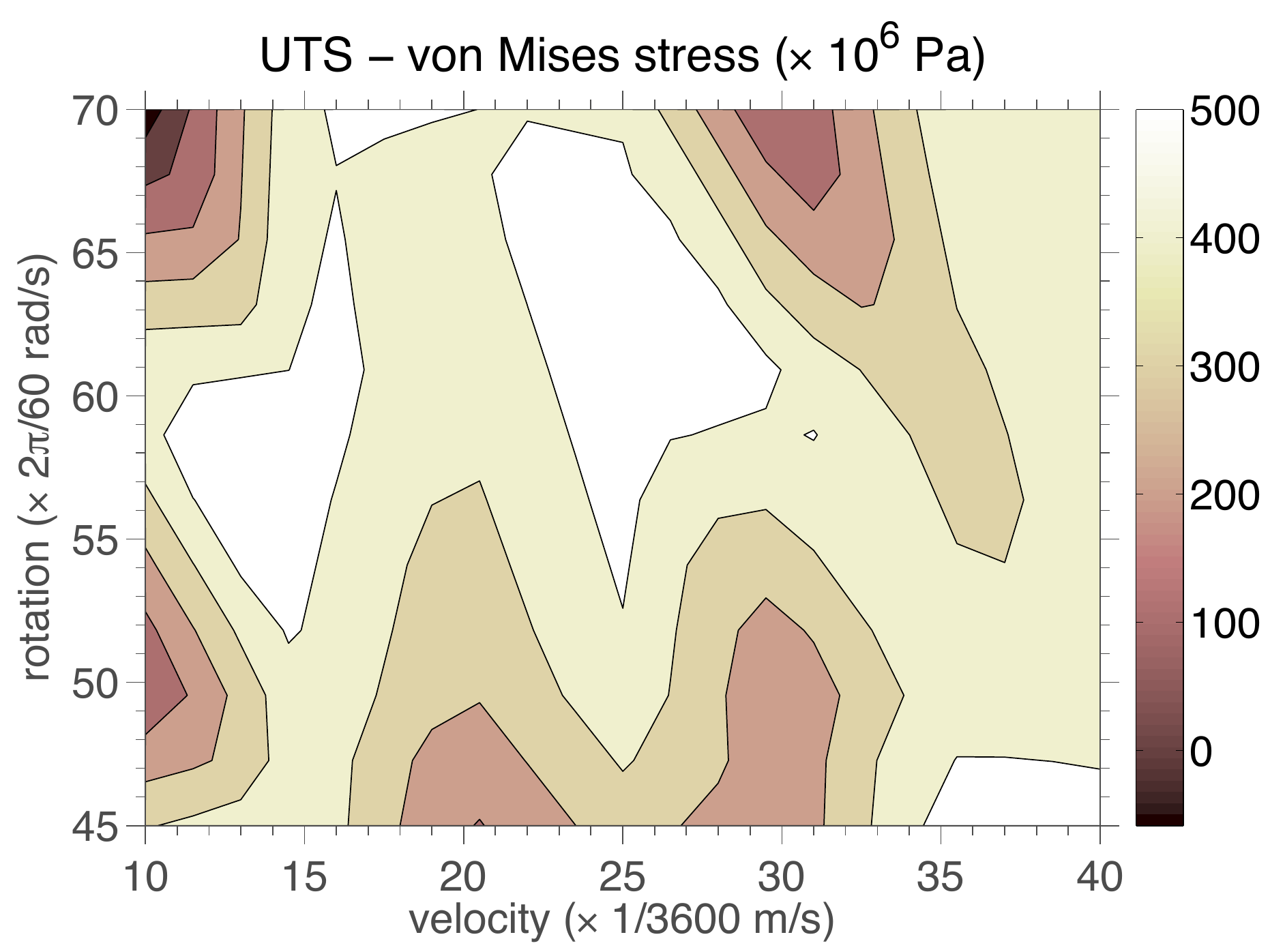}
	\caption{Illustration of maximum von Mises stress contour plot, for an
	``operating window" defined by $1/360~m/s \leq V_0 \leq 1/90~m/s$
	and $3\pi/2~rad/s \leq \Omega \leq 7\pi/3~rad/s$.}
	\label{sigvm_func_fig}
\end{figure}

The contour map of constraint (\ref{von_mises_stress_determ}),
is shown in Figure~\ref{sigvm_func_fig}. From the way 
(\ref{von_mises_stress_determ}) is written, the Mises criterion is not satisfied
when the function is negative, which occurs in a ``small neighborhood" of
the upper left corner of the rectangle that defines the ``operating window".
It is noted that all other points respect the material structural limits.
In this way, then, the ``operating window" \emph{admissible} region
consists of all points that satisfy the constraint.

\begin{figure}
	\centering
	\includegraphics[scale=0.45]{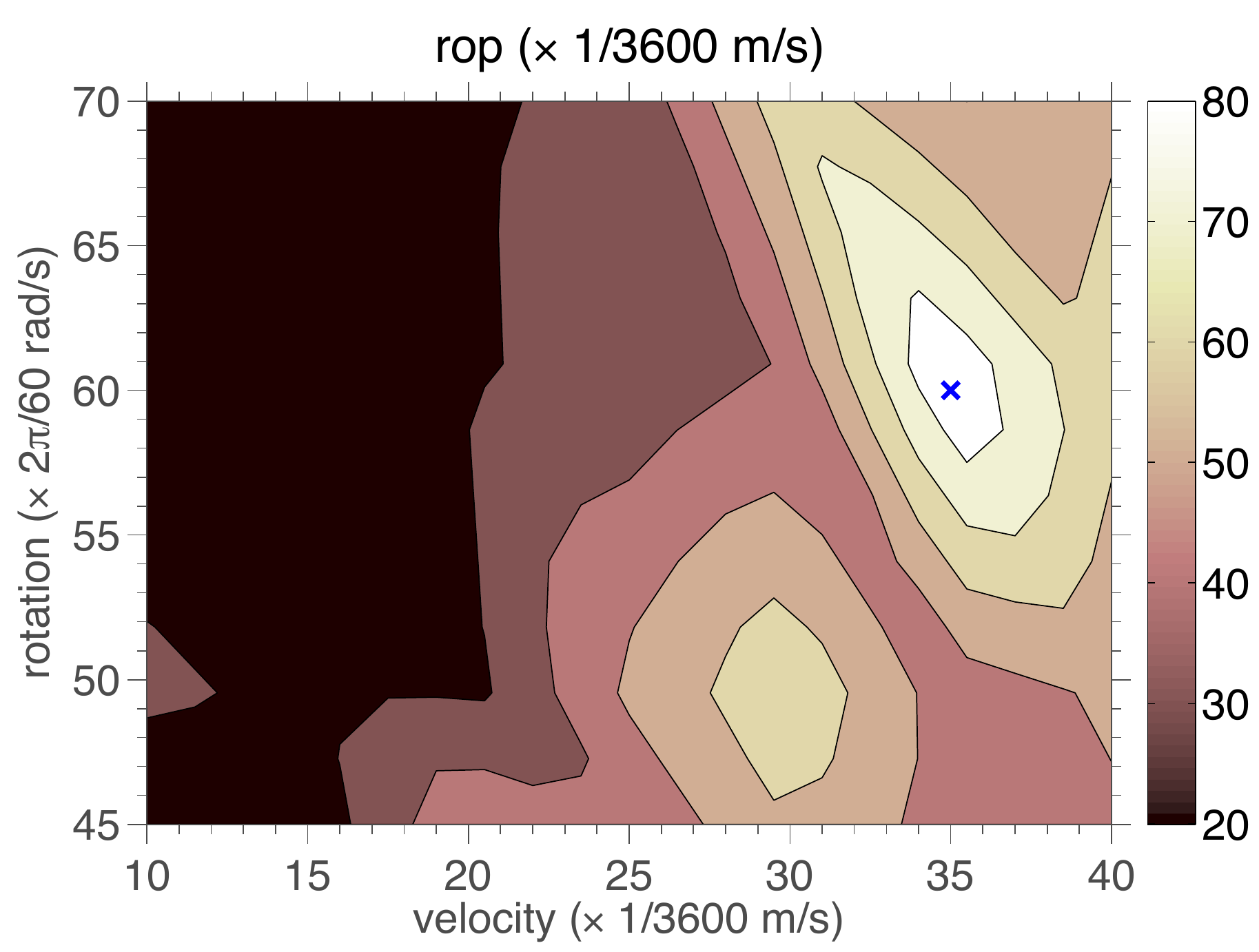}
	\caption{Illustration of rate of penetration function contour plot, for an
	``operating window" defined by $1/360~m/s \leq V_0 \leq 1/90~m/s$
	and $3\pi/2~rad/s \leq \Omega \leq 7\pi/3~rad/s$. The maximum is indicated
with a blue cross.}
	\label{rop_func_fig}
\end{figure}

In Figure~\ref{rop_func_fig} the reader can see the contour map of 
$\texttt{rop}$ function. Taking into account only points 
in the admissible region, the maximum of $\texttt{rop}$ occurs at
$(V_0,\Omega) = (7/720~m/s,2\pi~rad/s)$, which is indicated on the graph
with a blue cross. This point corresponds to a mean rate of penetration,
during the time interval analyzed, approximately equal to $90$ ``meters per hour".

It is worth remembering that the definition of $\texttt{rop}$ uses
temporal mean of $\dot{u}_{bit}(t)$ positive part. In such
a way, it is not surprising to find the maximum value of $\texttt{rop}$ much
higher than the corresponding velocity, $V_0$ imposed on the column left end.
This occurs because, by taking only the function positive part,
the rate of penetration value increases.

\begin{figure}
	\centering
	\includegraphics[scale=0.45]{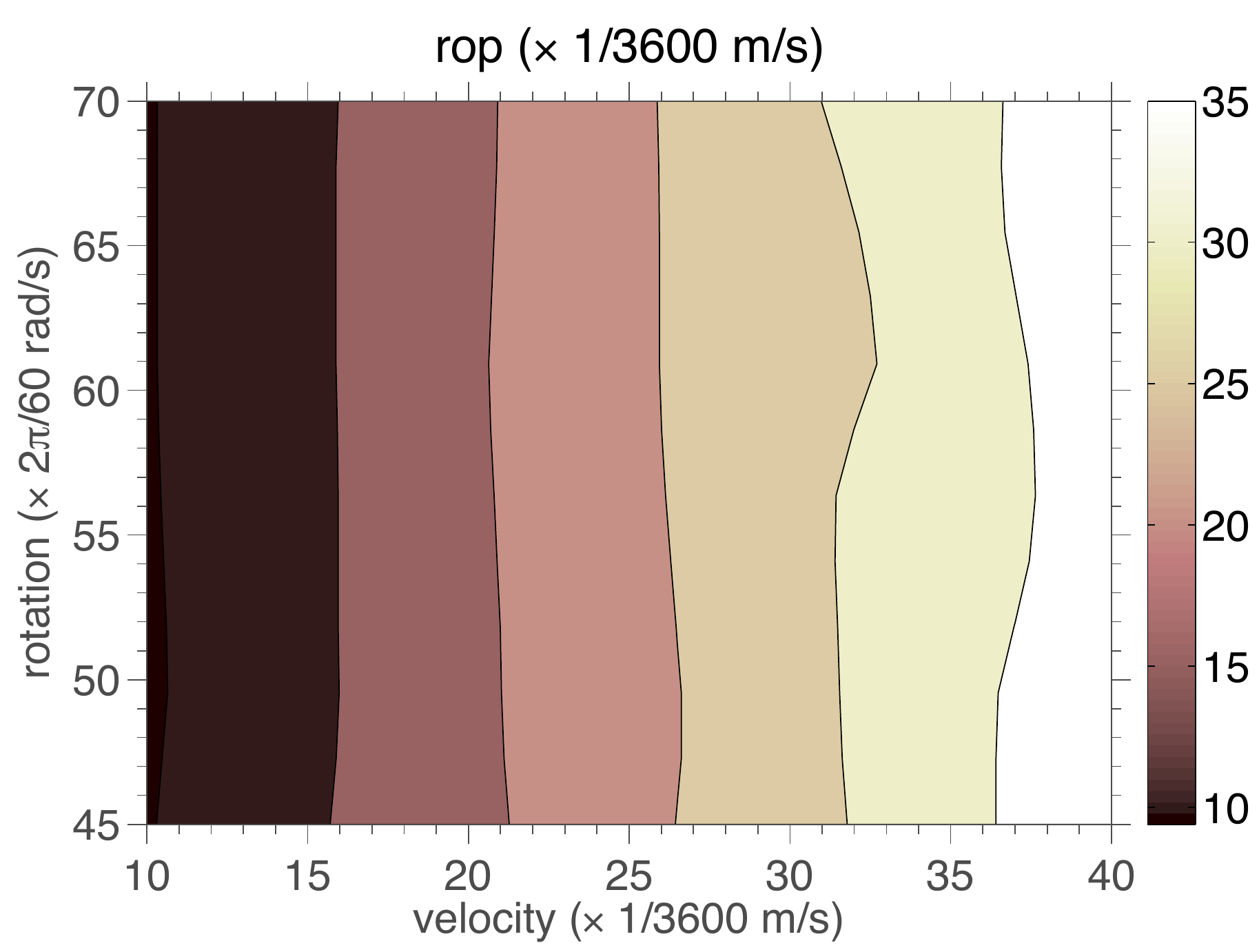}
	\caption{Illustration of the contour plot of the rate of penetration function,
	with an alternative definition, for an ``operating window" defined by $1/360~m/s \leq V_0 \leq 1/90~m/s$
	and $3\pi/2~rad/s \leq \Omega \leq 7\pi/3~rad/s$.}
	\label{rop2_func_fig}
\end{figure}

To see how significant is the inclusion of $\dot{u}_{bit}(t)$ positive part in
the definition of $\texttt{rop}$, the reader can see Figure~\ref{rop2_func_fig}.
This figure shows the same information as Figure~\ref{rop_func_fig}, i.e., the contour map
of $\texttt{rop}$ function, but now considering $\dot{u}_{bit}(t)$ instead of
$\dot{u}_{bit}^{+}(t)$ in the definition of $\texttt{rop}$. Note that, in comparison
with the contour map of Figure~\ref{rop_func_fig}, lower values for the function levels
are observed, and these values now are closer to $V_0$.
Furthermore, the topology of contour lines change, so that no local extreme point
can be seen isolated. This example shows the importance of considering
$\dot{u}_{bit}^{+}(t)$ in the definition of $\texttt{rop}$.

\subsection{Probabilistic analysis of the dynamics}

For the probabilistic analysis of the dynamic system
a parametric approach is used, where the random 
parameters distributions are constructed according to the
procedure presented in section~\ref{prob_mod_data_uncert}.
In this case, the random variables of interest are characterized
by the mean values
$\mean{\randvar{\bbalpha}_{\BR}} = 400~1/m/s$,
$\mean{\randvar{\Gamma}_{\BR}} = 30 \times 10^{3}~N$, and
$\mean{\randvar{\bbmu}_{\BR}} = 0.4$, and the
dispersion factors
$\delta_{\randvar{\bbalpha}_{\BR}} = 0.5\%$,
$\delta_{\randvar{\Gamma}_{\BR}} = 1\%$, and
$\delta_{\randvar{\bbmu}_{\BR}} = 0.5\%$.

To compute the parameters uncertainties propagation
through the model, MC method is employed. To analyze
the convergence of MC simulations, it is taken into consideration the map
$ n_s \in \N \mapsto \texttt{conv}_{\textsc{\tiny MC}} (n_{s}) \in \R $, being

\begin{equation}
		\texttt{conv}_{\textsc{\tiny MC}} (n_{s}) =
		\left(
		\frac{1}{n_{s}} \sum_{n=1}^{n_{s}}
		\int_{t=t_0}^{t_f} \norm{\randproc{q} (t, \SSpt_n)}^2~dt
		\right)^{1/2},
		\label{MC_conv_eq}
\end{equation}

\noindent
where $n_{s}$ is the number of MC realizations,
and $\norm{\cdot}$ denotes the standard Euclidean norm.
This metric allows one to evaluate the approximation 
$\randproc{q} (t, \SSpt_n)$ convergence in mean-square sense.
For further details the reader is encouraged to see \cite{soize2005p623}.

\begin{figure}[h!]
				\centering
				\includegraphics[scale=0.45]{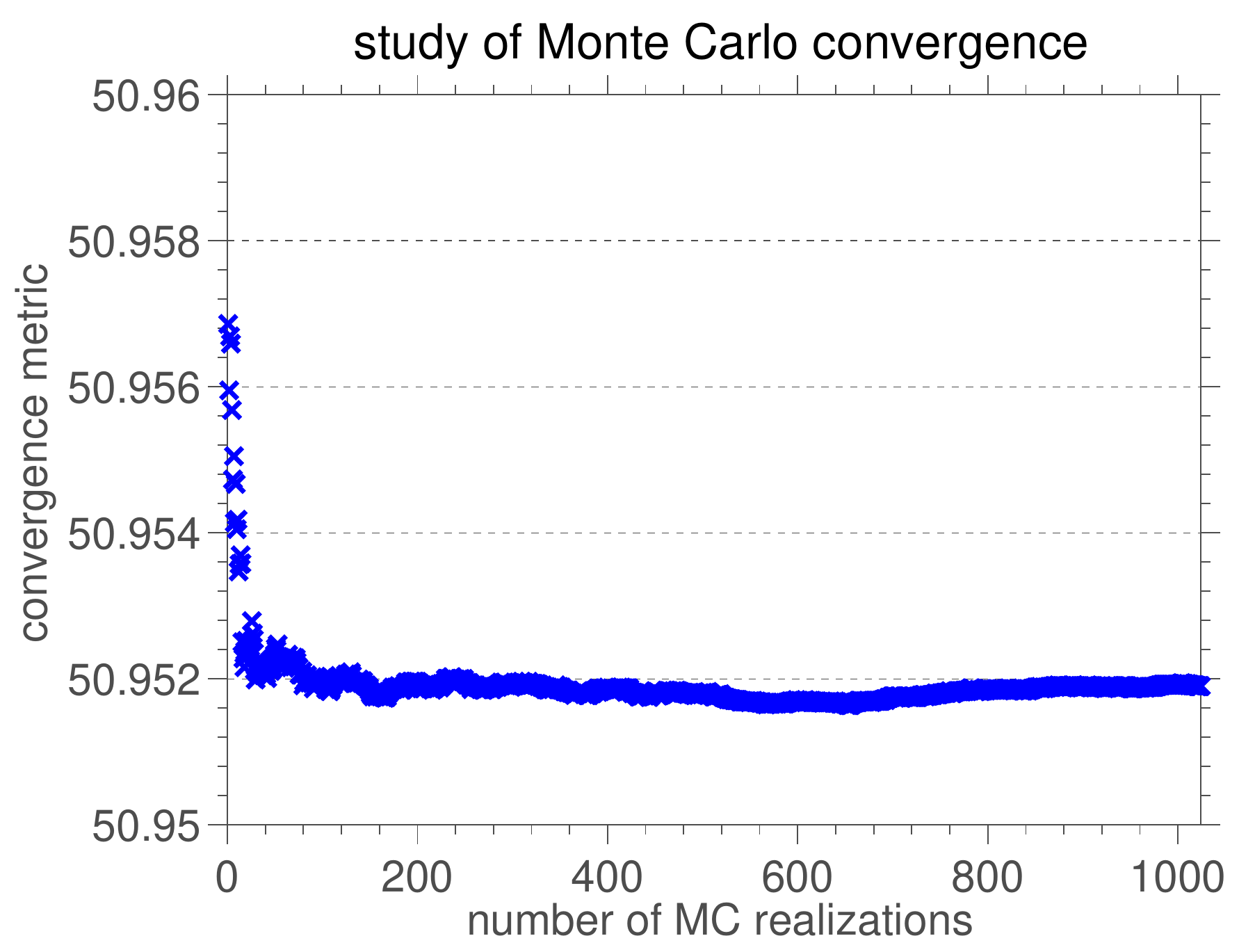}
				\caption{This figure illustrates the convergence metric of MC simulation as function
								of the number of realizations.}
				\label{MC_conv_fig}
\end{figure}

The evolution of $\texttt{conv}(n_{s})$ as function of
$n_{s}$ can be seen in Figure~\ref{MC_conv_fig}. Note that for
$n_{s}=1024$ the metric value has reached a steady value.
In this sense, if something is not stated otherwise,
all the stochastic simulations that follows in this work
use $n_{s}=1024$.

\begin{figure}
	\centering
	\includegraphics[scale=0.45]{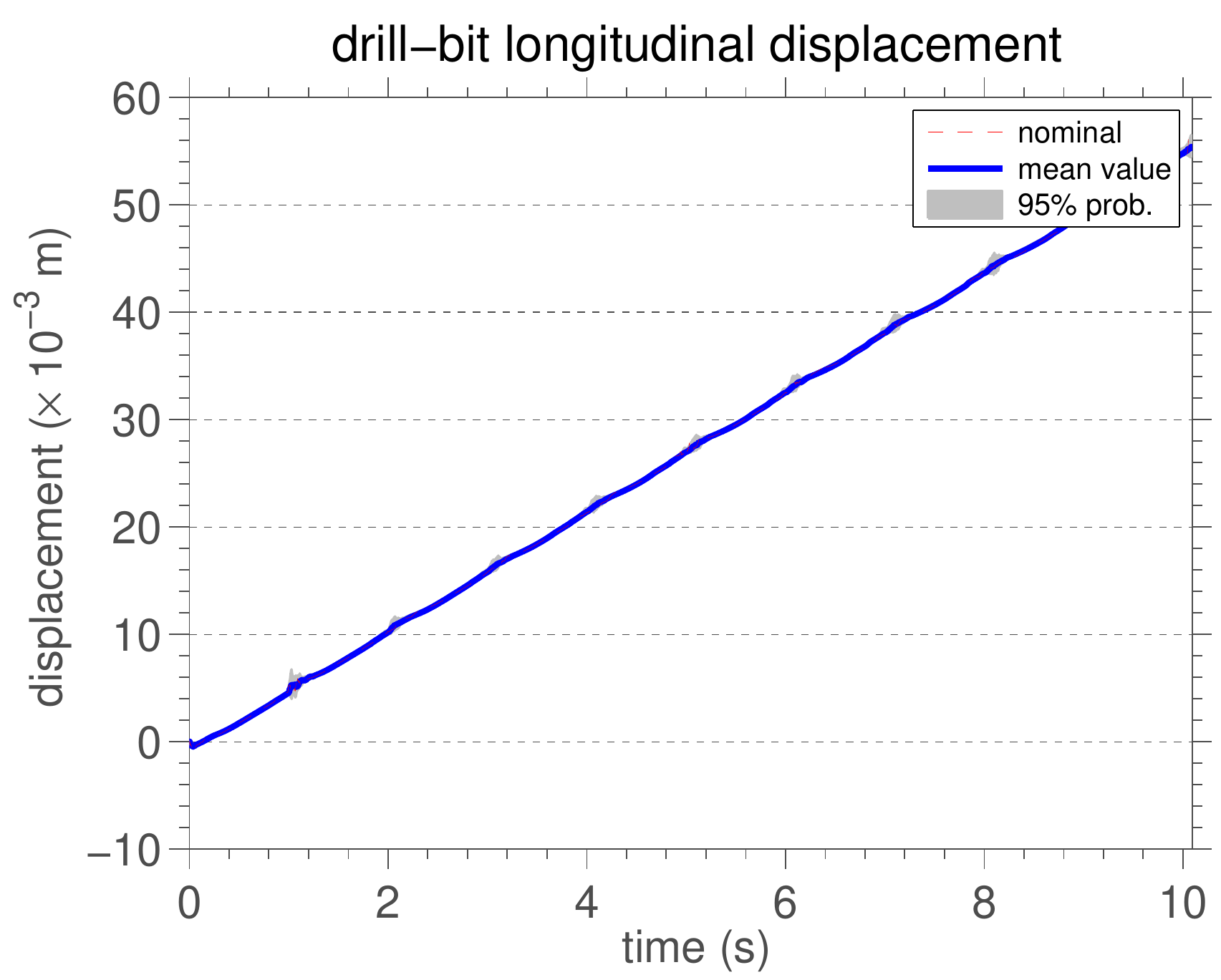}\\
	\vspace{2mm}
	\includegraphics[scale=0.45]{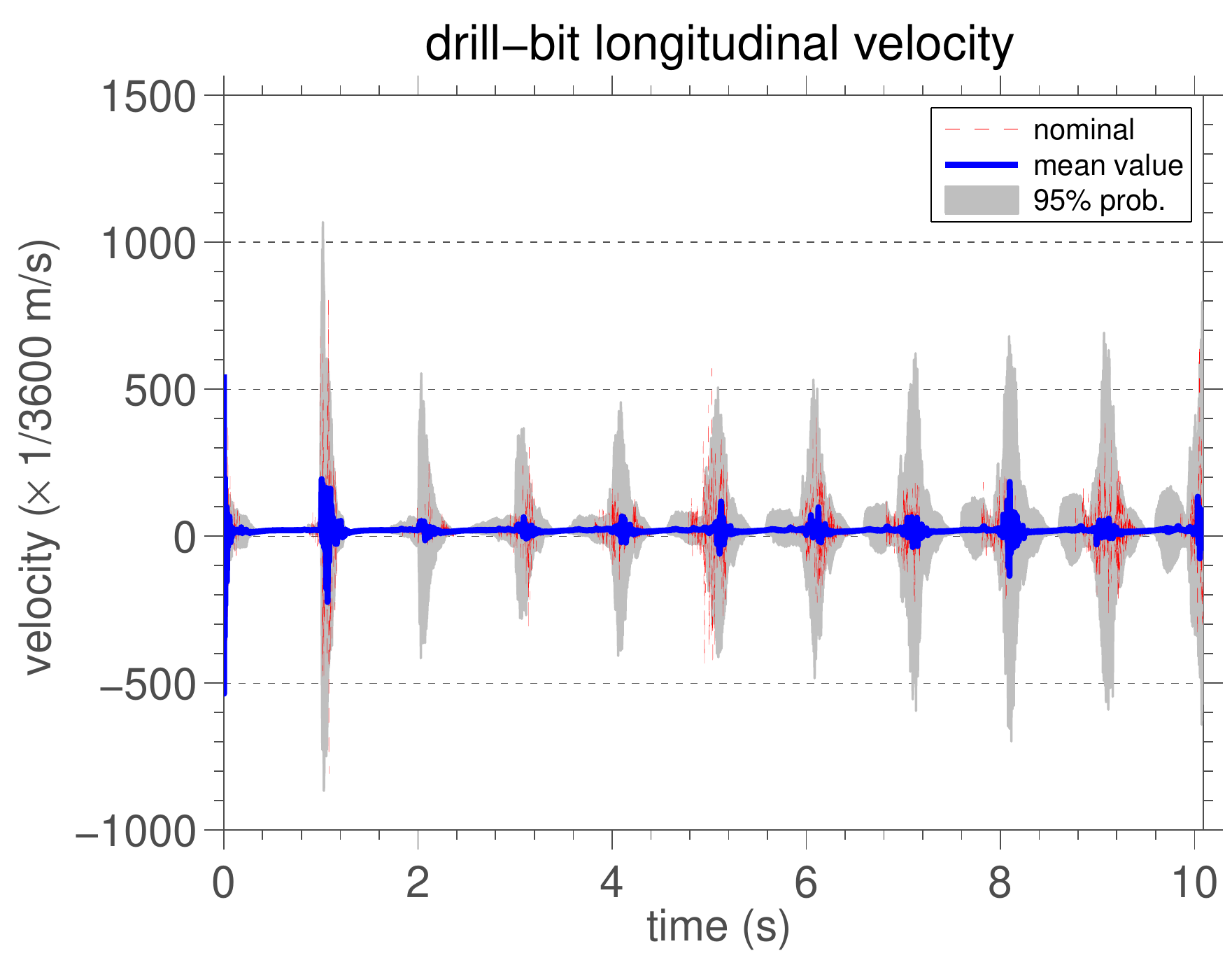}
	\caption{Illustration of the nominal model (red line), the mean value (blue line), and the 95\% probability envelope (grey shadow)
	for the drill-bit longitudinal displacement (top) and velocity (bottom).}
	\label{drill-bit_disp_velo_stoch}
\end{figure}

An illustration of the mean value (blue line), and a confidence band
(grey shadow), wherein a realization of the stochastic dynamic system has
95\% of probability of being contained, for drill-bit longitudinal
displacement and velocity is shown in Figure~\ref{drill-bit_disp_velo_stoch}.
For sake of reference, the deterministic model, which the numerical results were
presented earlier, is also shown and called nominal model (red line).
It is observed that for the displacement, mean value and 
nominal model are very similar. Meanwhile, for the velocity, mean value presents 
oscillations that are correlated with the nominal model, but with very different amplitudes.
Regarding the confidence band, there are significant amplitudes in the instants 
that corresponds to the fluctuation packages, and negligible amplitudes
in the other moments.

\begin{figure}
	\centering
	\includegraphics[scale=0.45]{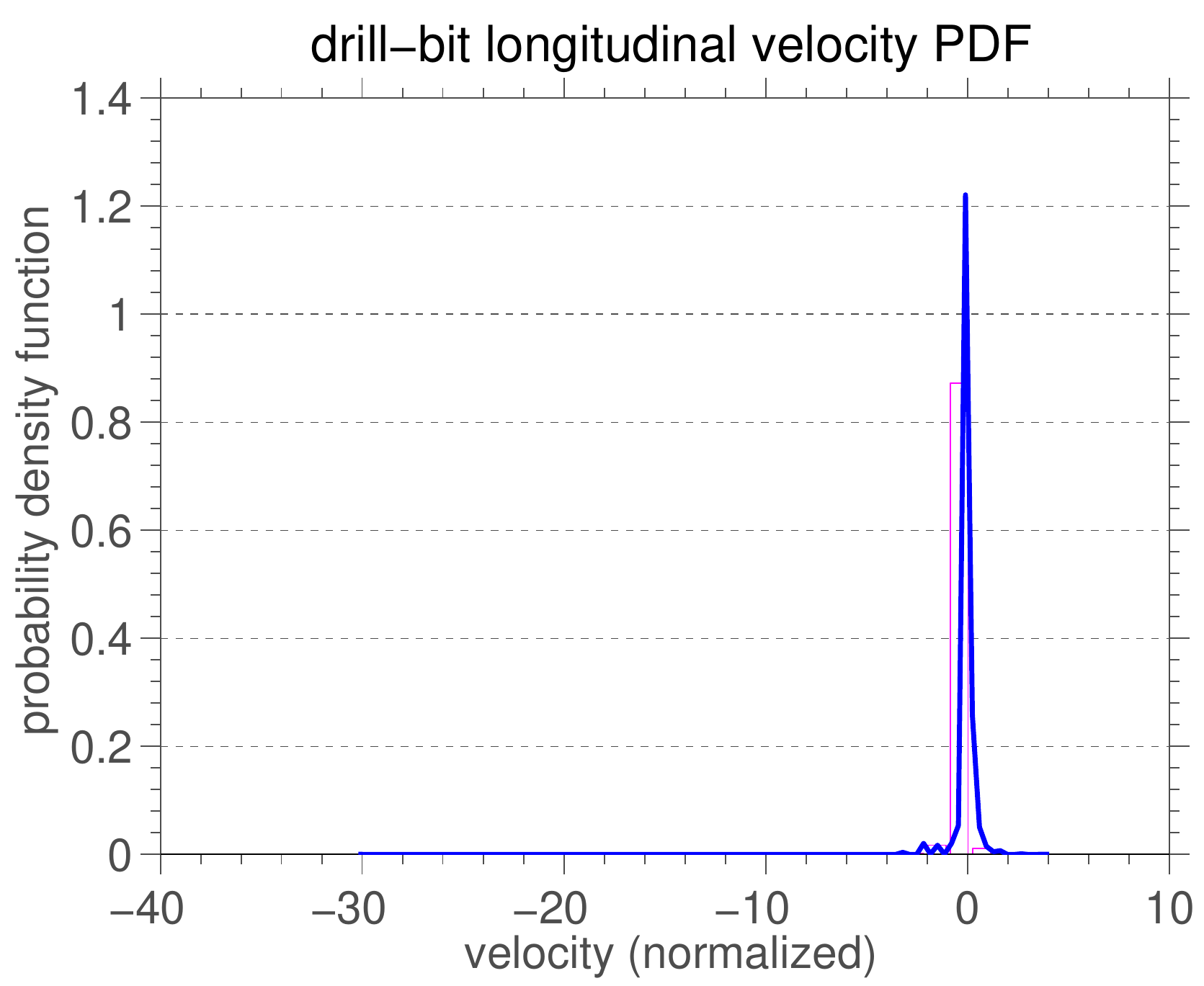}
				\caption{Illustration of the normalized probability density function of drill-bit longitudinal velocity.}
				\label{drill-bit_velo_pdf}
\end{figure}

Fixing time in $t=10~s$, it is possible to analyze the behavior of
drill-bit longitudinal velocity through its normalized PDF, which
is presented in Figure~\ref{drill-bit_velo_pdf}. In this context normalized
means a distribution of probability with zero mean and unit standard deviation.
It is observed an unimodal behavior, with maximum value occurring in a
neighborhood of the mean value, with small dispersion around this position.

\begin{figure}
	\centering
	\includegraphics[scale=0.45]{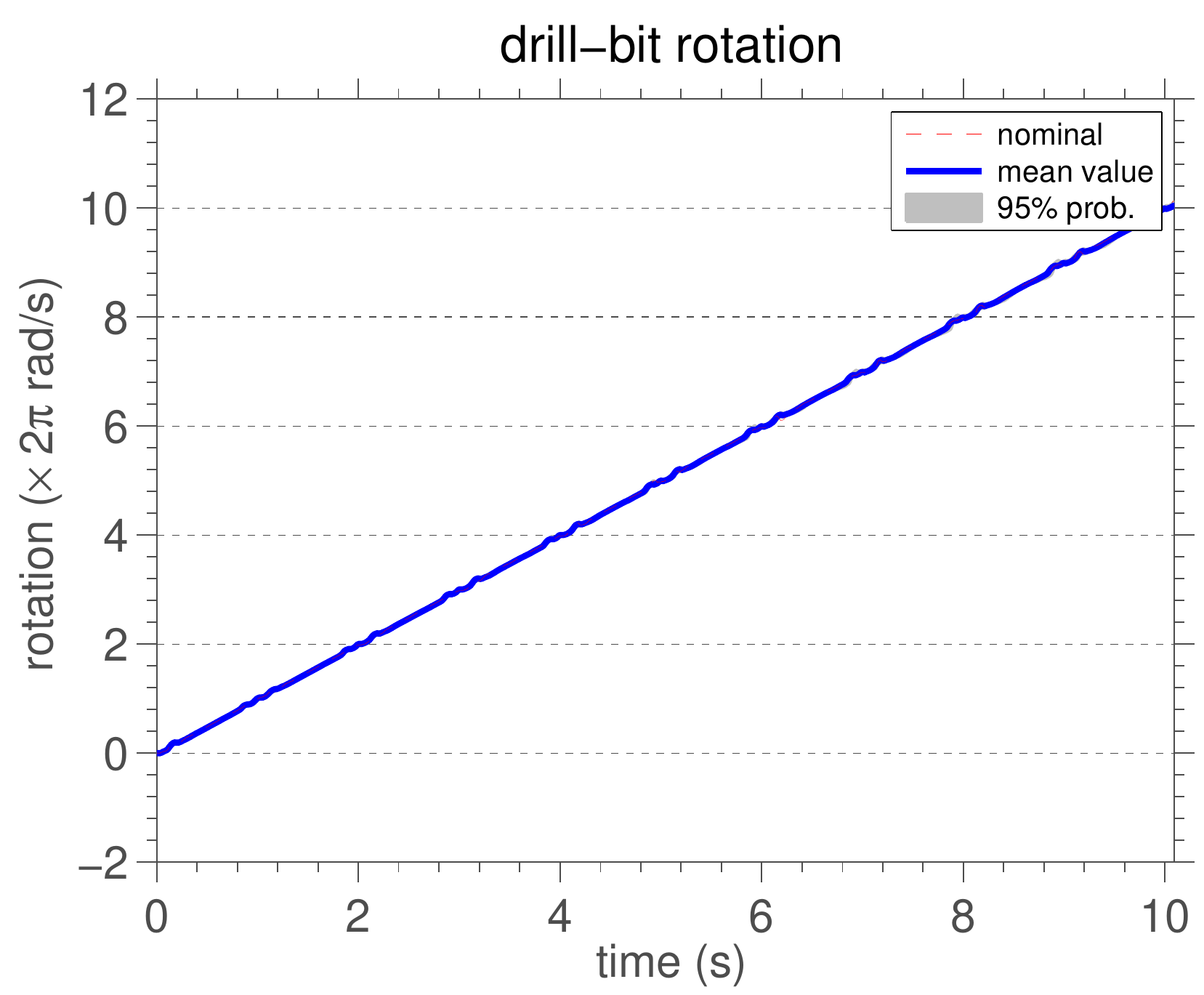}\\
	\vspace{2mm}
	\includegraphics[scale=0.45]{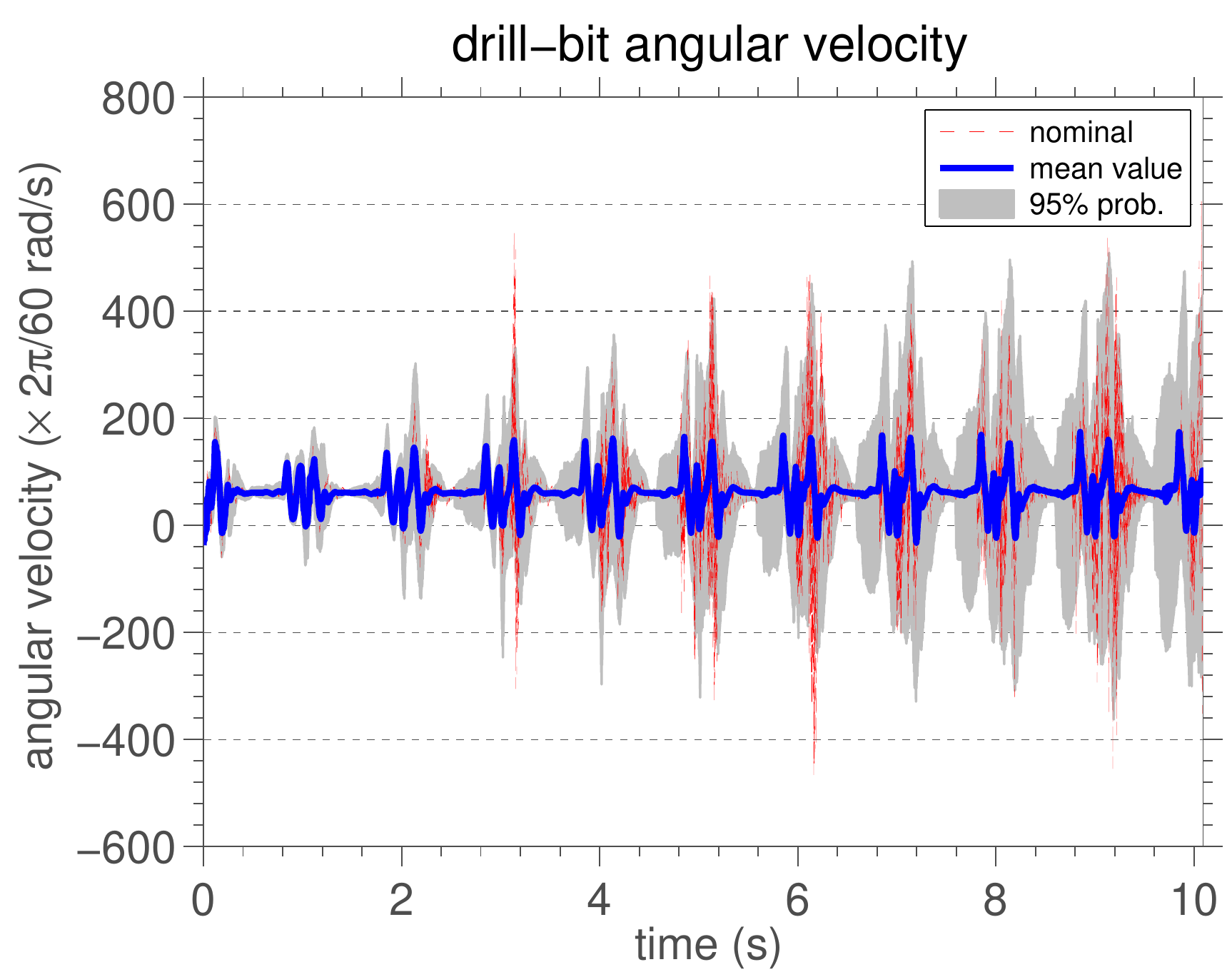}
	\caption{Illustration of the nominal model (red line), the mean value (blue line), and the 95\% probability envelope (grey shadow)
	for the drill-bit rotation (top) and angular velocity (bottom).}
	\label{drill-bit_rot_angvelo_stoch}
\end{figure}

In Figure~\ref{drill-bit_rot_angvelo_stoch}, the reader can see the nominal model,
the mean value, and the 95\% probability envelope of drill-bit rotation and 
angular velocity. A good agreement between rotation nominal model
and mean value is observed, and the confidence band around it
is negligible. On the other hand, with respect to the angular velocity, it is possible to
see discrepancies in the nominal model and mean value amplitudes.
These differences occur in the instants when the system is subject to shocks,
as in the case of drill-bit longitudinal velocity. The band of uncertainty shows that
the dispersion around mean value increases with time, due to accumulation of 
uncertainties, but also in reason of the impacts, once its amplitude increases a lot
near the instants where the mean value presents large fluctuations, i.e., the instants
which are correlated with the beam and borehole wall impacts.

\begin{figure}
	\centering
	\includegraphics[scale=0.45]{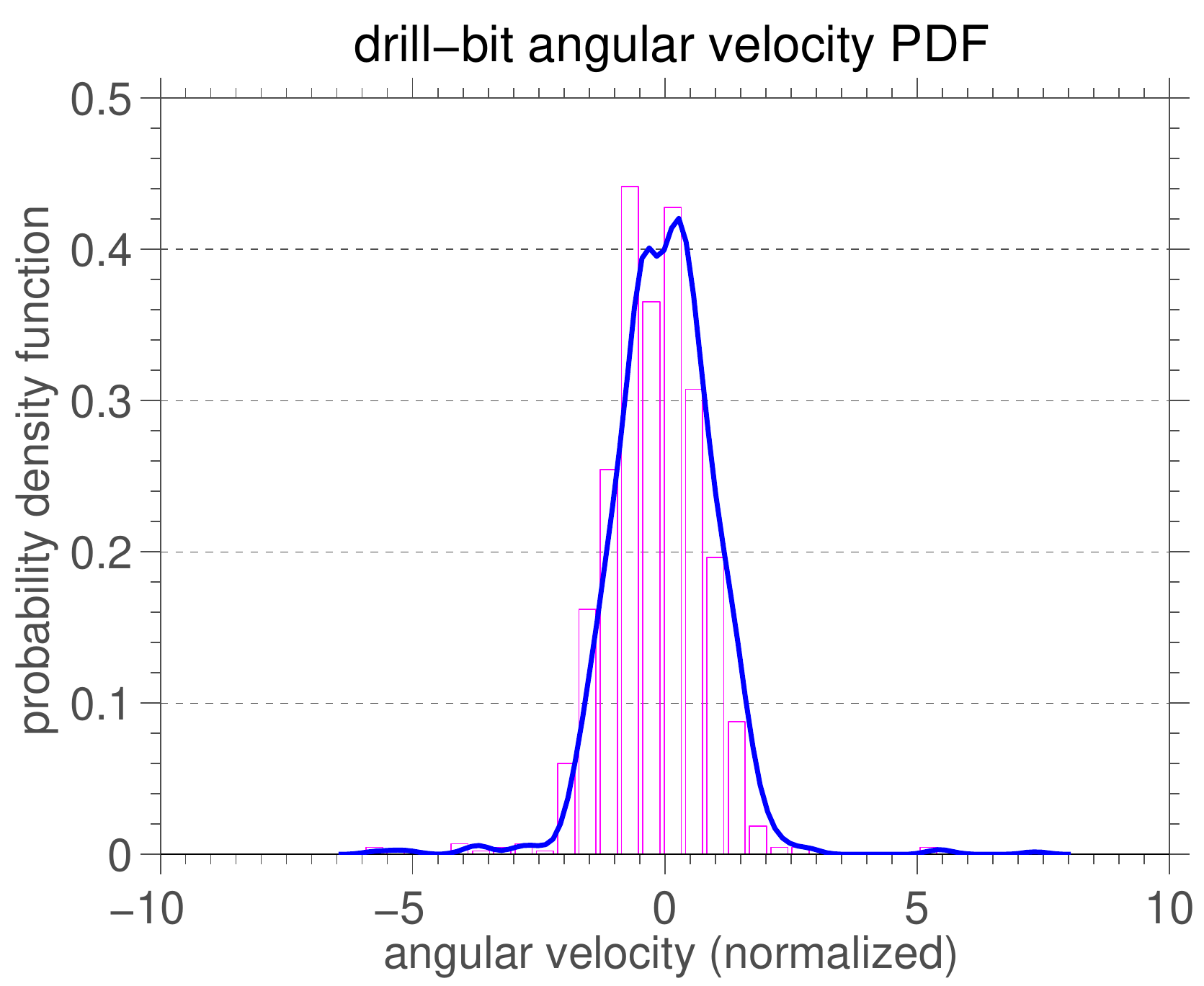}
				\caption{Illustration of the normalized probability density function of drill-bit angular velocity.}
				\label{drill-bit_angvelo_pdf}
\end{figure}

For $t=10~s$, the reader can see the normalized PDF of drill-bit angular velocity
in Figure~\ref{drill-bit_angvelo_pdf}. It is noted again an unimodal behavior, with
maximum again near the mean value. But now a large dispersion around the mean
can be seen.

\begin{figure}
	\centering
	\includegraphics[scale=0.45]{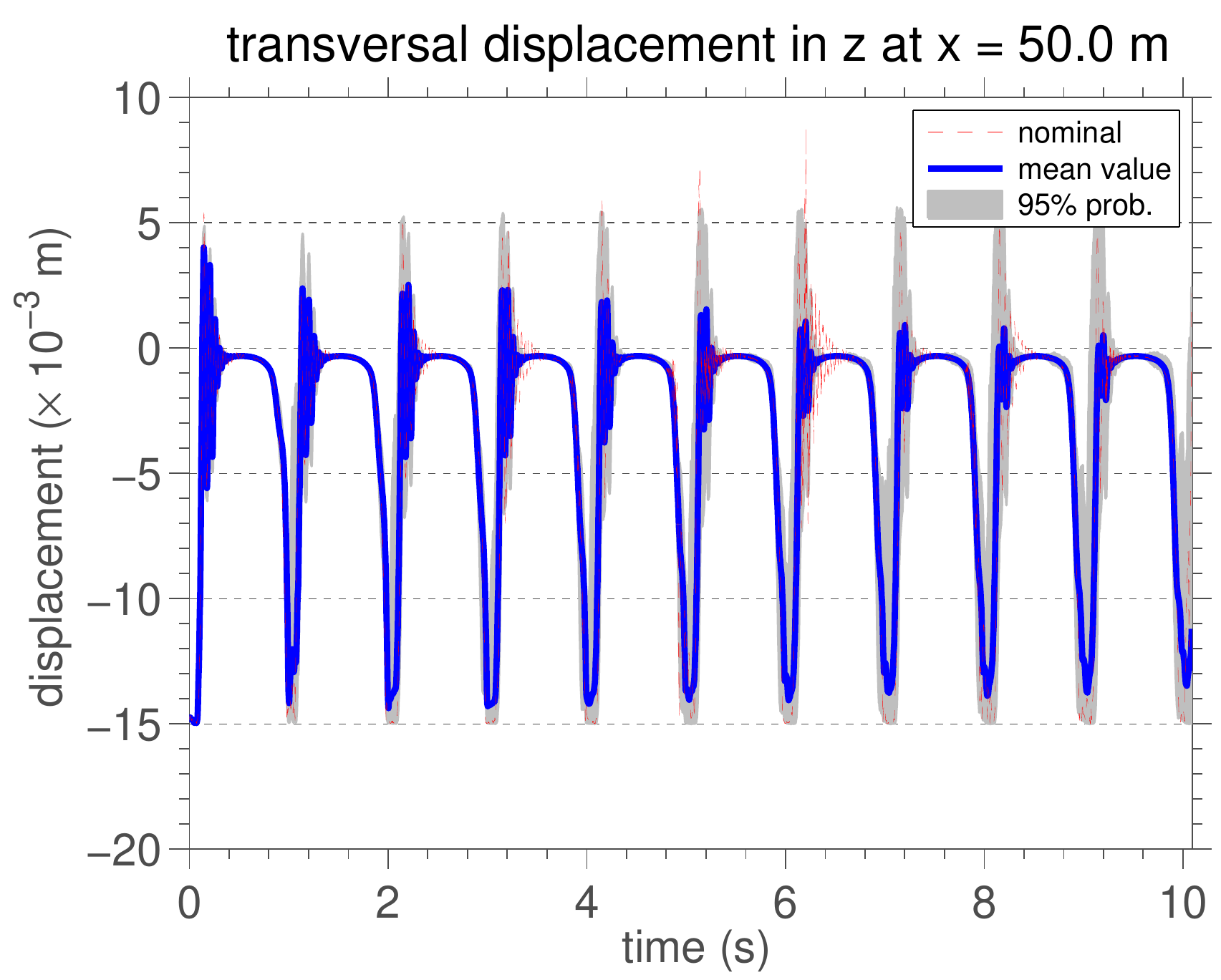}\\
	\vspace{2mm}
	\includegraphics[scale=0.45]{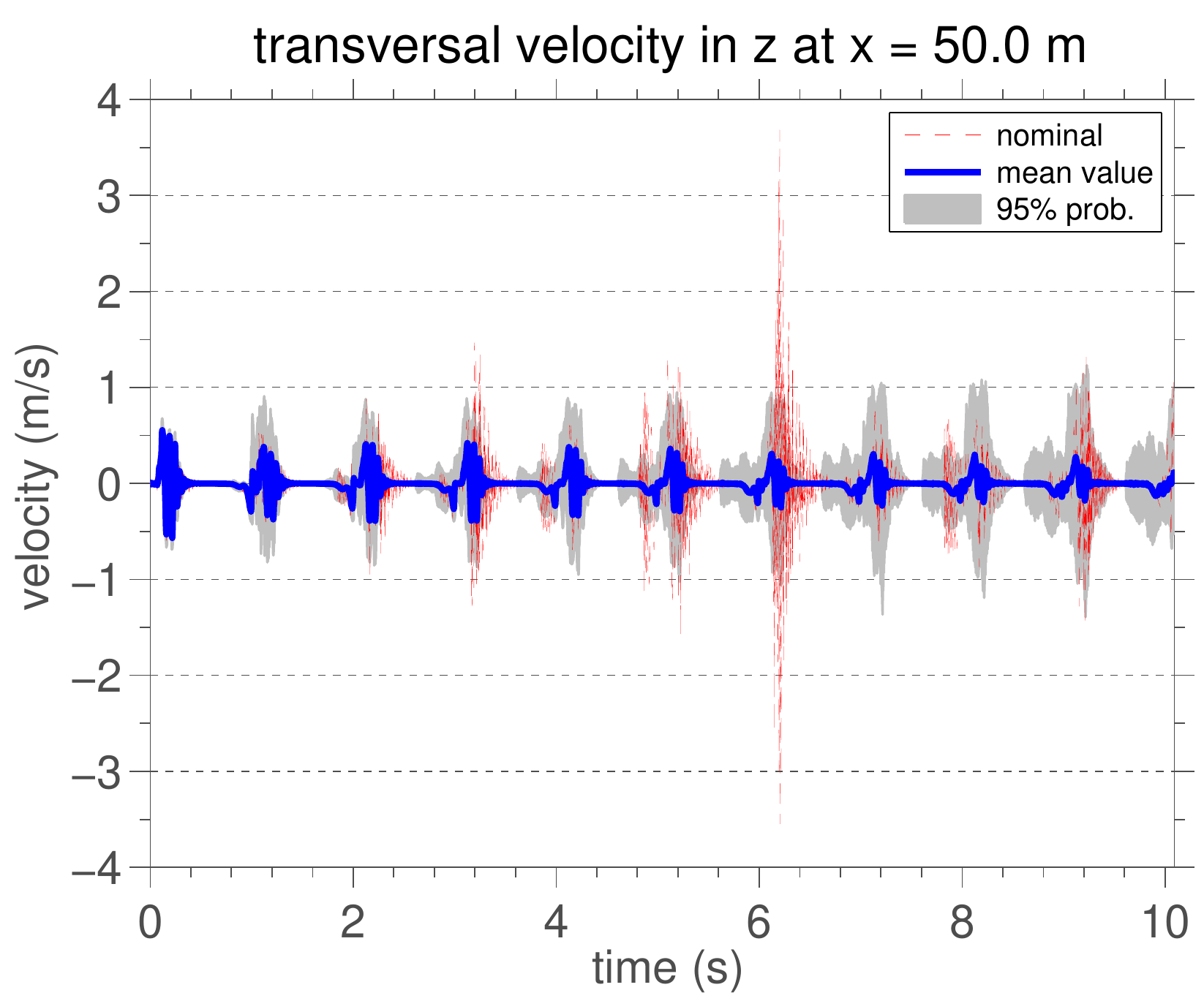}
	\caption{Illustration of the nominal model (red line), the mean value (blue line), and the 95\% probability envelope (grey shadow)
	for the beam transversal displacement (top) and velocity in z (bottom) at $x=50~m$.}
	\label{drill-bit_disp_velo_z_stoch}
\end{figure}

Moreover, in Figure~\ref{drill-bit_disp_velo_z_stoch} it is shown
the nominal model, the mean value, and the 95\% probability envelope of
beam transversal displacement and velocity in z at $x=50~m$.
Here the mean values of both, velocity and displacement, present correlation
with the nominal models. Indeed, both present discrepancies in the
oscillation amplitudes, especially the velocity, that are more pronounced, 
as before, in the instants wherein the system is subject to impacts.
The confidence bands present meaningful amplitudes, what evidentiates
a certain level of dispersion around the means, which are more significant,
as expected, at the instants of impact.

\begin{figure}
	\centering
	\includegraphics[scale=0.45]{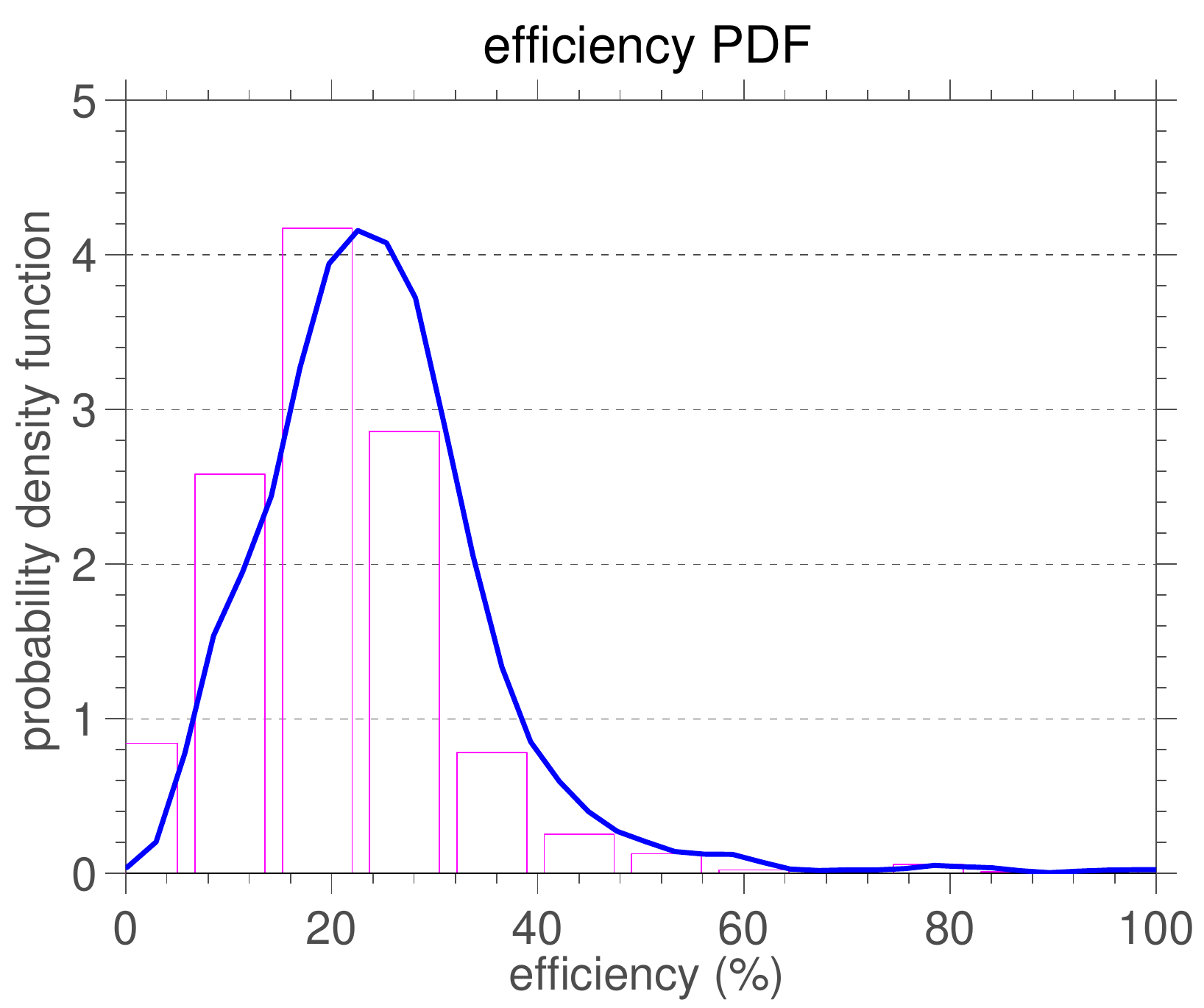}
				\caption{Illustration of the probability density function of the drilling process efficiency.}
				\label{Eff_pdf}
\end{figure}

The PDF of drilling process efficiency function is shown in
Figure~\ref{Eff_pdf}. One can observe a unimodal distribution with the
maximum around 16\% and wide dispersion between 0 and 40\%,
declining rapidly to negligible values outside this range.

\begin{figure}
	\centering
	\includegraphics[scale=0.45]{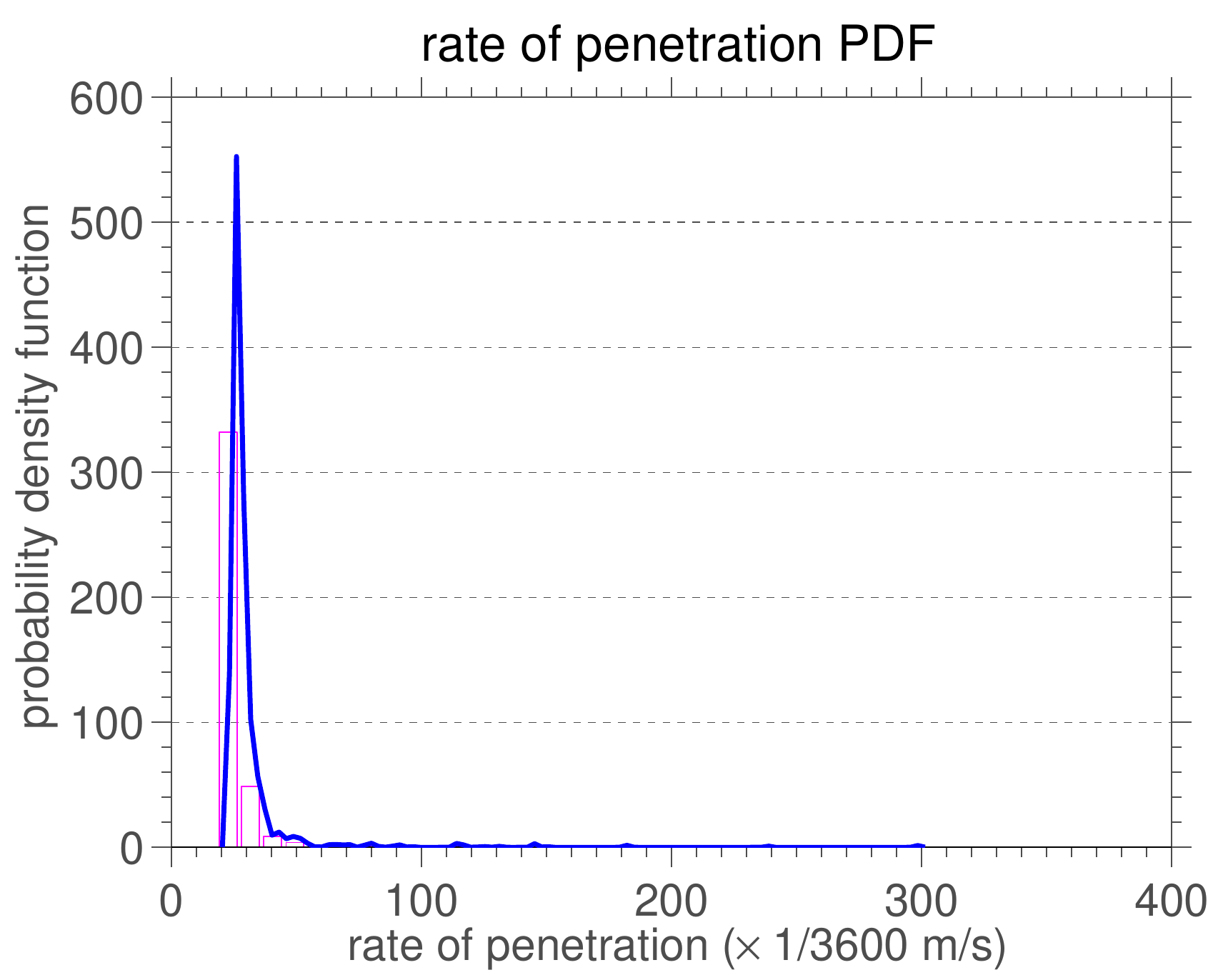}
				\caption{Illustration of the probability density function of the rate of penetration function.}
				\label{rop_pdf}
\end{figure}

Finally, in Figure~\ref{rop_pdf} one can see the PDF of drillstring rate of
penetration function. One notes an unimodal behavior in a narrow range between
20 and 50 ``meters per hour", with the maximum around 30 ``meters per hour".

\subsection{Robust optimization of drillstring rate of penetration}

To improve the level of confidence of drilling process optimization,
uncertainties intrinsic to the problem should be taken into account.
This leads to a robust optimization problem,
i.e, optimization under uncertainty where the range of random parameters
are known, but not necessarily their distribution
\cite{beyer2007p3190,schueller2008p2,capiez-lernout2008p021001,
capiez-lernout2008p021001_130,capiez-lernout2008p1774,
soize2008p2955,ben-tal2009}.

Taking into account the uncertainties, through the parametric approach presented
in section~\ref{prob_mod_data_uncert}, drill-bit velocity becomes the stochastic
process $\randproc{U}_{bit}(t,\SSpt)$, so that the random rate of penetration
is defined by

\begin{equation}
		\randproc{ROP}(V_0,\, \Omega,\SSpt) =
		\frac{1}{t_f - t_0} \int_{t=t_0}^{t_f} \dot{\randproc{U}}_{bit}^{+}(t,\SSpt) \, dt.
		\label{rop_stoch}
\end{equation}

In the robust optimization problem, who plays the role of objective function
is the expected value of the random variable $\randproc{ROP}(V_0,\, \Omega,\SSpt)$, i.e.,
$\expval{\randproc{ROP}(V_0,\, \Omega,\SSpt)}$.

Regarding the restriction imposed by the von Mises criteria, now the equivalent
stress is the random field $\randproc{\bbsigma}_{VM} (V_0,\, \Omega,\, x,\, t,\SSpt)$,
so that the inequality is written as

\begin{equation}
		\texttt{UTS} - \underset{t_0 \leq t \leq t_f}{\underset{0 \leq x \leq L}{\max}} \left\{\randproc{\bbsigma}_{VM} (V_0,\, \Omega,\, x,\, t,\SSpt) \right\} \geq 0.
		\label{von_mises_stress_stoch}
\end{equation}

However, the robust optimization problem considers as restriction 
a probability of the event defined by inequality (\ref{von_mises_stress_stoch}),

\begin{equation}
\resizebox{.98\hsize}{!}{$
		\PM \left\lbrace \texttt{UTS} - \underset{t_0 \leq t \leq t_f}{\underset{0 \leq x \leq L}{\max}} \left\{\randproc{\bbsigma}_{VM} (V_0,\, \Omega,\, x,\, t,\SSpt) \right\} \geq 0 \right\rbrace \geq 1 - P_{risk}$,
}
		\label{constraint_stoch}
\end{equation}

\noindent
where $0 < P_{risk} < 1$ is the risk percentage acceptable to the problem.

A robust optimization problem very similar to this one,
in the context of a vertical drillstring dynamics,
is considered in \cite{ritto2010p415}.

To solve this robust optimization problem it is employed a trial
strategy which discretizes the ``operating window" in a structured
grid of points and then evaluates the objective function
$\expval{\randproc{ROP}(V_0,\, \Omega,\SSpt)}$ and the
probabilistic constraint (\ref{constraint_stoch}) in these points.

Accordingly, it is considered the same ``operating window" used
in the deterministic optimization problem solved above, i.e.,
$1/360~m/s \leq V_0 \leq 1/90~m/s$ and
$3\pi/2~rad/s \leq \Omega \leq 7\pi/3~rad/s$, in addition to
$\texttt{UTS} = 650 \times 10^{6}~Pa$ and $P_{risk} = 10\%$.
Each MC simulation in this case used 128 realizations to compute
the propagation of uncertainties.

\begin{figure}[h!]
	\centering
	\includegraphics[scale=0.45]{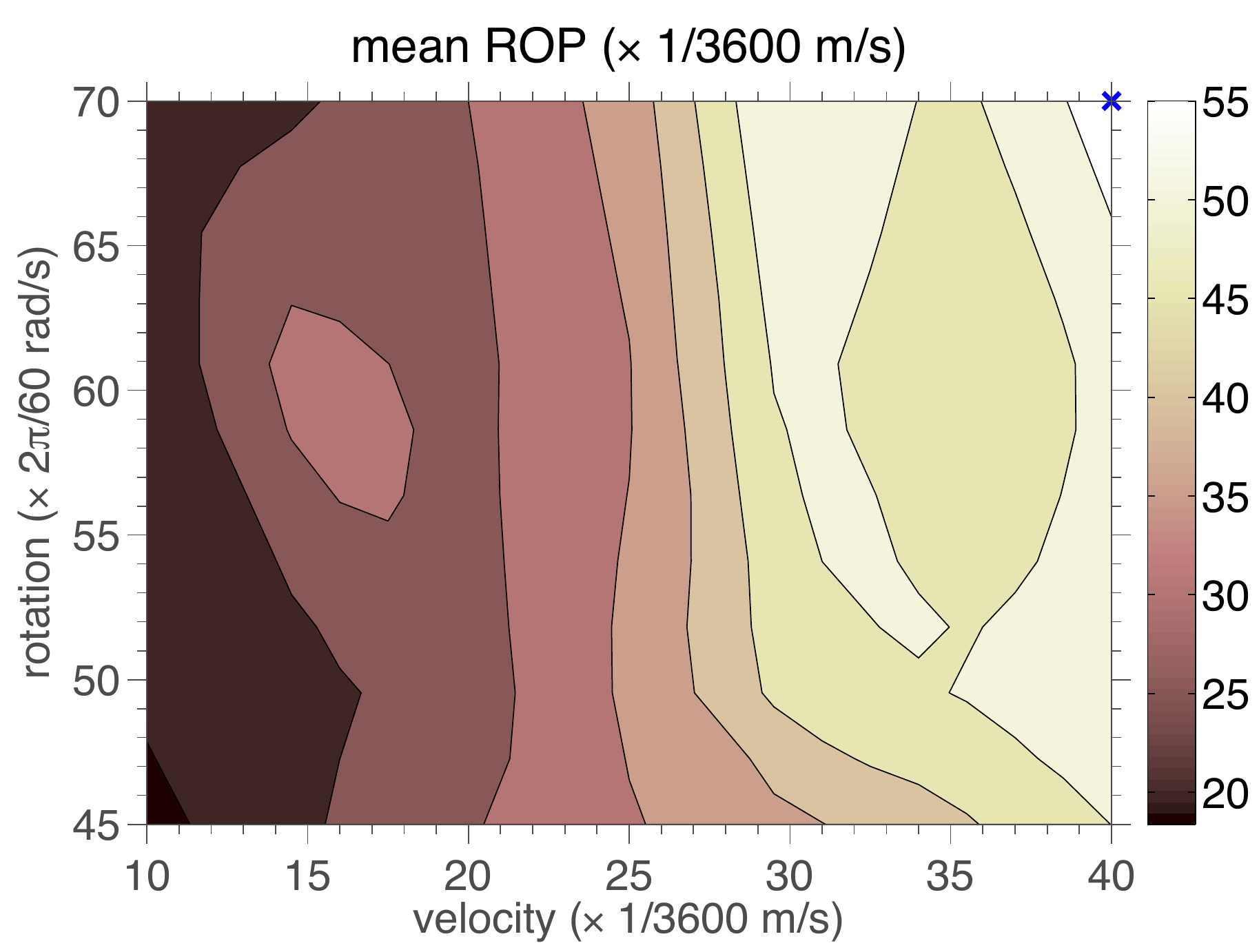}
	\caption{Illustration of the contour plot of the mean rate of penetration function, for an
	``operating window" defined by $1/360~m/s \leq V_0 \leq 1/90~m/s$
	and $3\pi/2~rad/s \leq \Omega \leq 7\pi/3~rad/s$. The maximum is indicated
with a blue cross in the upper right corner.}
	\label{rop_robust_func_fig}
\end{figure}

Concerning the simulation results, the probabilistic constraint (\ref{constraint_stoch})
is respected in all grid points that discretize the ``operating window". Thus, the
admissible region of robust optimization problem is equal to the ``operating window".
In what follows, the contour map of function $\expval{\randproc{ROP}(V_0,\, \Omega,\SSpt)}$
can be see in Figure~\ref{rop_robust_func_fig}. Note that the maximum, which is
indicated on the graph with a blue cross, occurs at the point
$(V_0,\Omega) = (1/90~m/s,7\pi/3~rad/s)$. This point is located in the boundary
of the admissible region, in the upper right corner, and corresponds to a expected value
of the mean rate of penetration, during the time interval analyzed, approximately equal to
$58$ ``meters per hour".

This result says that, in the ``operating window" considered here,
increasing drillstring rotational and translational velocities provides 
the most robust strategy to maximize its ROP into the soil. 
This is in some ways an intuitive result, but is at odds with the result
of the deterministic optimization problem, which provides another strategy
to achieve optimum operating condition.



\section{Concluding remarks}
\label{concl_remarks}

A model was developed in this work to describe the nonlinear dynamics
of horizontal drillstrings. The model uses a beam theory, with effects of
rotatory inertia and shear deformation, which is capable of reproducing
large displacements that the beam undergoes. This model also considers
the friction and shock effects due to transversal impacts, as well as,
the force and torque induced by bit-rock interaction.

It should be emphasized that the paper not only proposes a mechanical analysis 
of this complex dynamic system, but, through this analysis, also provides a formulation 
for this type of system, and a methodology, that can be reused to other rotating slender 
mechanical systems subjected to friction and shock effects.

Numerical simulations showed that the mechanical system of interest
has a very rich nonlinear dynamics, which reproduces complex phenomena
such as bit-bounce, stick-slip, and transverse impacts.
The study also indicated that large velocity fluctuations
observed in the phenomena of bit-bounce and stick-slip are
correlated with the transverse impacts, i.e., with the number of shocks
per unit time which the system is subjected. Also, the mechanical impacts
cause the beam to assume complex spatial configurations, which are formed
by flexural modes associated to high natural frequencies.

A study aiming to maximize drilling process efficiency, varying drillstring
velocities of translation and rotation was presented. The optimization strategy
used a trial approach to seek for a local maximum, which was located within
``operating window" and corresponds to an efficiency of approximately 16\%.

The probabilistic analysis of the nonlinear dynamics showed that,
with respect to the velocities, nominal model and mean value
of the stochastic model differ significantly. Furthermore, at the
instants which the system was subjected to mechanical impacts,
it was possible to see a more pronounced dispersion around the mean value.
Regarding the velocities probability distributions, it was noticed a
unimodal behavior essentially.

Two optimizations problems, one deterministic and one robust,
where the objective was to maximize drillstring rate of penetration into
the soil, respecting its structural limits, were formulated and solved.
The solutions of these problems provided two different strategies to
optimize the ROP.


Finally, it sounds stressing that the mathematical model 
used in this work has not gone through any process of experimental validation. 
This is because experimental data for this type of system is difficult to be obtained, 
and to construct an experimental apparatus in real scale is virtually impossible.
An interesting proposal for future work would be the construction
of an experimental test rig, in reduced scale, that emulates the main aspects
of a real drillstring. The model used in this study could be validated,
following, for instance, the methodology presented in \cite{batou2009p559},
with the aid of experimental measurements taken from this reduced apparatus.


\appendix
\section{Geometric nonlinearly force coefficients}
\label{nonlinear_gammas}

This appendix presents the coefficients which appears in
geometric nonlinearity force of Eq.(\ref{NLforce_op_se}).
For the sake of saving space, in the following lines it is used
the abbreviations: $S_{\theta_{x}} = \sin{\theta_{x}}$, and
$C_{\theta_{x}} = \cos{\theta_{x}}$.

\begin{eqnarray}
		\label{NLforce_gamma1}
		\Gamma_{1} & = &
		   E \, I_{4} \left(1 + u' \right) \left( v' \, \theta_{y}' + w' \, \theta_{z}' \right) S_{\theta_{x}} \, \theta_{x}' ~ + ~ \\ \nonumber
		& &
		   E \, I_{4} \left(1 + u' \right) \left( v' \, \theta_{z}' - w' \, \theta_{y}' \right) C_{\theta_{x}} \, \theta_{x}' ~+~ \\ \nonumber
		& &
           k_{s} \, G \, A \left(1 + u' \right) \left( \theta_{z} \, v' - \theta_{y} \, w' \right) S_{\theta_{x}} ~-~ \\ \nonumber
		& &
           k_{s} \, G \, A \left(1 + u' \right) \left( \theta_{y} \, v' + \theta_{z} \, w' \right) C_{\theta_{x}},
\end{eqnarray}

\begin{eqnarray}
		\label{NLforce_gamma2}
		\Gamma_{2} & = &
           k_{s} \, G \, I_{4} \left( \theta_{y} \left( \theta_{y}'^{\,2} + \theta_{z}'^{\,2} \right) -
           									   \theta_{x}' \, \theta_{z}'
                    	 				\right) ~ + ~ \\ \nonumber
		& &
           k_{s} \, G \, A \left( - w' + u' \, \theta_{y} \left( 2 + u' \right) \right) ~-~ \\ \nonumber
		& &
           k_{s} \, G \, A \left(1 + u' \right) \left( v' \, S_{\theta_{x}} - w' \, C_{\theta_{x}} \right),
\end{eqnarray}

\begin{eqnarray}
		\label{NLforce_gamma3}
		\Gamma_{3} & = &
           k_{s} \, G \, I_{4} \left( \theta_{z} \left( \theta_{y}'^{\,2} + \theta_{z}'^{\,2} \right) +
           									   \theta_{x}' \, \theta_{y}'
                    	 				\right) ~ + ~\\ \nonumber
		& &                    	 				
           k_{s} \, G \, A \left( v' + u' \, \theta_{z} \left( 2 + u' \right) \right)  ~-~ \\ \nonumber
		& &                    	 				
           k_{s} \, G \, A \left(1 + u' \right) \left( w' \, S_{\theta_{x}} + v' \, C_{\theta_{x}} \right),
\end{eqnarray}

\begin{eqnarray}
		\label{NLforce_gamma4}
		\Gamma_{4} & = &
		E \, A \left( \frac{1}{2} \left( 1 +  u' \right) \left( v'^{\,2} +  w'^{\,2} \right) +
						   \frac{1}{2} u'^{\,2} \left(3 + u' \right)
				  \right) ~ + ~ \\ \nonumber
		& &
           E \, I_{4} \left(
	           						S_{\theta_{x}} \left( v' \, \theta_{z}' - w' \, \theta_{y}' \right) -
		   							C_{\theta_{x}} \left( v' \, \theta_{y}' + w' \, \theta_{z}' \right)
							\right) \theta_{x}' ~ + ~\\ \nonumber
		& &
           E \, I_{4} \left( 1 + u' \right) \left( \theta_{x}'^{\,2} +
           														\frac{3}{2}\left(\theta_{y}'^{\,2} + \theta_{z}'^{\,2} \right)
           												\right) ~ + ~\\ \nonumber
		& &
			k_{s} \, G \, A \left(
											C_{\theta_{x}} \left( \theta_{y} \, w' - \theta_{z} \, v' \right) -
											S_{\theta_{x}} \left( \theta_{y} \, v' + \theta_{z} \, w'  \right)
                     			   \right) ~ + ~\\ \nonumber
		& &
        	k_{s} \, G \, A \left(1 + u' \right) \left( \theta_{y}^{2} + \theta_{z}^{2} \right),
\end{eqnarray}

\begin{eqnarray}
		\label{NLforce_gamma5}
		\Gamma_{5} & = &
		E \, A \left( u' + \frac{1}{2} \left( u'^{\,2} + v'^{\,2} +  w'^{\,2} \right)
				  \right) v' ~ + ~\\ \nonumber
		& &
           E \, I_{4} \left(  2 \, \theta_{x}'^{\,2} +
           						  \frac{1}{2} \left( \theta_{y}'^{\,2} + \theta_{z}'^{\,2} \right)
                    	 \right) v' ~ + ~\\ \nonumber
		& &
           E \, I_{4} \left(1 + u' \right) \left( \theta_{z}' S_{\theta_{x}} - \theta_{y}' C_{\theta_{x}} \right) \theta_{x}'  ~ + ~\\ \nonumber
		& &
           k_{s} \, G \, A \left(1 + u' \right) \left( \theta_{z} - \theta_{y} \, S_{\theta_{x}} - \theta_{z} \, C_{\theta_{x}} \right),
\end{eqnarray}

\begin{eqnarray}
		\label{NLforce_gamma6}
		\Gamma_{6} & = &
		E \, A \left( u' + \frac{1}{2} \left( u'^{\,2} + v'^{\,2} +  w'^{\,2} \right)
			      \right) w' ~ + ~\\ \nonumber
		& &
           E \, I_{4} \left( 2 \, \theta_{x}'^{\,2} +
           						\frac{1}{2} \left( \theta_{y}'^{\,2} + \theta_{z}'^{\,2} \right)
                    	  \right) w' ~ + ~\\ \nonumber
		& &
           E \, I_{4} \left(1 + u' \right)  \left( - \theta_{y}' \, S_{\theta_{x}} - \theta_{z}' \, C_{\theta_{x}} \right) \theta_{x}' ~ + ~\\ \nonumber
		& &
           k_{s} \, G \, A \left(1 + u' \right) \left(  - \theta_{y} + \theta_{y} \, C_{\theta_{x}} - \theta_{z} \, S_{\theta_{x}} \right),
\end{eqnarray}

\begin{eqnarray}
		\label{NLforce_gamma7}
		\Gamma_{7} & = &
			E \, I_{4} \left( u'^{\,2} + 2 \left( u' + v'^{\,2} + w'^{\,2} \right) \right) \theta_{x}'  ~ + ~\\ \nonumber
		& &
			E \, I_{4} \left(1 + u' \right) \left( v' \, \theta_{z}' - w' \, \theta_{y}' \right) S_{\theta_{x}} ~ - ~ \\ \nonumber
		& &
			E \, I_{4} \left(1 + u' \right) \left( v' \, \theta_{y}' + w' \, \theta_{z}' \right) C_{\theta_{x}} ~ + ~ \\ \nonumber
		& &
           E \, I_{6} \left( 4 \, \theta_{x}'^{\,2} +
           						   2 \left( \theta_{y}'^{\,2} + \theta_{z}'^{\,2} \right)
                         \right) \theta_{x}' ~ + ~\\ \nonumber
		& &		
           k_{s} \, G \, A \left( \theta_{z} \, \theta_{y}' - \theta_{y} \, \theta_{z}' \right),
\end{eqnarray}

\begin{eqnarray}
		\label{NLforce_gamma8}
		\Gamma_{8} & = &
            		 E \, I_{4} \left( 3 \, u' +
            		 						\frac{1}{2} \left( 3 \, u'^{\,2} + v'^{\,2} + w'^{\,2} \right)
			    	 \right) \theta_{y}' ~ + ~\\ \nonumber
           & &
            		 E \, I_{4} \left(1 + u' \right) \left( - w' \, S_{\theta_{x}} - v' \, C_{\theta_{x}} \right) \theta_{x}' ~ + ~\\ \nonumber
           & &
           E \, I_{6} \left( 2 \, \theta_{x}'^{\,2} +
           							\frac{3}{2} \left( \theta_{y}'^{\,2} + \theta_{z}'^{\,2} \right)
                         \right) \theta_{y}' ~ + ~ \\ \nonumber
		& &
		k_{s} \, G \, I_{4} \left( \theta_{z} \, \theta_{x}' + \theta_{y}' \left( \theta_{y}^2 + \theta_{z}^2 \right) \right),
\end{eqnarray}

\noindent
and

\begin{eqnarray}
		\label{NLforce_gamma9}
		\Gamma_{9} & = &
         		 E \, I_{4} \left( 3 \, u' +
         		 						\frac{1}{2} \left( 3 \, u'^{\,2} + v'^{\,2} + w'^{\,2} \right)
			    	 \right) \theta_{z}'  ~ + ~\\ \nonumber
           & &
         		 E \, I_{4} \left(1 + u' \right) \left( v' \, S_{\theta_{x}} - w' \, C_{\theta_{x}} \right) \theta_{x}' ~ + ~\\ \nonumber
        & &
           E \, I_{6} \left( 2 \theta_{x}'^{\,2} +
           						 \frac{3}{2} \left( \theta_{y}'^{\,2} + \theta_{z}'^{\,2} \right)
                         \right) \theta_{z}' ~ + ~ \\ \nonumber
		& &
		k_{s} \, G \, I_{4} \left( - \theta_{y} \, \theta_{x}' + \theta_{z}' \left( \theta_{y}^2 + \theta_{z}^2 \right) \right).
\end{eqnarray}


\begin{acknowledgements}
The authors are indebted to Brazilian agencies CNPq, CAPES, and FAPERJ, and
French agency COFECUB for the financial support given to this \mbox{research}.
The first author is grateful for the institutional support received from 
PUC-Rio and Universit\'{e} Paris-Est to carry out this work.
\end{acknowledgements}

\bibliographystyle{spbasic}      
\bibliography{bibliography}

\end{document}